%% file: moarref-sharma-tropp-mckeon-JFM2013.tex
\documentclass[10pt]{jfm}
\usepackage{graphicx}
\usepackage{float}
\usepackage{rotating}
\usepackage{epstopdf}
\AppendGraphicsExtensions{.tif}

\usepackage{multirow}

\input{commands}

\usepackage{amsfonts,amssymb,latexsym,color,amsmath,pifont,epsfig,mathdots,mathtools,natbib}
\usepackage{setspace}
\usepackage{subfigure}
\usepackage{url}

\newcommand{\cS}{{\cal S}}

\newcommand{\cO}{{\cal O}}

\newcommand{\non}{\nonumber}
\newcommand{\ds}{\displaystyle}

\newcommand{\hf}{\hat{f}}
\newcommand{\mrd}{\mathrm{d}}
\newcommand{\mre}{\mathrm{e}}
\newcommand{\mri}{\mathrm{i}}

\newcommand{\fvec}{{\bf f}}

\newcommand{\bu}{{\bf u}}
\newcommand{\bU}{{\bf U}}

\newcommand{\bzeta}{\mbox{\boldmath$\zeta$}}
\newcommand{\bphi}{\mbox{\boldmath$\phi$}}
\newcommand{\bpsi}{\mbox{\boldmath$\psi$}}

\newcommand{\p}{\partial}

\providecommand{\abs}[1]{\lvert#1\rvert}
\providecommand{\norm}[1]{\lVert#1\rVert}
\newcommand{\tc}{\textcolor}

\newcommand{\bjm}[1]{{{\color{black}#1}}}
\newcommand{\moa}[1]{{{\color{black}#1}}}
\newcommand{\moar}[1]{{{\color{black}#1}}}

\title{\moa{Model-based scaling of the streamwise energy density in high-Reynolds number turbulent channels}}

\shorttitle{Model-based scaling of the streamwise energy density}

\author{Rashad Moarref$^{1}$, Ati S.\ Sharma$^{2}$, \\[0.15cm] Joel A.\ Tropp$^{3}$ \and Beverley J.\ McKeon$^{1}$}

\affiliation{
$^{1}$Graduate Aerospace Laboratories, California Institute of Technology, CA 91125, USA
\\[0.1cm]
$^{2}$Engineering and the Environment, University of Southampton, SO17 1BJ, UK
\\[0.1cm]
$^{3}$Computing \& Mathematical Sciences, California Institute of Technology, CA 91125, USA
}

\shortauthor{R.\ Moarref, A.\ S.\ Sharma, J.\ A.\ Tropp, \& B.\ J.\ McKeon}

\begin{document}

\maketitle

    \begin{abstract}

We study the Reynolds number scaling \moa{and the geometric self-similarity} of a gain-based, low-rank approximation to turbulent channel flows, determined by the resolvent formulation of \cite{mcksha10}, in order to obtain a description of the streamwise turbulence intensity from direct consideration of the Navier-Stokes equations. Under this formulation, the velocity field is decomposed into propagating waves (with single streamwise and spanwise wavelengths and wave speed) whose wall-normal shapes are determined from the principal singular function of the corresponding resolvent operator. \moa{Using the accepted scalings of the mean velocity in wall-bounded turbulent flows, we establish that the resolvent operator} admits three classes of wave parameters that induce universal behavior with Reynolds number on the low-rank model, and which are consistent with scalings proposed throughout the wall turbulence literature. \moa{In addition, it was shown that a necessary condition for geometrically self-similar resolvent modes is the presence of a logarithmic turbulent mean velocity. Under the practical assumption that the mean velocity consists of a logarithmic region, we identify the scalings that constitute hierarchies of self-similar modes that are parameterized by the critical wall-normal location where the speed of the mode equals the local turbulent mean velocity.} For the rank-1 model subject to broadband forcing, the integrated streamwise energy density takes a universal form which is consistent with the dominant near-wall turbulent motions.  When the shape of the forcing is optimized to enforce matching with results from direct numerical simulations at low turbulent Reynolds numbers, further similarity appears. Representation of these weight functions using similarity laws enables prediction of the Reynolds number and wall-normal variations of the streamwise energy intensity at high Reynolds numbers (${Re}_\tau \approx 10^3 - 10^{10}$). Results from this low-rank model of the Navier-Stokes equations compare favorably with experimental results in the literature.

    \end{abstract}

\section{Introduction}
\label{sec.introduction}

Understanding the behavior of wall-bounded turbulent flows at high Reynolds numbers has tremendous technological implications, for example, in air and water transportation. This problem has received significant attention over the last two decades especially in the light of full-field flow information revealed by direct numerical simulations (DNS) at relatively small Reynolds numbers and high-Reynolds number experiments. Notwithstanding the recent developments, the highest Reynolds numbers that are considered in DNS are an order of magnitude smaller than experiments, which are in turn conducted at Reynolds numbers that are typically two orders of magnitude smaller than most applications. This creates a critical demand for model-based approaches that describe and predict the behavior of turbulent flows at technologically relevant Reynolds numbers.

Wall turbulence has been the topic of several reviews; see, for example,~\cite{rob91,adr07} for structure of coherent motions,~\cite{gadban94} for turbulence statistics and scaling issues,~\cite{pan01} for self-sustaining turbulence mechanisms, and~\cite{kle10,marmckmonnagsmisre10,smimckmar11} for the latest findings and main challenges in examining high-Reynolds number wall turbulence. In the present study, special attention is paid to scaling, universality, \moa{and geometric self-similarity} of the turbulent energy spectra at high Reynolds numbers. We also note that the energy spectra exhibit clear signatures of coherent turbulent motions such as the near-wall streaks, the large-scale motions (LSMs), and the very large-scale motions (VLSMs).

\subsection{Overview of dominant coherent motions}
\label{sec.structures}

In the interests of giving a brief overview of the energetically dominant coherent motions in wall turbulence, we will review three classes of structure. The near-wall system of quasi-streamwise streaks and counter-rotating vortices with streamwise length and spanwise spacing of approximately $1000$ and $100$ inner (viscous) units, centered at approximately $15$ inner units above the wall, has been well-studied. These ubiquitous features of wall turbulence are responsible for large production of turbulent kinetic energy~\citep{klireyschrun67,smimet83}.

Another commonly-observed feature of turbulent flows is the hairpin vortex. In low Reynolds number flows, at least, packets of hairpin vortices have been observed to extend from the wall to the edge of the boundary layer and constitute LSMs~\citep{heaban81,adrmeitom00,adr07}, with streamwise extent approximately $2$-$3$ outer units (channel half-height, pipe radius, or the boundary layer thickness).

VLSMs have been observed to reside in the logarithmic region of the turbulent mean velocity, with lengths of approximately $10$-$15$ outer units in boundary layers and up to $30$ outer units in channels and pipes~\cite[see, for example,][]{kimadr99,baladr07,monstewilcho07}. The emergence of VLSMs was originally attributed to alignment of LSMs~\citep{kimadr99}. However,~\cite{smimckmar11} concluded that this is unlikely since the detached LSMs are located at a farther distance from the wall than the VLSMs and the attached LSMs have much smaller width than VLSMs and are convected at different speeds. Recently, the correlation between the envelope of small scale activity and the large scale velocity signal (identified via filtering in spectral space), which has been interpreted as an amplitude modulation of the small scales, has been investigated in detail, see e.g. ~\cite{hutmar07-PTRS,mathutmar09,matmonhutmar09,chumck09,hutmonganngmar11}.

\subsection{Overview of scaling issues}
\label{sec.scaling}

In spite of recent advances in understanding the structure of wall turbulence, the Reynolds number scaling of the turbulent energy spectra and the energy intensities remains an open area of research. The main experimental obstacle is maintaining the necessary spatial resolution for measurement accuracy while achieving the high Reynolds numbers required for large separation between the small and large turbulent scales. For example, the available experiments are performed at relatively small friction Reynolds numbers, ${Re}_\tau \approx \cO (10^4)$, with a notable exception of the atmospheric surface layer measurements of, e.g., \cite{metkle01} (${Re}_\tau \approx \cO (10^6)$) that are in turn generally contaminated by surface roughness effects. Most high-Reynolds number experiments suffer from spatial resolution issues in the inner region~\cite[see, for example,][]{hutnicmarcho09}.

Significant experimental effort has been devoted to determining the behavior of the streamwise energy intensity at high Reynolds numbers since it dominates the turbulent kinetic energy and is easier to measure relative to the wall-normal and spanwise velocities. It is understood that both small and large scales contribute to the streamwise energy intensity~\citep{metkle01,markun03,hutmar07-JFM,marmathut10}. It is well-known that a region of the streamwise wavenumber spectrum scales with inner units; \cite{marmathut10} showed by filtering that the contribution of such scales to the streamwise energy intensity, and therefore by extension also the streamwise spectrum, is universal, i.e. independent of Reynolds number. On the other hand, the large motions have been shown to scale in outer units \citep{kimadr99}; \cite{mathutmar09} proposed that the corresponding peak in streamwise intensity occurs close to the geometric mean of the limits of the logarithmic region in the turbulent mean velocity. The amplitude of this energetic peak increases with Reynolds number and has a footprint down to the wall~\citep{hutmar07-PTRS}. \moa{Using data from experiments of canonical wall-bounded turbulent flows,~\cite{alforlseg12} proposed a composite profile for the streamwise turbulence intensity and showed the possibility for emergence of an outer peak at high Reynolds numbers.} Note however, that available data are not sufficiently well-resolved to determine unequivocally the Reynolds number scaling of either the inner or outer peaks of the streamwise energy intensity~\cite[see, for example,][]{marmathut10}.

Theoretical approaches also offer insight into the scaling of the spectrum with increasing Reynolds numbers, originating with the attached eddy concepts described by \cite{tow76}. \moa{These eddies are attached in the sense that their height scales with their distance from the wall, and they are geometrically self-similar since their wall-parallel length scales are proportional to their height.~\cite{percho82} developed these ideas to include hierarchies of geometrically self-similar attached eddies in the logarithmic region of the turbulent mean velocity. They systematically predicted that if the population density of the attached eddies inversely decreases with their height, both the turbulent mean velocity and the wall-parallel energy intensities exhibit logarithmic dependence with the distance from the wall. The logarithmic behavior of the mean velocity and the streamwise energy intensity was recently confirmed using high-Reynolds number experiments~\citep{marmonhulsmi13}. However, the attached eddy hypothesis does not predict the exact shape of the eddies or their evolution in time.}

Subsequent works by Perry and co-authors extended the attached eddy formulation beyond the logarithmic region;~\cite{markun03} used empirical scaling arguments concerning the effective forcing of the outer turbulence on the viscous region to propose a similarity expression for the streamwise energy intensity that is valid throughout the zero pressure boundary layer. Recently, \cite{marmathut10-science} outlined an observationally-based, predictive formulation for the variation of the streamwise turbulent intensity up to the geometric mean of the logarithmic region based on consideration of the correlation between large and small scales.~\moa{Most recently,~\cite{mizjim13} used DNS to show that self-similarity of the velocity fluctuations is sufficient and seemly important for reproducing a logarithmic profile in the mean velocity. They also observed that the logarithmic region can be maintained independent of the near-wall dynamics.}

\subsection{Review of previous model-based approaches}
\label{sec.previous}

We seek in this work a description of the streamwise turbulence intensity for all wall-normal locations arising from direct consideration, and modeling, of the Navier-Stokes equations (NSE). There has been much work in this vein, highlighting several important features of the NSE. We provide a brief review of the most relevant literature here.

The critical role of linear amplification mechanism in promoting and maintaining turbulent flows was highlighted in direct numerical simulations of~\cite{kimlim00}. \moa{In addition, it was shown that nonlinearity plays an important role in regenerating the near-wall region of turbulent shear flows through a self-sustaining process~\citep{hamkimwal95,wal97,schhus02}. More recently, significant effort has been directed at identification and analysis of \emph{exact} solutions of the NSE, such as traveling waves and periodic orbits, see e.g.~\cite{wal03,wedker04} and the review paper by~\cite{ker05}.} 

It is understood that high sensitivity of the laminar flow to disturbances provides alternative paths to transition that bypass linear instability; see, for example,~\cite{schhen01}.~\cite{tretrereddri93} showed that the high flow sensitivity is related to non-normality of the coupled Orr-Sommerfeld and Squire operators; see also~\cite{sch07}. These operators are coupled in the presence of mean shear and spanwise-varying fluctuations. Physically, as originally explained by~\cite{lan75}, a large streamwise disturbance is induced on the flow in response to lift-up of a fluid particle by the wall-normal velocity such that its wall-parallel momentum is conserved.

Even in linearly stable flows, the high sensitivity can result in large transient responses, meaning that the energy of certain initial perturbations significantly grows before eventual decay to zero~\citep{gus91,kli92,butfar92,schhen94,redhen93}. In addition, the high sensitivity is responsible for high energy amplification, meaning that the velocity fluctuations achieve a large variance at the steady state for the flow subject to zero-mean stochastic disturbances~\citep{farioa93,bamdah01,jovbamJFM05}. The dominant structures that emerge from the above transient growth and energy amplification analyses are reminiscent of the streamwise streaks observed at the early stages of transition to turbulence~\citep{matalf01}. They are characterized by infinitely long spanwise-periodic regions of high and low streamwise velocity associated with pairs of counter-rotating streamwise vortices that are separated by approximately $3.5$ outer units.

It is believed that the NSE linearized around the turbulent mean velocity are stable for all Reynolds numbers~\citep{mal56,reytie67}. Early model-based approaches extended the aforementioned sensitivity analyses of the laminar flow to the turbulent channel flow and found dominance of streamwise streaks that are spaced by $3$ outer units, which is approximately the same as in the laminar flow. In addition to the outer-scaled dominant structures,~\cite{butfar93,farioa93a} showed that the largest transient response over an eddy turnover time of $80$ inner units, associated with the near-wall cycle, is obtained for initial perturbations that are infinitely long and have the same spanwise spacing as the near-wall streaks, i.e. $100$ inner units. The same streamwise and spanwise lengths were obtained in flows subject to stochastic disturbances over a coherence time of $90$ inner units~\citep{farioa98}.

\cite{reyhus72-3} put forward a modified linear model to account for the effect of background Reynolds stresses on the velocity fluctuations. They proposed to augment the molecular viscosity by the turbulent eddy viscosity that is required to maintain the mean velocity. This model yields two local optima for the structures with largest transient growth~\citep{alajim06,pujgarcosdep09} and energy amplification~\citep{hwacos10} without the need for confining the optimization time. These peaks correspond to streamwise-elongated structures with a spacing of $80$ inner units and $3$-$4$ outer units and are in fair agreement with the spacing of near-wall streaks and the very large-scale motions in real turbulent flows. \moa{The geometric similarity of the optimal transient response to initial perturbations and the optimal responses to harmonic and stochastic forcings was highlighted by~\cite{hwacos10} using the linearized NSE with turbulent eddy viscosity. These authors found that the streamwise constant optimal responses scale with the spanwise wavelength in the wall-normal direction for spanwise wavelengths between the inner- and outer-scaled regions.}

An exact representation of the NSE was introduced by~\cite{mcksha10} in which (i) a set of linear sub-systems describe extraction of energy from the mean velocity at individual wavenumbers/frequencies; and (ii) the only source of coupling between these sub-systems is the conservative nonlinear interaction of their outputs, that determines both the input to the sub-systems and the turbulent mean velocity. At its heart is the ability to analyze the flow of energy from the mean velocity to all the velocity scales and identify the essential linear amplification and nonlinear redistribution mechanisms that drive the turbulent flow. The input-output relationship of the linear sub-systems can be described by transfer functions whose low-rank nature in the wall-normal direction enables significant simplification of their analysis.

One of the main differences between the formulation of~\cite{mcksha10} and other input-output analyses of laminar and turbulent flows~\cite[see, for example,][]{jovbamJFM05,hwacos10} is parameterization of the waves with wave speed rather than temporal frequency. The latter approaches showed that the globally optimal transient growth and energy amplification takes place for zero streamwise wavenumber and temporal frequency. Selecting the wave speed, as emphasized by~\cite{mcksha10}, (i) enables systematical search for both locally (in wall-normal direction) and globally optimal wave shapes and parameters; (ii) removes the ambiguity about the wave speed corresponding to the globally optimal waves by determining the limit of the ratio between zero streamwise wavenumber and temporal frequency; and (iii) distinguishes between non-normality and critical behavior as the main linear amplification mechanisms.

\cite{mcksha10} showed that the principal forcing and response directions associated with the linear sub-systems are consistent with the dominant response shapes in real turbulent pipe flows. In addition, the low-dimensional and sparse feature of the resulting model enables development and utilization of compressive sampling techniques for analyzing the turbulent flow dynamics~\citep{boushatromck13}. This formulation has also proven useful for pre- and post-diction of experimental observations in turbulent pipe flow~\citep{shamck13,mckshajac13}.

\subsection{Paper outline}
\label{sec.contributions}

In this paper, we identify the Reynolds number scaling of a low-rank approximation to turbulent channel flow and utilize it for predicting the streamwise energy intensity at high Reynolds numbers. Our development is outlined as follows: In~\S~\ref{sec.low-rank}, we briefly review the resolvent formulation, highlight its low-rank nature, and show that a rank-1 approximation captures the characteristics of the most energetic modes of real turbulent channels. The stage is set for studying the energy density of fluctuations using a minimum number of assumptions by considering a rank-1 model in the wall-normal direction subject to broadband forcing in the wall-parallel directions and time. Furthermore, a summary of the computational approach for determining the rank-1 model is provided.

Three classes of wave parameters for which the low-rank approximation of the resolvent exhibits universal behavior (independence) with Reynolds number are identified in~\S~\ref{sec.Re-scaling-theory}. The requirement for universality highlights the role of wave speed in distinguishing these classes. Each class of waves is characterized by a unique range of wave speeds and a unique spatial scaling that emerge from the resolvent. For the rank-1 model subject to broadband forcing, we reveal the universal streamwise energy densities, and show that the peaks of these energy densities roughly agree with the most energetic turbulent motions, i.e. the near-wall streaks, the VLSMs, and the LSMs.

In~\S~\ref{sec.energy-density-weighted}, we show that the streamwise energy density of the rank-1 model with broadband forcing can be optimally weighted as a function of wave speed to match the intensity of simulations at low turbulent Reynolds numbers. The weight functions are then formulated using similarity laws which, in conjunction with the universal energy densities, enable prediction of the streamwise energy intensity at high Reynolds numbers. The paper is concluded in~\S~\ref{sec.conclusion} and limitations and several future directions are discussed.

\section{Low-rank approximation to channel flow}
\label{sec.low-rank}

An overview of the rationale for considering a low-rank approximation to turbulent channel flow is presented in this section. We follow the development of~\cite{mcksha10} for turbulent pipe flow, showing that equivalent results are obtained for channels and highlighting the new observations. 

\begin{figure}
        \begin{center}
        \begin{tabular}{cc}
        \subfigure[]{\includegraphics[height=2.6cm]
                {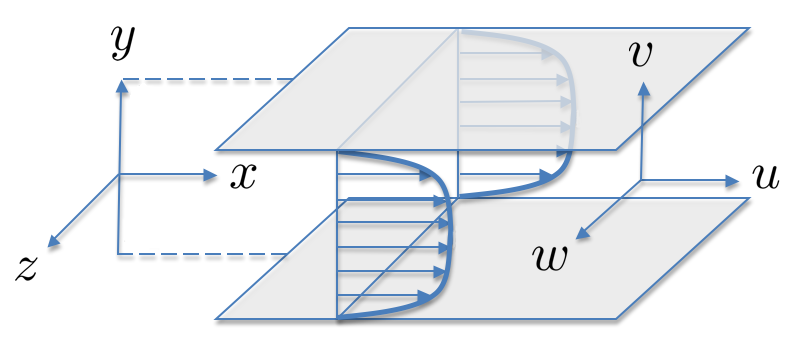}
        \label{fig.channel}}
        &
        \subfigure[]{\includegraphics[height=3cm]
                {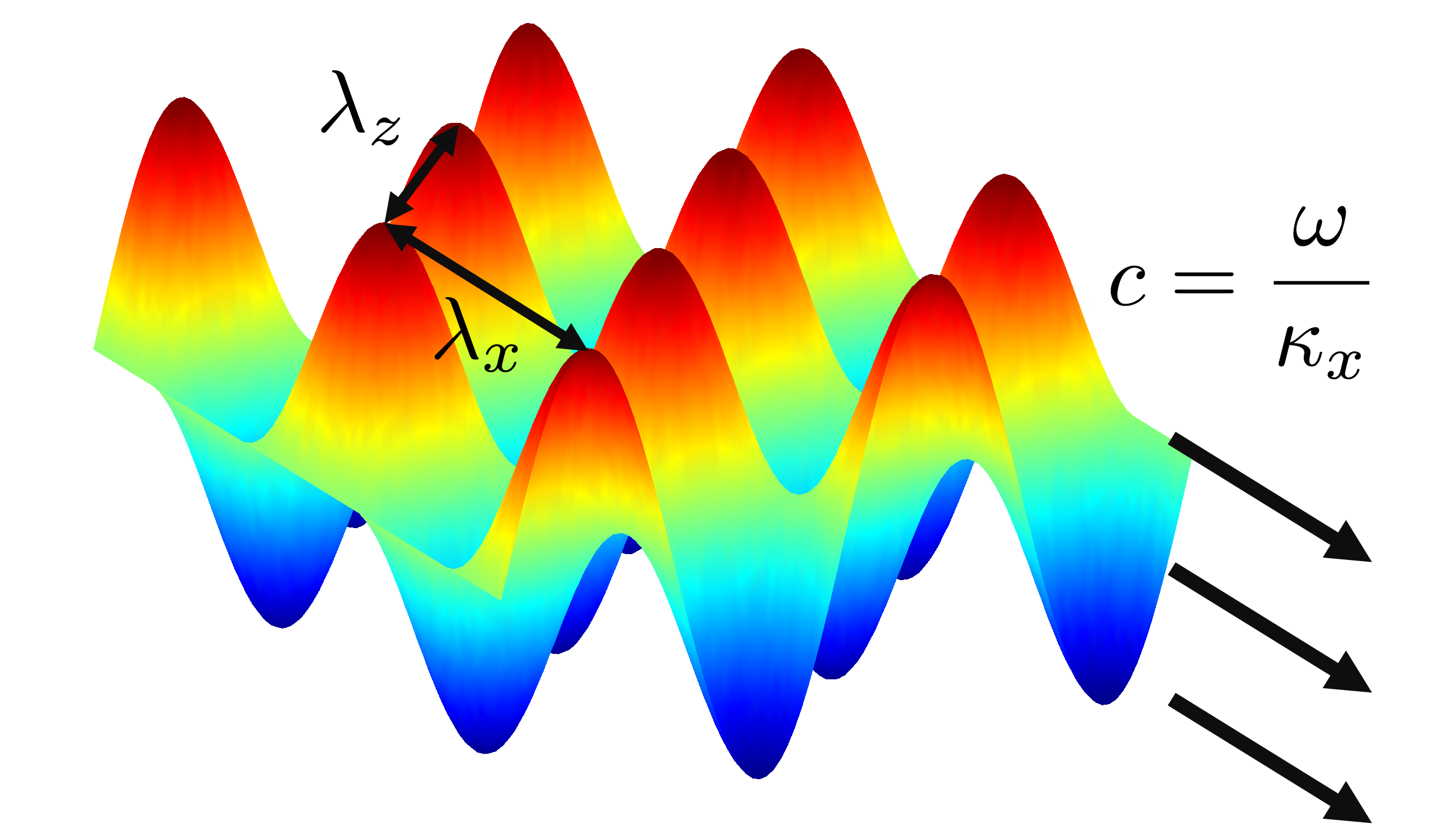}
        \label{fig.prop-waves}}
        \end{tabular}
        \end{center}
        \caption{(a) Pressure driven channel flow. (b) Schematic of a two-dimensional propagating wave with streamwise and spanwise wavelengths $\lambda_x$ and $\lambda_z$ and streamwise speed ${c}$.}
      \end{figure}

The pressure-driven flow of an incompressible Newtonian fluid is governed by the nondimensional NSE and the continuity constraint
	\be
	\ba{l}
	\bu_t
	\, + \,
	(\bu \cdot \nabla) \bu
	\, + \,
	\nabla P
	\; = \;
	(1/{Re}_\tau) \Delta \bu,
	\\[0.15cm]
	\nabla \cdot \bu
	\; = \;
	0,
	\ea
	\label{eq.NS}
	\ee
where $\bu (x,y,z,t)$ is the velocity vector, $P (x,y,z,t)$ is the pressure, $\nabla$ is the gradient operator, and $\Delta = \nabla \cdot \nabla$ is the Laplacian. The streamwise and spanwise directions, $x$ and $z$, are infinitely long, the wall-normal direction is finite, $0 \leq y \leq 2$, and $t$ denotes time; see figure~\ref{fig.channel} for the geometry. The subscript $t$ represents temporal derivative, e.g. $\bu_t = \p \bu/ \p t$. The Reynolds number ${Re}_\tau = u_\tau h/\nu$ is defined based on the channel half-height $h$, kinematic viscosity $\nu$, and friction velocity $u_\tau = \sqrt{\tau_w/\rho}$, where $\tau_w$ is the shear stress at the wall, and $\rho$ is the density. Velocity is normalized by $u_\tau$, spatial variables by $h$, time by $h/u_\tau$, and pressure by $\rho u_\tau^2$. The spatial variables are denoted by $^+$ when normalized by the viscous length scale $\nu/u_\tau$, e.g. $y^+ = {Re}_\tau y$.

\subsection{Decomposition in homogenous directions}
\label{sec.prop-waves}

The velocity is decomposed using the Fourier transform in the homogenous directions and time
	\be
	\bu (x,y,z,t)
	\; = \;
	\ds{
	\iiint_{-\infty}^{\infty}
	}
	\,
	\hat{\bu} (y; \kappa_x, \kappa_z, \omega)\,
	\mre^{\mri
	(
	\kappa_x x
	\, + \,
	\kappa_z z
	\, - \,
	\omega t
	)
	}
	\mrd \kappa_x \,
	\mrd \kappa_z \,
	\mrd \omega,
	\label{eq.prop-waves}
	\ee
where $~\hat{}~$ denotes a variable in the transformed domain, and the triplet $(\kappa_x, \kappa_z, \omega)$ is the streamwise and spanwise wavenumbers and the temporal (angular) frequency. \moa{The Fourier basis is optimal in the homogeneous wall-parallel directions. It is also an appropriate basis in time under stationary conditions.} For any $(\kappa_x, \kappa_z, \omega) \neq 0$, $\hat{\bu} (y; \kappa_x, \kappa_z, \omega)$ represents a propagating wave with streamwise and spanwise wavelengths $\lambda_x = 2\pi/\kappa_x$ and $\lambda_z = 2\pi/\kappa_z$ and speed ${c} = \omega/\kappa_x$ in the streamwise direction; see figure~\ref{fig.prop-waves} for an illustration. Some special cases include standing waves ($c = 0$), infinitely long waves ($\kappa_x = 0$), and infinitely wide waves ($\kappa_z = 0$). In this study, we emphasize the eminent role of wave speed, a factor that was highlighted by~\cite{mcksha10} while being predominantly neglected in the previous studies, in determining the classes of propagating waves that are universal with Reynolds number.

The turbulent mean velocity $\bU(y) = [\,U (y)~0~0\,]^T = \hat{\bu} (y; 0,0,0)$ corresponds to $(\kappa_x, \kappa_z, \omega) = 0$ and is assumed to be known. \moa{Note that our main results, i.e. the identified scalings in~\S~\ref{sec.Re-scaling-theory}, rely on the accepted scales of the turbulent mean velocity and, otherwise, do not depend on the exact shape of $U$.}~\cite{mcksha10} avoided the closure problem for the mean velocity by using $U (y)$ obtained in pipe flow experiments, but note that the resolvent formulation could be used to determine the mean velocity profile, a topic of ongoing work \citep[see][]{mckshajac13}. Here, we use a semi-empirical turbulent viscosity model, originally proposed for pipe flow~\citep{mal56,ces58} and extended to channel flow~\citep{reytie67}, to determine $U (y)$
	\be
	\ba{rcl}
	U (y)
	&\!\! = \!\!&
	{Re}_\tau
	\, \ds{\int_{0}^{y}} \,
	\dfrac{1 - \xi}{1 + \nu_{T} (\xi)} \, \mrd \xi,
	\\[0.05cm]
	\nu_{T} (y)
	&\!\!\! = \!\!\!&
	\dfrac{1}{2}
	\,
	\Bigg\{
	1
	+
	\Bigg(
	\dfrac{\kappa {Re}_\tau}{3} \,
	\big(2y - y^2\big) \,
	\big(3 - 4y + 2 y^2\big) \,
	\bigg\{1 - \mre^{\big(\abs{y-1} - 1\big) \dfrac{{Re}_\tau}{\alpha}}\bigg\} \,
	\Bigg)^2
	\Bigg\}^{1/2}
	-
	\dfrac{1}{2},
	\ea
	\label{eq.U-nuT}
	\ee
where $\nu_T$ is normalized by $\nu$, and the parameters $\alpha$ and $\kappa$ appear in the van Driest's wall law and the von K{\'a}rm{\'a}n log law. These parameters are obtained by minimizing the deviation between $U (y)$ in~(\ref{eq.U-nuT}) and the DNS-based turbulent mean velocity profile. The $\alpha$ and $\kappa$ obtained for ${Re}_\tau = 186$, $547$, and $934$~\citep{moajovJFM12} suggest that both of these values converge for large ${Re}_\tau$. We take $\alpha = 25.4$ and $\kappa = 0.426$ for all Reynolds numbers and note that these values are optimized for ${Re}_\tau = 2003$~\citep{alajim06,pujgarcosdep09}.
		
\begin{figure}
    \begin{center}
    \includegraphics[width=0.65\columnwidth]
    {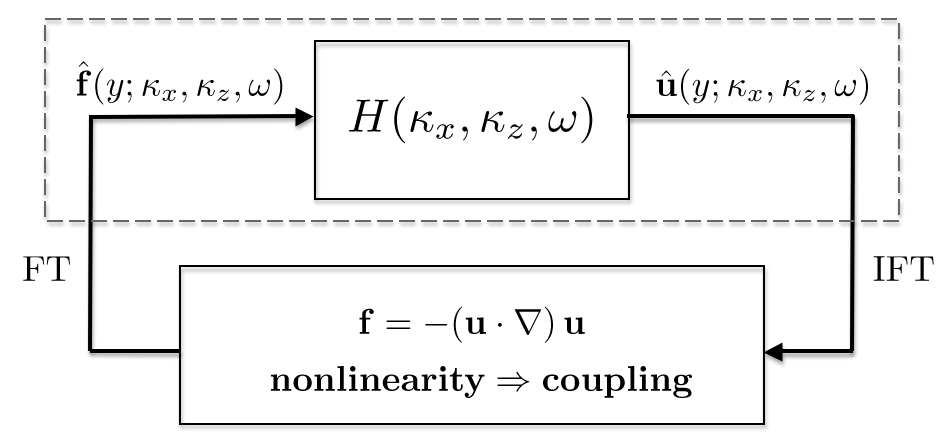}
    \end{center}
    \caption{For any triplet $(\kappa_x, \kappa_z, \omega)$, the operator $H (\kappa_x, \kappa_z, \omega)$ maps the forcing $\hat{\fvec}$ to the response $\hat{\bu}$. The different wavenumbers are coupled via the quadratic relationship between $\fvec (x,y,z,t)$ and $\bu (x,y,z,t)$. FT and IFT stand for Fourier transform and inverse Fourier transform, respectively. The input-output map (shown with the dashed rectangle) is the main focus of the present study.}
    \label{fig.block-diag}
    \end{figure}

Following~\cite{mcksha10}, the convective nonlinearity in~(\ref{eq.NS}) is considered as a forcing term ${\fvec} = -(\bu \cdot \nabla) \bu$ that drives the velocity fluctuations, see also figure~\ref{fig.block-diag}. For any $(\kappa_x, \kappa_z, \omega) \neq 0$, an equation for velocity fluctuations $\hat{\bu} (y; \kappa_x, \kappa_z, \omega) = [\, \hu~\hv~\hw \,]^T$ around the turbulent mean velocity is obtained by substituting~(\ref{eq.prop-waves}) in~(\ref{eq.NS}), and using the orthonormality of the complex exponential functions
	\be
	\ba{l}
	-\mri \omega \hat{\bu}
	\, + \,
	(\bU \cdot \nabla) \hat{\bu}
	\, + \,
	(\hat{\bu} \cdot \nabla) \bU
	\, + \,
	\nabla \hat{p}
	\, - \,
	(1/{Re}_\tau) \Delta \hat{\bu}
	\; = \;
	\hat{\fvec},
	\\[0.15cm]
	\nabla \cdot \hat{\bu}
	\; = \;
	0.
	\ea
	\label{eq.NS-prop-waves}
	\ee
Here, $\nabla = [\,\mri \kappa_x~\p_y~\mri \kappa_z\,]^T$, $\Delta = \p_{yy} - \kappa^2$ with $\kappa^2 = \kappa_x^2 + \kappa_z^2$, and

	\be
	\ba{l}
	\hat{\fvec} (y; \kappa_x, \kappa_z, \omega)
	\, = \,
	[\,\hf_1~\hf_2~\hf_3\,]^T
	\; = \;
	\\[0.2cm]
	-
	\hspace{-0.4cm}
	\ds{
	\iiint\limits_{
	{\small
	\ba{c}
	(\kappa_x', \kappa_z', \omega') \neq (0, 0, 0)
	\\
	(\kappa_x', \kappa_z', \omega') \neq (\kappa_x, \kappa_z, \omega)
	\ea
	}
	}
	}
	\hspace{-0.6cm}
	\left(
	\hat{\bu} (y; \kappa_x', \kappa_z', \omega')
	\cdot
	\nabla
	\right)
	\hat{\bu} (y; \kappa_x - \kappa_x', \kappa_z - \kappa_z', \omega - \omega')\,
	\mrd \kappa_x' \,
	\mrd \kappa_z' \,
	\mrd \omega'.
	\ea
	\label{eq.f}
	\ee

\cite{mcksha10} implicitly accounted for the continuity constraint by projecting the velocity field onto the divergence-free basis of~\cite{mestre03}. Here, we use a standard choice of wall-normal velocity $\hv$ and wall-normal vorticity $\hat{\eta} = \mri \kappa_z \hat{u} - \mri \kappa_x \hat{w}$ as the state variables, $\hat{\bzeta} (y; \kappa_x, \kappa_z, \omega) = [\,\hat{v}~~\hat{\eta}\,]^T$, to eliminate the pressure term and the continuity constraint from~(\ref{eq.NS-prop-waves}) and obtain
	\be
	\ba{rcl}
	-\left(
	\mri \omega I \, + \, A (\kappa_x, \kappa_z)
	\right)
	\hat{\bzeta} (y; \kappa_x, \kappa_z, \omega)
	&\!\! = \!\!&
	C^{\dagger} (\kappa_x, \kappa_z)
	\,
	\hat{\fvec} (y; \kappa_x, \kappa_z, \omega),
	\\[0.15cm]
	\hat{\bu} (y; \kappa_x, \kappa_z, \omega)
	&\!\! = \!\!&
	C (\kappa_x, \kappa_z)
	\,
	\hat{\bzeta} (y; \kappa_x, \kappa_z, \omega).
	\ea
	\label{eq.evolution}
	\ee
Here, $A$ is the state operator, $C$ maps the state vector to the velocity vector, and the adjoint of $C$ (denoted by $C^{\dagger}$) maps the forcing vector to the state vector. $A$, $C$, and $C^\dagger$ are operators in $y$ and parameterized by $\kappa_x$ and $\kappa_z$
	\be
	\ba{rcl}
	A
	&\!\! = \!\!&
	\tbt
	{
	\Delta^{-1}
	\left(
	(1/{Re}_\tau) \Delta^2
	\, + \,
	\mri \kappa_x \,
	(U'' \, - \, U \Delta)
	\right)
	}
       {0}
       {-\mri \kappa_z U'}
       {
       (1/{Re}_\tau) \Delta
       \, - \,
       \mri \kappa_x U
       },
	\\[0.3cm]
       C
       &\!\! = \!\!&
	\dfrac{1}{\kappa^2}
       \,
       \thbt
       {\mri \kappa_x \p_y}{-\mri \kappa_z}
       {\kappa^2}{0}
       {\mri \kappa_z \p_y}{\mri \kappa_x},
       ~~
       C^{\dagger}
	\; = \;
	\tbth
	{-\mri \kappa_x \Delta^{-1} \p_y}
       {\kappa^2 \Delta^{-1}}
	{-\mri \kappa_z \Delta^{-1} \p_y}
	{\mri \kappa_z}
	{0}
	{-\mri \kappa_x},
	\ea
       \label{eq.A-C}
       \ee
where $\Delta^2 = \p_{yyyy} - 2 \kappa^2 \p_{yy} + \kappa^4$, and the prime denotes differentiation in $y$, e.g. $U' (y) = \mrd U/\mrd y$. The input-output relationship between $\hat{\fvec}$ and $\hat{\bu}$ is obtained upon elimination of $\hat{\bzeta}$ from~(\ref{eq.evolution})
	\be
	\ba{c}
	\hat{\bu} (y; \kappa_x, \kappa_z, \omega)
	\, = \,
	H (\kappa_x, \kappa_z, \omega) \, \hat{\fvec} (y; \kappa_x, \kappa_z, \omega),
	\\[0.2cm]
	H (\kappa_x, \kappa_z, \omega)
	\, = \,
	C (\kappa_x, \kappa_z) R_A (\kappa_x, \kappa_z, \omega) C^{\dagger} (\kappa_x, \kappa_z),
	\ea
	\label{eq.H}
	\ee
where $R_A (\kappa_x, \kappa_z, \omega) = -(\mri \omega I \, + \, A (\kappa_x, \kappa_z))^{-1}$ is the resolvent of $A$
	\be
	\ba{rcl}
	R_A
	&\!\! = \!\!&
	\tbt
	{
	\Delta^{-1}
	\left(
	\mri \kappa_x \,
	( (U - {c}) \Delta \, - \, U'')
	\, - \,
	(1/{Re}_\tau) \Delta^2
	\right)
	}
       {0}
       {\mri \kappa_z U'}
       {
       \mri \kappa_x (U - {c})
       \, - \,
       (1/{Re}_\tau) \Delta
       }^{-1}.
	\ea
       \label{eq.RA}
       \ee
As illustrated in figure~\ref{fig.block-diag}, the only source of coupling between propagating waves with different wavenumbers is the quadratic dependence of $\fvec (x,y,z,t)$ on $\bu (x,y,z,t)$. For any wavenumber triplet, the input-output map from $\hat{\fvec}$ to $\hat{\bu}$ (shown by the dashed rectangle) represents a sub-system of the full NSE.

\subsection{Decomposition in the wall-normal direction}
\label{sec.svd}

The transfer function $H (\kappa_x, \kappa_z, \omega)$ provides a large amount of information about the input-output relationship between $\hat{\fvec}$ and $\hat{\bu}$. Following the gain analysis of~\cite{mcksha10}, we use the Schmidt (singular value) decomposition to provide a wall-normal basis based on the most highly amplified forcing and response directions:
	\be
	\ba{rcl}
	\hat{\bu} (y; \kappa_x, \kappa_z, \omega)
	&\!\! = \!\!&
	H (\kappa_x,\kappa_z,\omega) \, \hat{\fvec} (y; \kappa_x,\kappa_z,\omega)
	\\[0.1cm]
	&\!\! = \!\!&
	\ds{\sum_{j = 1}^{\infty}}
	\,
	\sigma_j (\kappa_x,\kappa_z,\omega) \, a_j (\kappa_x,\kappa_z,\omega)
	\,
	\hat{\bpsi}_j (y; \kappa_x, \kappa_z,\omega),
	\\[0.4cm]
	a_j (\kappa_x,\kappa_z,\omega)
	&\!\! = \!\!&
	\ds{\int_{-1}^{1}}
	\hat{\bphi}_j^* (y; \kappa_x, \kappa_z,\omega)
	\,
	\hat{\fvec} (y; \kappa_x, \kappa_z,\omega)
	\,
	\mrd y,
	\ea
	\label{eq.svd}
	\ee
where $\sigma_1 \geq \sigma_2 \geq \cdots > 0\,$ denote the singular values of $H$, and the singular functions $\hat{\bphi}_j = [\,\hat{f}_{1j}~\hat{f}_{2j}~\hat{f}_{3j}\,]^T$ and $\hat{\bpsi}_j = [\,\hat{u}_j~\hat{v}_j~\hat{w}_j\,]^T$ are respectively the forcing and response directions corresponding to $\sigma_j$. \moa{In principle, there are infinite number of singular values/modes because the wall-normal coordinate is continuous. For the discretized equation, the total number of singular values/modes is twice the number of grid points in $y$ since the resolvent operator $R_A$ in~(\ref{eq.RA}) acts on a vector of two functions in $y$. As highlighted by~\cite{mcksha10}, the singular value decomposition effectively demonstrates that there are a limited number of relatively highly-amplified modes within this total number of modes. Throughout this paper, we consistently refer to $\hat{\bpsi}_j$ by \emph{the resolvent mode}, and distinguish it from the real turbulent flow that, under stationary conditions, can be represented by a weighted sum of the resolvent modes. The latter is denoted by \emph{the weighted mode}. Note that the resolvent modes were denoted by response modes in~\cite{mcksha10,mckshajac13,shamck13}.}

While the singular values of $H$ are unique, additional treatment is necessary to obtain unique singular functions. Unlike in a pipe, the singular values come in pairs due to the wall-normal symmetry in the channel (which reflects itself in the resolvent operator); see, for example, figure~\ref{fig.sigma1to20}. \moa{For the modes with smaller streamwise and spanwise wavelengths than the channel half-height, the singular values come in equal pairs. Therefore, any linear combination of the corresponding singular functions represents a legitimate singular function. For example, if the symmetric and anti-symmetric modes are denoted by $\psi_s$ and $\psi_a$ where $|\psi_s| = |\psi_a|$, the singular function given by $\psi_d = \psi_s - \psi_a$ is zero in one half of the channel and twice $\psi_s$ in the other half. Clearly, $\psi_d$ is also a singular function of the transfer function with the same singular value as $\psi_s$ and $\psi_a$. Physically, this means that the modes with lengths and widths smaller than the channel half-height exhibit the potential to independently evolve in either halves of the channel provided that they are forced with a forcing (e.g. disturbance) that is present only in one half of the channel. On the other hand, for the modes with larger wavelengths than the channel half-height, the paired singular values are different and the singular modes are either symmetric or anti-symmetric in the opposite halves of the channel. Physically, these modes represent convective global phenomena meaning that they cannot take place independently in the opposite halves of the channel. They convect with the same magnitude in the opposite halves of the channel even though they can be of the same or opposite phases.}

When the paired singular values are different, we obtain unique singular functions, modulo a complex multiplicative constant of unit magnitude, by imposing an orthonormality constraint on them
	\be
	\ds{\int_{-1}^{1}}
	\hat{\bphi}_j^* (y; \kappa_x, \kappa_z,\omega)
	\,
	\hat{\bphi}_k (y; \kappa_x, \kappa_z,\omega)
	\,
	\mrd y
	\, = \,
	\ds{\int_{-1}^{1}}
	\hat{\bpsi}_j^* (y; \kappa_x, \kappa_z,\omega)
	\,
	\hat{\bpsi}_k (y; \kappa_x, \kappa_z,\omega)
	\,
	\mrd y
	\, = \,
	\delta_{jk},
	\label{eq.orthonormal}
	\ee
where $\delta$ denotes the Kronecker delta. In the case where the paired singular values are equal, we impose a symmetry/anti-symmetry constraint on the singular functions in addition to the above orthonormality constraint. In other words, the corresponding singular functions assume the same magnitude throughout the channel while being in phase in one half of the channel and out of phase in the other half.

In this study, we select the unknown multiplicative constant (after orthonormalization) such that $\hat{u}_j (y_{\mathrm{max}}; \kappa_x, \kappa_z, \omega)$ is a real number at the wall-normal location $y_{\mathrm{max}}$ where the absolute value of $\hat{u}_j$ is the largest. This choice places the maximum of $u_j (x,y,z,t; \kappa_x, \kappa_z, \omega)$ at the origin $x = z = t = 0$. The channel symmetries in the streamwise and spanwise directions can be used to obtain $u_j$, $v_j$, and $w_j$ in the physical domain
	\be
	\ba{rcl}
	u_j (x,y,z,t; \kappa_x, \kappa_z, \omega)
	&\!\! = \!\!&
	4
	\cos (\kappa_z z)
	\,
	\mbox{Re} \left(
	\hat{u}_j (y; \kappa_x, \kappa_z, \omega) \,\mre^{\mri (\kappa_x x - \omega t)}
	\right),
	\\[0.15cm]
	v_j (x,y,z,t; \kappa_x, \kappa_z, \omega)
	&\!\! = \!\!&
	4
	\cos (\kappa_z z)
	\,
	\mbox{Re} \left(
	\hat{v}_j (y; \kappa_x, \kappa_z, \omega) \,\mre^{\mri (\kappa_x x - \omega t)}
	\right),
	\\[0.15cm]
	w_j (x,y,z,t; \kappa_x, \kappa_z, \omega)
	&\!\! = \!\!&
	-4
	\sin (\kappa_z z)
	\,
	\mbox{Im} \left(
	\hat{w}_j (y; \kappa_x, \kappa_z, \omega) \,\mre^{\mri (\kappa_x x - \omega t)}
	\right),
	\ea
	\non
	\ee
where $\mbox{Re}$ and $\mbox{Im}$ denote the real and imaginary parts of a complex number. The representation of the forcing directions in the physical domain is obtained using similar expressions.

    \begin{figure}
    \begin{center}
    \begin{tabular}{cc}
    $u_1$
    &
    $u_1 (\mbox{color}); v_1, w_1 (\mbox{arrows})$
    \\[-0.05cm]
    \includegraphics[width=0.5\columnwidth]
    {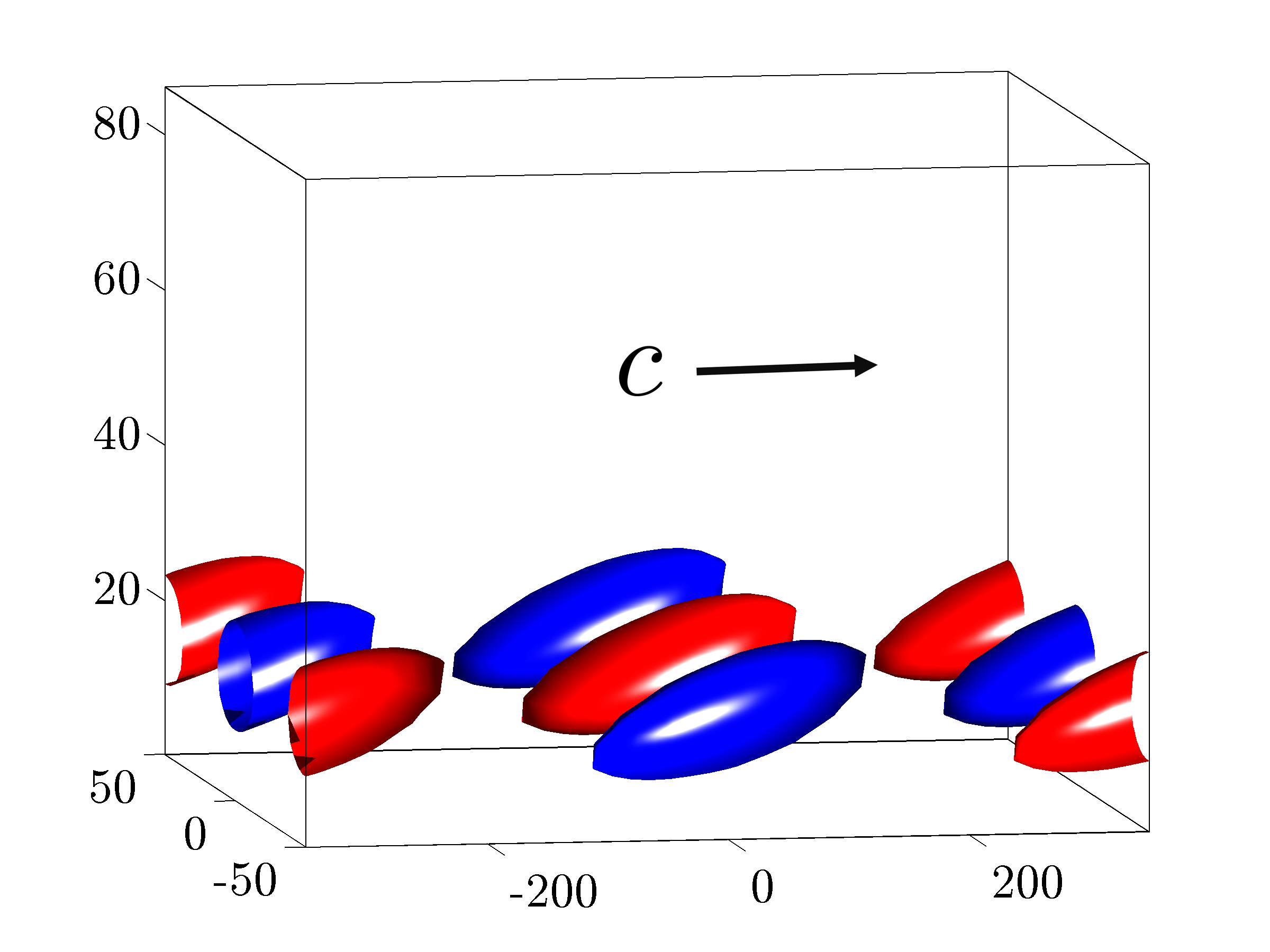}
    &
    \hskip-0.4cm
    \includegraphics[width=0.5\columnwidth]
    {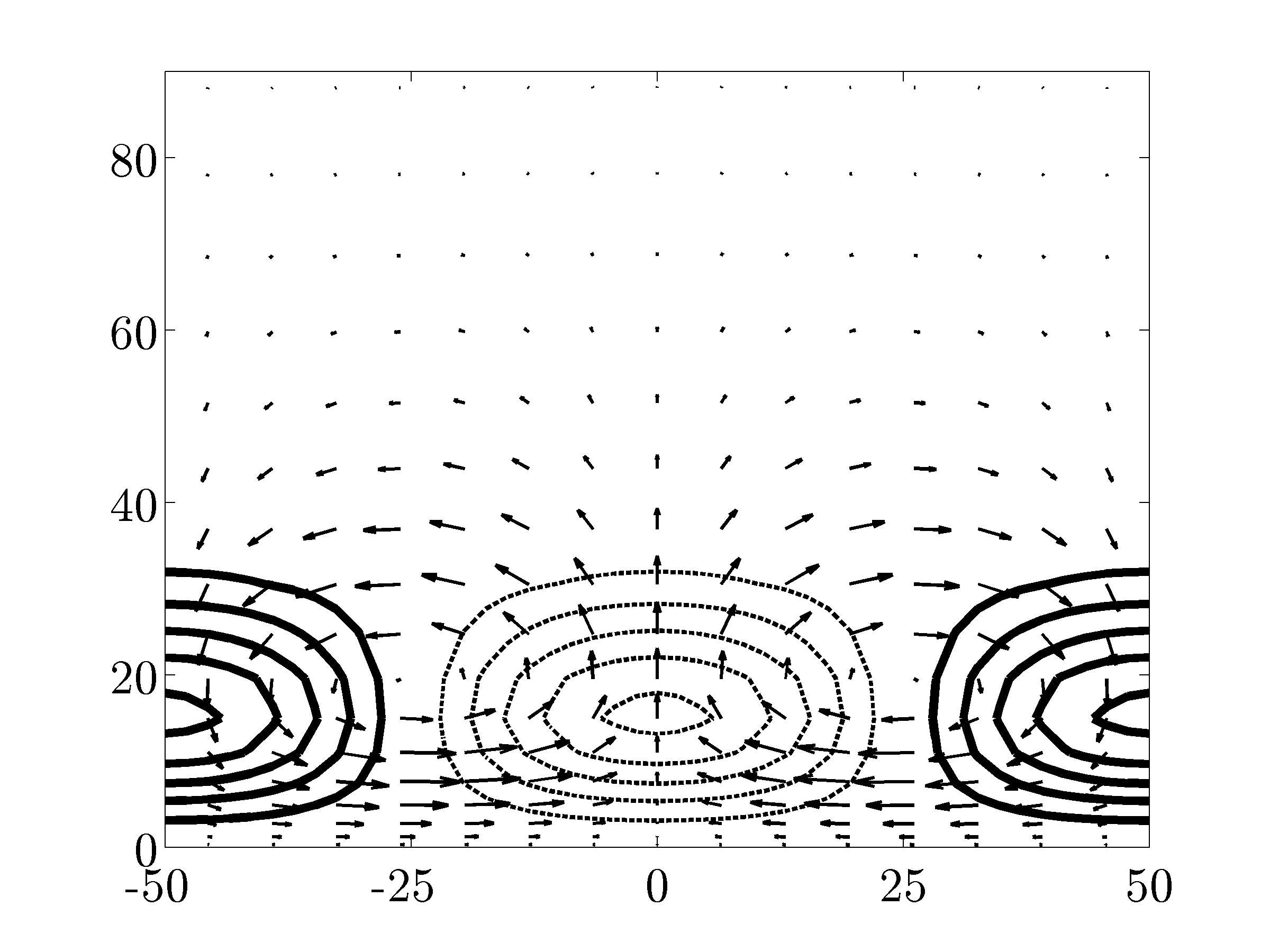}
    \end{tabular}
    \begin{tabular}{c}
    \\[-3.4cm]
    \begin{tabular}{c}
    \hskip-6.8cm
    \tc{black}{$y^+$}
    \end{tabular}
    \\[1.9cm]
    \begin{tabular}{c}
    \hskip-6.5cm
    \tc{black}{$z^+$}
    \end{tabular}
    \\[0.1cm]
    \begin{tabular}{c}
    \hskip0.8cm
    \tc{black}{$x^+$}
    \hskip5.4cm
    \tc{black}{$z^+$}
    \end{tabular}
    \end{tabular}
    \end{center}
    \caption{The principal velocity response $\bpsi_1 (x,y,z,t; \kappa_x, \kappa_z, c) = [\,u_1~v_1~w_1\,]^T$ for $\lambda_x^+ = 700$, $\lambda_z^+ = 100$, ${c} = 10$, and ${Re}_\tau = 10000$ at $t = 0$. (Left) The isosurfaces of the streamwise velocity at $60 \%$ of its maximum; (Right) The streamwise velocity (contours) and the spanwise and wall-normal velocity (arrows) at $x^+ = \lambda_x^+/2$. The contours in (b) represent positive (thick solid) and negative (thin dashed) values from $3$ to $15$ with increments of $3$.}
    \label{fig.u1-v1-w1}
    \end{figure}

From the singular value decomposition~(\ref{eq.svd}) and the orthonormality constraints~(\ref{eq.orthonormal}) it follows that if the forcing is aligned in the $\hat{\bphi}_j$ direction with unit energy, the response is aligned in the $\hat{\bpsi}_j$ direction with energy $\sigma_j^2$. Consequently, the forcing and response directions with the largest gain correspond to the principal singular functions $\hat{\bphi}_1$ and $\hat{\bpsi}_1$. For any $(\kappa_x, \kappa_z, \omega)$, the singular functions of $H$ should be thought of as propagating waves in the physical domain. \moa{In the rest of the paper, the resolvent modes are characterized by $c$ instead of $\omega$ and we note that prescribing any two of $\kappa_x$, $\omega$, and $c$ leads to the other.}

Equivalent near-wall structures to those reported for pipe flows by~\cite{mcksha10,mckshajac13} are obtained for channel flows. For example, the principal singular function $\bpsi_1 (x,y,z,t; \kappa_x, \kappa_z, \omega) = [\,u_1~v_1~w_1\,]^T$ for the propagating wave corresponding to the energetic near-wall cycle ($\lambda_x^+ = 700$, $\lambda_z^+ = 100$, ${c} = U(y^+ = 15) = 10$) for ${Re}_\tau = 10000$ is shown in figure~\ref{fig.u1-v1-w1}. The streamwise component of these structures contains regions of fast- and slow- moving fluids that are aligned in the streamwise direction, slightly inclined to the wall, and are sandwiched between counter-rotating vortical motions in the cross-stream plane.

    \begin{figure}
    \begin{center}
    \begin{tabular}{cc}
    $\sigma_j$
    &
    $(\sigma_1^2+\sigma_2^2)/(\sum_{j=1}^{\infty} \sigma_j^2)$
    \\[-0.15cm]
    \subfigure{\includegraphics[width=0.5\columnwidth]
    {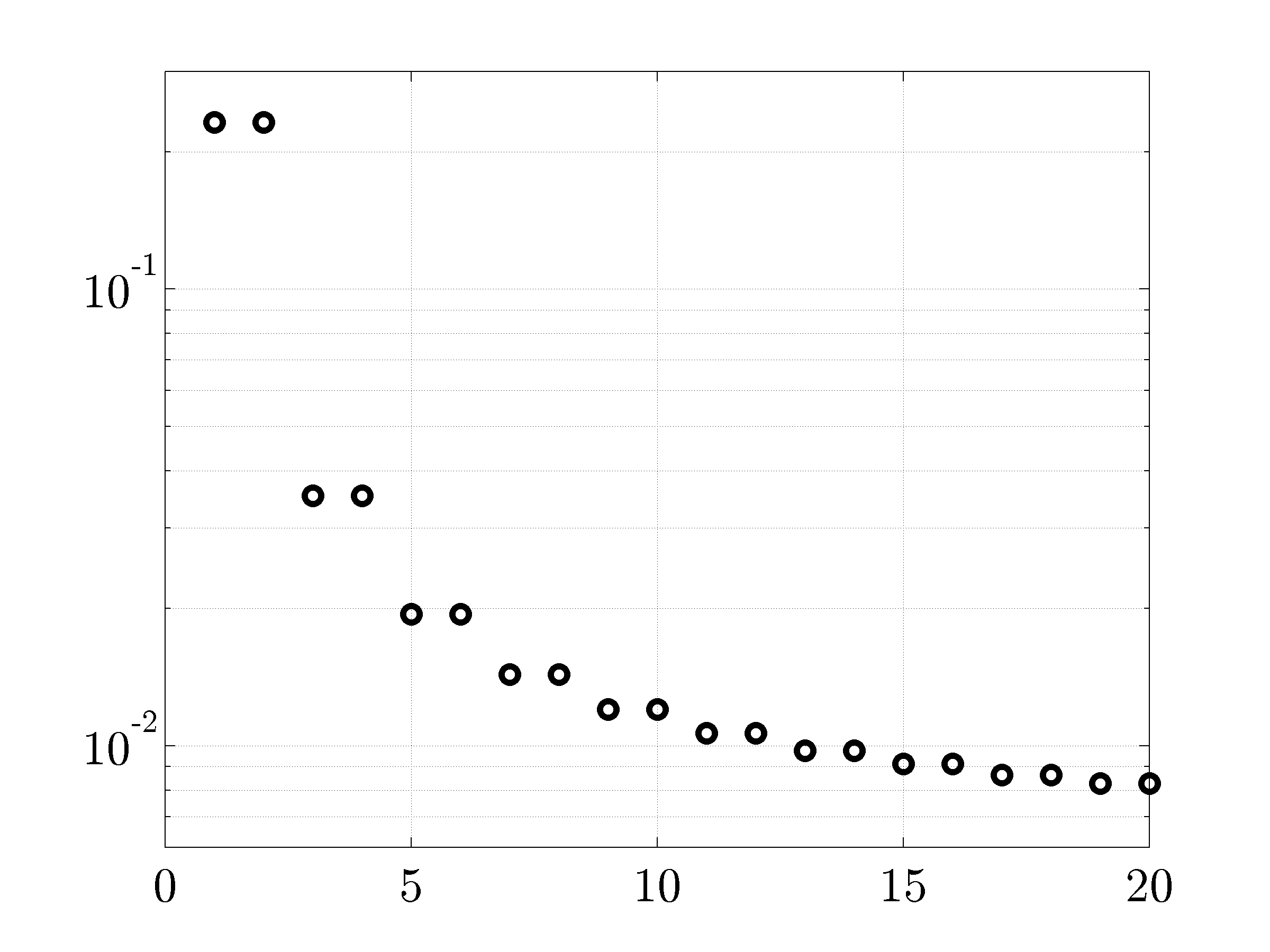}
    \label{fig.sigma1to20}}
    &
    \hskip-0.4cm
    \subfigure{\includegraphics[width=0.5\columnwidth]
    {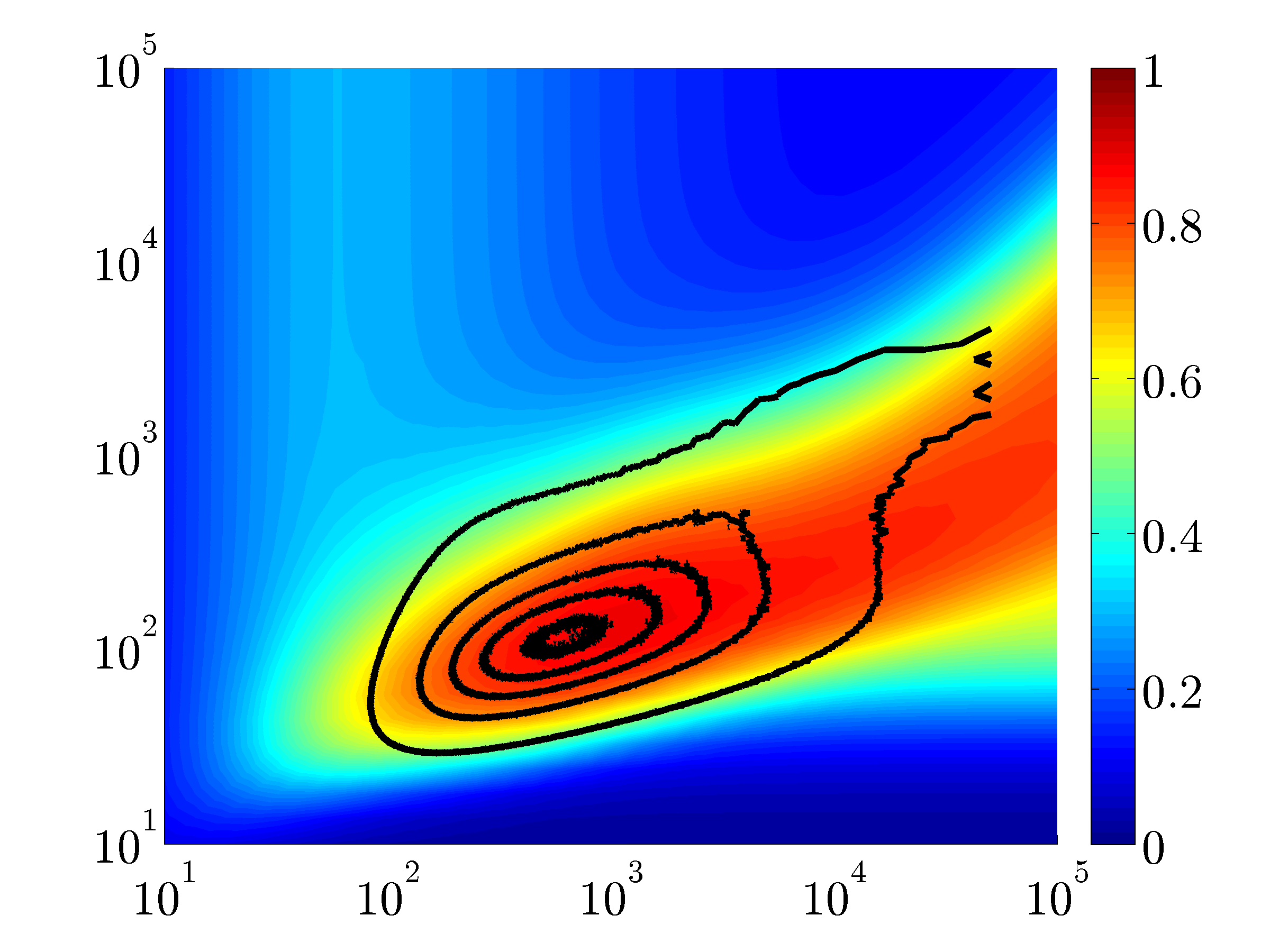}
    \label{fig.sigma1-vs-DNS-yp15}}
    \\[0.1cm]
    $(a)$
    &
    $(b)$
    \end{tabular}
    \begin{tabular}{c}
    \\[-4cm]
    \begin{tabular}{c}
    \hskip0.5cm
    \tc{black}{$\lambda_z^+$}
    \end{tabular}
    \\[2.3cm]
    \begin{tabular}{c}
    \hskip0.3cm
    \tc{black}{$j$}
    \hskip6.1cm
    \tc{black}{$\lambda_x^+$}
    \end{tabular}
    \end{tabular}
    \\[0.2cm]
    \begin{tabular}{cc}
    $(\sigma_1^2+\sigma_2^2)/(\sum_{j=1}^{\infty} \sigma_j^2)$
    &
    $(\sigma_1^2+\sigma_2^2)/(\sum_{j=1}^{\infty} \sigma_j^2)$
    \\[-0.15cm]
    \subfigure{\includegraphics[width=0.5\columnwidth]
    {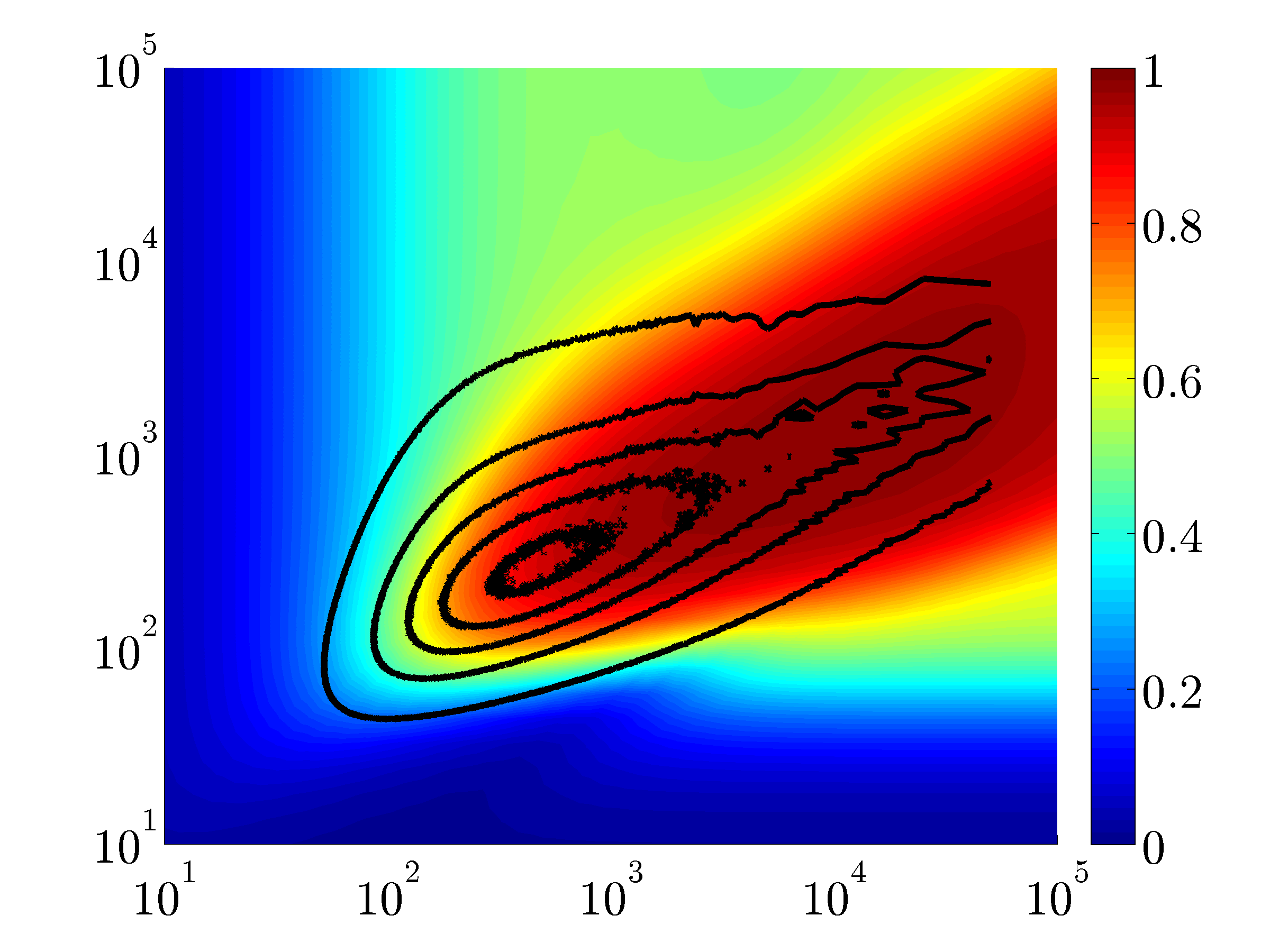}
    \label{fig.sigma1-vs-DNS-yp100}}
    &
    \hskip-0.4cm
    \subfigure{\includegraphics[width=0.5\columnwidth]
    {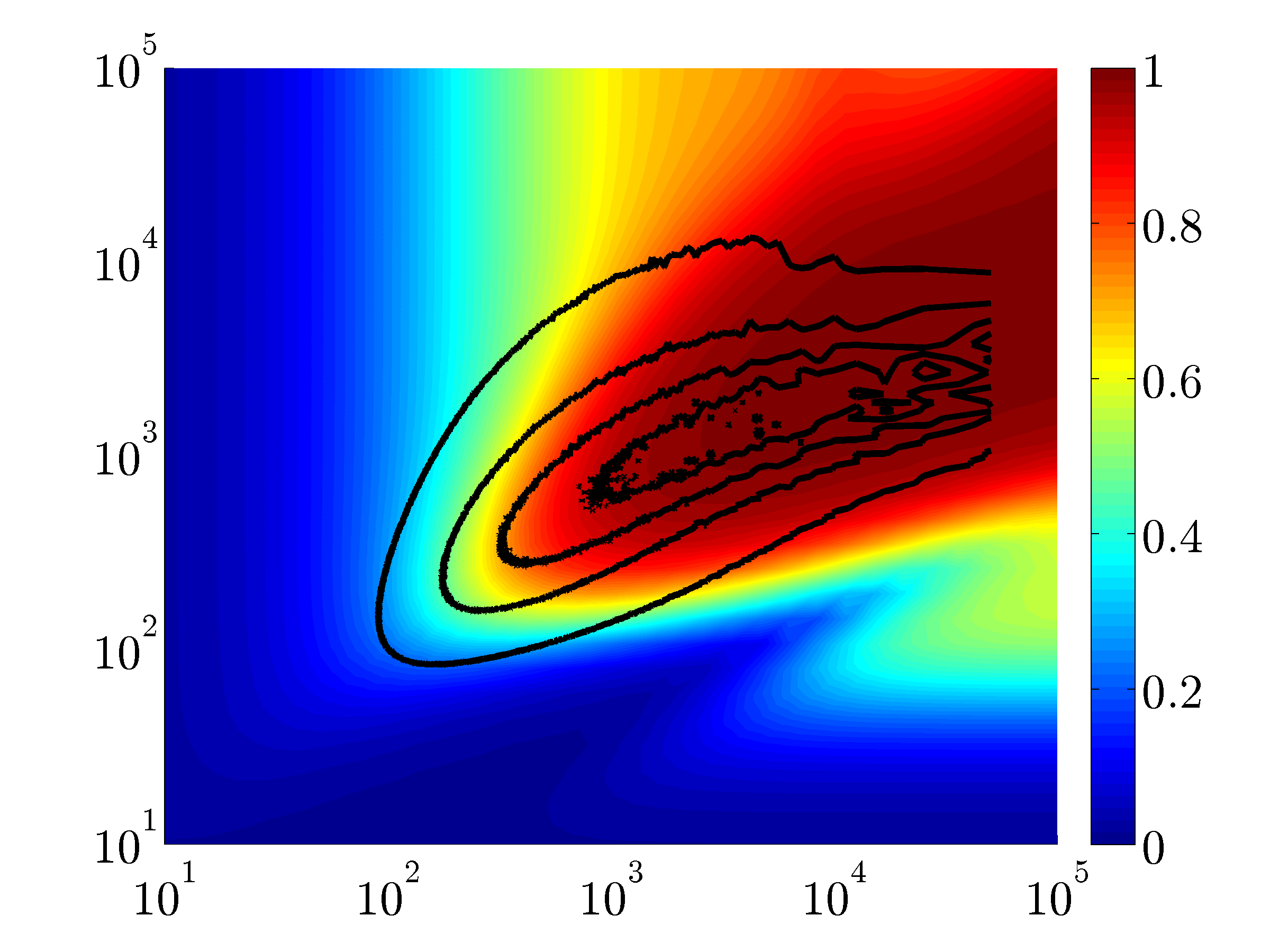}
    \label{fig.sigma1-vs-DNS-y0p2}}
    \\[0.1cm]
    $(c)$
    &
    $(d)$
    \end{tabular}
    \begin{tabular}{c}
    \\[-4cm]
    \begin{tabular}{c}
    \hskip-6.2cm
    \tc{black}{$\lambda_z^+$}
    \hskip6.1cm
    \tc{black}{$\lambda_z^+$}
    \end{tabular}
    \\[2.3cm]
    \begin{tabular}{c}
    \hskip0.3cm
    \tc{black}{$\lambda_x^+$}
    \hskip6.1cm
    \tc{black}{$\lambda_x^+$}
    \end{tabular}
    \end{tabular}
    \end{center}
    \caption{(a) The twenty largest singular values of $H$ for $\lambda_x^+ = 700$, $\lambda_z^+ = 100$, ${c} = 10$, and ${Re}_\tau = 2003$.
    (b)-(d): The color plots show the energy that is contained in the largest two response modes relative to the total response for different streamwise and spanwise wavelengths and (b) ${c} = U(y^+ = 15)$; (c) ${c} = U(y^+ = 100)$; and (d) ${c} = U(y = 0.2)$. The black contours show the turbulent kinetic energy spectrum from the DNS of~\cite{hoyjim06} at the corresponding critical wall-normal locations (b) $y^+ = 15$; (c) $y^+ = 100$; and (d) $y = 0.2$. The contours represent $10 \%$ to $90 \%$ of the maximum energy spectrum at each wall-normal location with increments of $20 \%$.}
    \label{fig.low-rank}
    \end{figure}

\subsection{Low-rank nature of $H$}
\label{sec.low-rank-H}

The operator $H$, acting on functions of $y$, can be described as low-rank if a significant portion of its response to a broadband forcing in $y$ is captured by projection on the first few response directions.~\cite{mcksha10} highlighted the low-rank nature of $H$ for turbulent pipe flow. Figure~\ref{fig.sigma1to20} shows the first twenty singular values of $H$ for $\lambda_x^+ = 700$, $\lambda_z^+ = 100$, and ${c} = 10$ in turbulent channel flow with ${Re}_\tau = 2003$. We see that the largest pair of singular values is approximately one order of magnitude larger than the other singular values.

The energetic contribution of the $k$\/-th direction $\hat{\bpsi}_k$ to the total response in the model subject to broadband forcing in $y$ with fixed $\lambda_x$, $\lambda_z$, and $c$ is quantified by $\sigma_k^2/(\sum_{j=1}^{\infty} \sigma_j^2)$. Figures~\ref{fig.sigma1-vs-DNS-yp15}-\ref{fig.sigma1-vs-DNS-y0p2} highlight the low-rank nature of $H$ by showing that the first two principal response directions $\hat{\bpsi}_1$ and $\hat{\bpsi}_2$ contribute to more than $80 \%$ of the total response over a large range of wall-parallel wavelengths (red region) for wave speeds ${c} = U(y^+ = 15)$, $U(y^+ = 100)$, $U(y = 0.2)$, and ${Re}_\tau = 2003$.  \moa{The relevance of studying the low-rank approximation of $H$ is further emphasized by noting that the most energetic wavenumbers from the DNS of~\cite{hoyjim06} (contours) coincide with the wavenumbers and critical wave speeds for which $H$ is low-rank. We note that the streamwise velocity has the largest contribution to the kinetic energy. Even though the shapes of the two-dimensional wall-normal and spanwise spectra may be significantly different from the streamwise spectrum, the contours corresponding to $70 \%$ of the maximum in all spectra (not shown) lie within the region where the contribution of the largest two singular values is more than $50 \%$.}

\subsection{Rank-1 model subject to broadband forcing}
\label{sec.rank-1}

In the present study, we consider a rank-1 model by only keeping the most energetic forcing and response directions corresponding to $\sigma_1$ and show that significant understanding of the scaling of wall turbulence can be obtained using this simple model. This is motivated by the observation in~\S~\ref{sec.low-rank-H} that the operator $H$ is essentially a directional amplifier. In other words, we expect to see the principal singular response of $H$ in real turbulent flows provided that the principal forcing direction is present in the nonlinear forcing term. \moa{Even though the resolvent modes corresponding to $\sigma_1$ and $\sigma_2$ comparably contribute to the total response, cf.~\S~\ref{sec.low-rank-H}, considering one of the resolvent modes is sufficient for capturing the wall-normal shape of the energy density. This is because the two resolvent modes are symmetric/anti-symmetric counterparts of each other and have the same magnitude. Therefore, accounting for both resolvent modes yields the same result as accounting for one resolvent mode.}

It is well-known that the streamwise energy spectrum can be divided into regions that scale in inner and outer variables~\cite[see, for example,][]{mormckjiasmi04}. Our objective is to explore the Reynolds number scaling of the streamwise energy density and predict the behavior of the streamwise turbulence intensity at high ${Re}_\tau$. We focus on the streamwise velocity because it dominates the kinetic energy density in turbulent flows. Similarly, the principal singular responses of $H$ that result in the largest energy amplification are dominated by their streamwise component, \moa{such that the proposed gain-based decomposition yields the streamwise velocity most accurately. This is in agreement with previous linear analyses of the global optimal responses, e.g.~\cite{alajim06,hwacos10}. We note that higher-order resolvent modes may have comparable or larger wall-normal and spanwise components relative to the streamwise velocity, studying of which is a subject of ongoing work.}

In order to use the least number of assumptions, we consider the case where the forcing $\hat{\fvec}$ equals the principal forcing direction $\hat{\bphi}_1$. Consequently, the forcing has unit energy for all wave parameters, meaning that it is broadband in $\kappa_x$, $\kappa_z$, and $c$. For the rank-1 model with broadband forcing, we define the premultiplied streamwise energy density of the principal response of $H$ by
	\be
	\ba{c}
	E_{uu} (y; \kappa_x, \kappa_z, {c})
	\, = \,
	\kappa_x^2 \kappa_z
	\left(
	\sigma_1 (\kappa_x,\kappa_z,c) \abs{u_1} (y; \kappa_x, \kappa_z, {c})
	\right)^2,
	\ea
	\label{eq.Euu}
	\ee
such that the premultiplied one-dimensional energy densities and the energy intensity are obtained by integrating $E_{uu} (y; \kappa_x, \kappa_z, {c})$ over the set of all wave parameters $\cS$, e.g.
	\be
	\ba{rcl}
	E_{uu} (y,{c})
	&\!\! = \!\!&
	\ds{
	\iint\limits_{\cS}
	}
	\,
	E_{uu} (y; \kappa_x, \kappa_z, {c})
	\,
	\mrd \log(\kappa_x)
	\,
	\mrd \log(\kappa_z),
	\\[0.25cm]
	E_{uu} (y)
	&\!\! = \!\!&
	\ds{
	\iiint\limits_{\cS}
	}
	\,
	E_{uu} (y; \kappa_x, \kappa_z, {c})
	\,
	\mrd \log(\kappa_x)
	\,
	\mrd \log(\kappa_z)
	\,
	\mrd {c},
	\ea
	\label{eq.Euu-1-0-D}
	\ee
and $E_{uu} (y,\kappa_x)$ and $E_{uu} (y,\kappa_z)$ are determined similarly.

The above formulation of the energy density is used in~\S~\ref{sec.Re-scaling-theory} to identify the contribution of confined subsets of wave parameters to the energy density. We establish that the energy density exhibits universal behavior with ${Re}_\tau$ for properly selected subsets of wave parameters. It is further shown that the emerging scales are consistent with those observed in experiments. In addition, the scales of energetically dominant waves roughly agree with the scales of dominant near-wall motions in real turbulent flows.

\subsection{Computational approach}
\label{sec.comput}

A pseudo-spectral method is used to discretize the differential operators in the wall-normal direction on a set of Chebyshev collocation points. This is implemented using the Matlab Differentiation Matrix Suite developed by~\cite{weired00}. Table~\ref{table.numerics} summarizes the selected range of wave parameters and their respective resolution in numerical computations. It has been verified that the excluded wave parameters are not energetically important and therefore do not change the results of the present study.

An efficient randomized scheme developed by~\cite{halmartro11} is utilized to compute the principal singular directions of $H$ for different Reynolds numbers and wave parameters. The accuracy and computation time depend on the decay of the singular values; a faster decay results in high accuracy or equivalently less computation time to reach the same accuracy. In addition, if the singular values are not well separated, the problem of computing the associated singular functions is badly conditioned, meaning that it is hard for any method to determine them very accurately. In this study, the above scheme approximately halves the total computation time relative to Matlab's {\verb svds } algorithm. This becomes increasingly important considering the three-dimensional wave parameter space that we need to explore and the large size of the discretized resolvent operator (twice the number of collocation points in $y$) at high Reynolds numbers. In addition, the randomized nature of this scheme enables its parallel implementation which makes it especially suitable for large-scale computations. Even though we have not used this feature in the present study, it may find use in designing turbulent flow control strategies, e.g. by means of spatially or temporally periodic actuations.

         \begin{table}
         \centering
         \begin{tabular}{ccccccccccc}
            \hspace{0.1cm}
            ${Re}_\tau$
            \hspace{0.1cm}
            &
            \hspace{0.1cm}
            $N_x$
            \hspace{0.1cm}
            &
            \hspace{0.1cm}
            $N_y$
            \hspace{0.1cm}
            &
            \hspace{0.1cm}
            $N_z$
            \hspace{0.1cm}
            &
            \hspace{0.1cm}
            $N_c$
            \hspace{0.1cm}
            &
            \hspace{0.1cm}
            $\lambda_{x,\mathrm{min}}^+$
            \hspace{0.1cm}
            &
            \hspace{0.1cm}
            $\lambda_{x,\mathrm{max}}$
            \hspace{0.1cm}
            &
            \hspace{0.1cm}
            $y_{\mathrm{min}}^+$
            \hspace{0.1cm}
            &
            \hspace{0.1cm}
            $\lambda_{z,\mathrm{min}}^+$
            \hspace{0.1cm}
            &
            \hspace{0.1cm}
            $\lambda_{z,\mathrm{max}}$
            \hspace{0.1cm}
            &
            \hspace{0.1cm}
            $U_{cl}$
            \hspace{0.1cm}
            \\[0.2cm] 
            $934$
            &
            $64$
            &
            $251$
            &
            $32$
            &
            $100$
            &
            $10$
            &
            $10^6$
            &
            $0.07$
            &
            $10$
            &
            $100$
            &
            $22.39$
            \\[0.2cm] 
            $2003$
            &
            $64$
            &
            $251$
            &
            $32$
            &
            $100$
            &
            $10$
            &
            $5\times10^5$
            &
            $0.15$
            &
            $10$
            &
            $50$
            &
            $24.02$
		 \\[0.2cm] 
            $3333$
            &
            $64$
            &
            $301$
            &
            $32$
            &
            $100$
            &
            $10$
            &
            $3\times10^5$
            &
            $0.18$
            &
            $10$
            &
            $30$
            &
            $25.22$
            \\[0.2cm] 
            $10000$
            &
            $64$
            &
            $401$
            &
            $32$
            &
            $100$
            &
            $10$
            &
            $10^5$
            &
            $0.3$
            &
            $10$
            &
            $10$
            &
            $27.81$
            \\[0.2cm] 
            $30000$
            &
            $80$
            &
            $601$
            &
            $40$
            &
            $100$
            &
            $10$
            &
            $3.3\times10^6$
            &
            $0.4$
            &
            $10$
            &
            $33$
            &
            $30.39$
         \end{tabular}
         \caption{Summary of the selected parameters in numerical computations at different Reynolds numbers. In the wall-normal direction, $N_y$ Chebyshev collocation points are used with $y_{\mathrm{min}}^+$ denoting the closest point to the wall. In the streamwise and spanwise directions, $N_x$ and $N_z$ logarithmically spaced wavelengths are used between $\lambda_{\mathrm{min}}^+$ and $\lambda_{\mathrm{max}}$. In addition, $N_c$ linearly spaced wave speeds are chosen between $c_{\mathrm{min}} = 2$ and $c_{\mathrm{max}} = U_{cl}$.}
         \label{table.numerics}
         \end{table}
	
\moa{
\section{Universal behavior of the resolvent}
\label{sec.Re-scaling-theory}
}

The formulation of~\S~\ref{sec.low-rank} facilitates analysis of the contribution of different wave parameters $(\kappa_x,\kappa_z,c)$ to the streamwise energy density. For the rank-1 model with broadband forcing, the energy density of each wave is determined from the principal singular values and singular functions of the transfer function $H$; see~(\ref{eq.Euu}). 
\moa{
In this section, we identify unique classes of wave parameters for which $E_{uu} (y; \kappa_x, \kappa_z, {c})$ exhibits either universal behavior with ${Re}_\tau$ or geometrically self-similar behavior with distance from the wall. Each class is characterized by a unique range of wave speeds and a unique scaling of the wall-normal coordinate and the wall-parallel wavelengths. These classes are inherent to the linear mechanisms in the NSE and are rigorously identified by analysis of the transfer function.
}

\moa{
\subsection{Requirement for universality of the resolvent modes}
\label{sec.self-sim-req}
}

We start by showing that a requirement for universal behavior is the wall-normal locality of the resolvent modes. This is done by examining the underlying operators in $H$, cf.~(\ref{eq.A-C})-(\ref{eq.RA}). We see that the difference between the turbulent mean velocity and the wave speed, $U(y) - {c}$, and its wall-normal derivatives, $U'(y)$ and $U''(y)$, appear as spatially-varying coefficients in $H$. Since the turbulent mean velocity scales differently with ${Re}_\tau$ in different wall-normal locations, only the resolvent modes that are sufficiently narrow in $y$ have the potential to be universal. This is because such resolvent modes are purely affected by a certain part of the mean velocity that scales uniquely with ${Re}_\tau$.

We next show that the resolvent modes corresponding to the energetically significant modes are in fact localized. As summarized by~\cite{lehguamck11}, the energetic contribution of structures with convection velocities less than $10 u_\tau$ and larger than the centerline velocity $U_{cl} = U (y = 1)$ is negligible in real turbulent flows. \moa{However, we are interested in determining the effect of a broader range of wave speeds on the energy density. Note that small values of $c$ result in small amplification because the corresponding singular values are small. In fact, it is shown in~\S~\ref{sec.energy-density-weighted} that including the modes with $c \lesssim 2$ does not improve the matching error between the model-based and DNS-based energy intensities.} This motivates defining a conservative subset of $\cS$, denoted by $\cS_{e}$, that includes all wall-parallel wavenumbers and the energetically important wave speeds
	\be
	\cS_e
	\, = \,
	\left\{
	(\kappa_x,\kappa_z,{c})
	~~|~~
	2 \leq {c} \leq U_{cl}
	\right\}.
	\label{eq.S_e}
	\ee

    \begin{figure}
    \begin{center}
    \begin{tabular}{cc}
    \subfigure{\includegraphics[width=0.5\columnwidth]
    {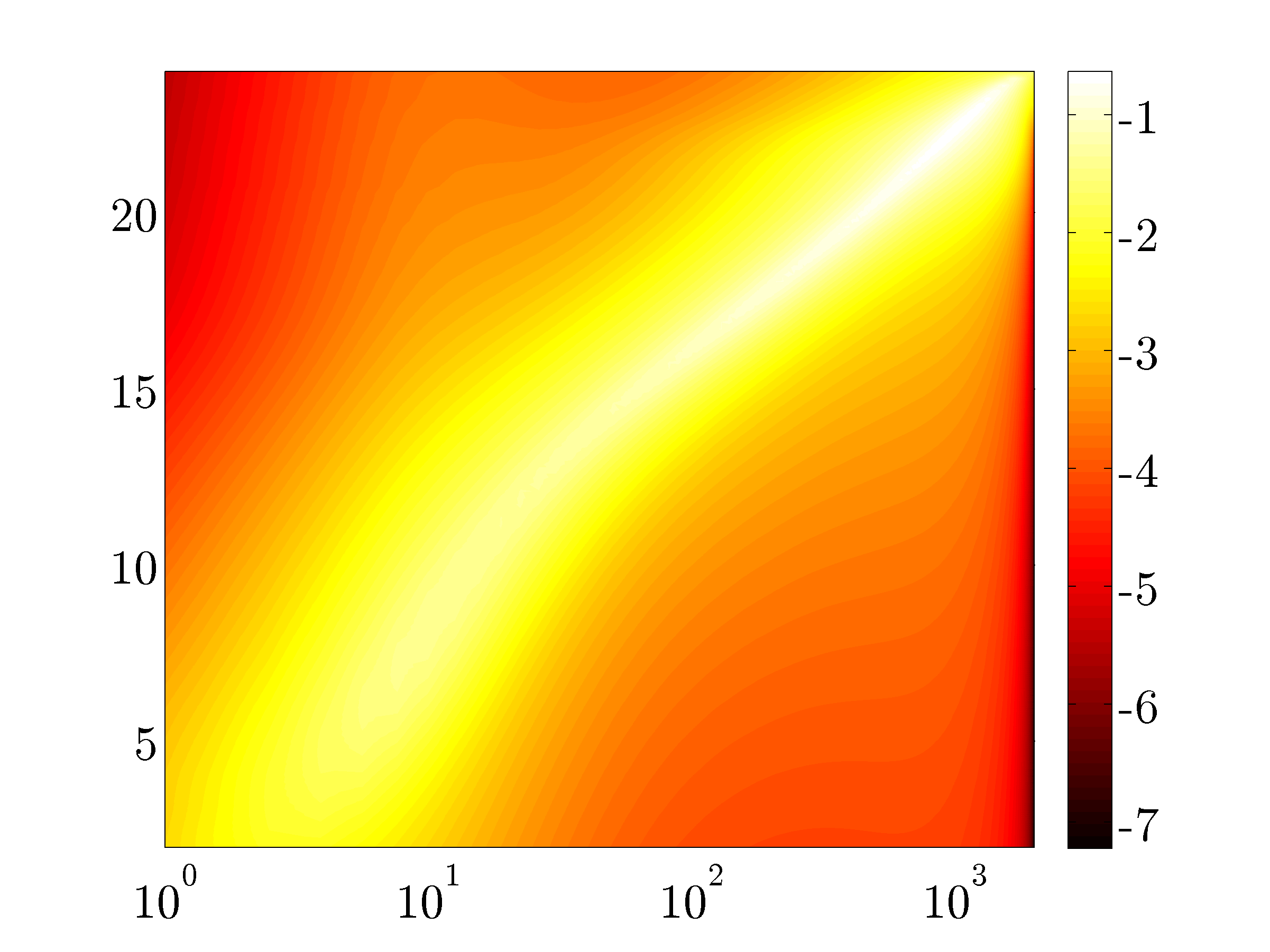}
    \label{fig.Euu-Rm2-kzkx2-vs-yp-Ux-R30000-sigma1-log10}}
    &
    \hskip-0.4cm
    \subfigure{\includegraphics[width=0.5\columnwidth]
    {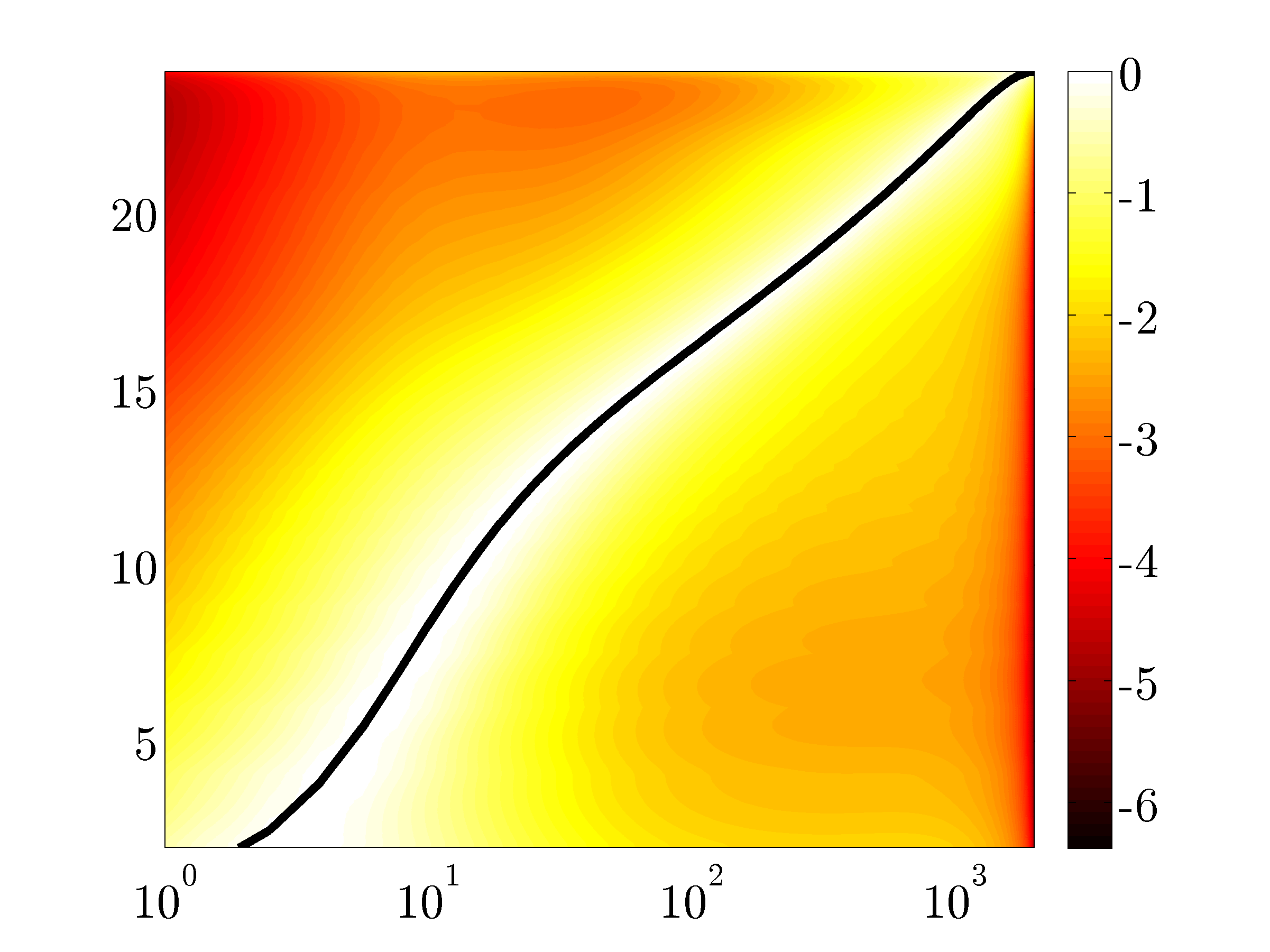}
    \label{fig.Euu-Rm2-kzkx2-vs-yp-Ux-R30000-sigma1-normalized-log10}}
    \\[0.15cm]
    $(a)$
    &
    $(b)$
    \end{tabular}
    \begin{tabular}{c}
    \\[-3.95cm]
    \begin{tabular}{c}
    \hskip-2.6cm
    \tc{black}{${c}$}
    \hskip6.45cm
    \tc{black}{${c}$}
    \hskip3.1cm
    \tc{black}{$U$}
    \end{tabular}
    \\[2.25cm]
    \begin{tabular}{c}
    \hskip0.5cm
    \tc{black}{$y^+$}
    \hskip6.4cm
    \tc{black}{$y^+$}
    \end{tabular}
    \end{tabular}
    \end{center}
    \caption{(Color online) (a) The one-dimensional energy density ${Re}_\tau^{-2} E_{uu} (y,{c})$ for $\cS = \cS_e$ and ${Re}_\tau = 2003$; and (b) the energy density normalized by its maximum value over all $y$ for fixed values of ${c}$. The colors are in logarithmic scale. The turbulent mean velocity is shown by the black curve in (b).}
    \label{fig.Euu-y-Up-1D}
    \end{figure}

Figure~\ref{fig.Euu-y-Up-1D} shows the one-dimensional energy density as a function of wave speed $E_{uu} (y,{c})$ for ${Re}_\tau = 2003$ and $\cS = \cS_{e}$. As evident from figure~\ref{fig.Euu-Rm2-kzkx2-vs-yp-Ux-R30000-sigma1-log10}, the energy density for a fixed ${c}$ is localized in a narrow wall-normal region; note that the colors are given in logarithmic scale. The localization is highlighted in figure~\ref{fig.Euu-Rm2-kzkx2-vs-yp-Ux-R30000-sigma1-normalized-log10} where $E_{uu} (y,{c})$ is normalized by its maximum value over $y$ for fixed values of $c$. We see that the largest energy amplification takes place in the vicinity of the critical wall-normal location where the turbulent mean velocity (thick black curve) equals the wave speed.~\cite{mcksha10} argued that emergence of critical layers is one of the three means of maximizing the Hilbert-Schmidt norm of $H$ (sum of squares of the singular values), i.e. by locally minimizing the term $U (y) - {c}$ that appears in the resolvent operator $R_A$ given in~(\ref{eq.RA}).

According to Taylor's frozen turbulence hypothesis~\citep{tay38}, the flow structures in boundary layers propagate downstream with a speed close to the local mean velocity. Consistent with this hypothesis, figure~\ref{fig.Euu-Rm2-kzkx2-vs-yp-Ux-R30000-sigma1-normalized-log10} shows that among all the waves with arbitrary streamwise and spanwise wavelengths at the wall-normal location $y$, the ones with critical speed ${c} = U (y)$ are the most highly amplified. This provides strong evidence for the importance of critical layers in amplification of flow disturbances. In addition, figure~\ref{fig.Euu-Rm2-kzkx2-vs-yp-Ux-R30000-sigma1-normalized-log10} shows that the scatter in the energetic wave speeds increases as the peak of energy density approaches the wall. This agrees with the practical observation that Taylor's hypothesis yields inaccurate energy spectra close to the wall; see, for example,~\cite{kimhus93,moncho09,deljim09,lehguamck11}.

\moa{
\subsection{Requirement for geometric self-similarity of the resolvent modes}
\label{sec.Req-self-similar-log}
}

\moa{We show that a necessary condition for existence of geometrically self-similar resolvent modes is the presence of a logarithmic region in the turbulent mean velocity. The boundary conditions in the inhomogeneous direction $y$, the wall-normal symmetry relative to the center plane, and the presence of $y$-dependent coefficients, e.g. $U(y) - c$, in the resolvent pose limitations on wall-normal scaling of the transfer function. As discussed later in~\S~\ref{sec.summary-R-scale}, the first two limitations are removed owing to the critical behavior of the resolvent modes, cf.~\S~\ref{sec.self-sim-req}, requiring that the resolvent modes have a zero support near the walls and the center plane. The third limitation concerns with scalability of $U(y) - {c}$, $U'(y)$, and $U''(y)$ in the resolvent, cf.~(\ref{eq.RA}), and reduces to identifying the necessary conditions under which 
	\be
	U(y) - c 
	\; = \;
	g_1 (y/y_c),
	\ee 
for some functions $U (y)$ and $g_1 (y)$ and some scale $y_c$ to be determined. Let the relationship between $c$ and $y_c$ be governed by $c = g_2 (y_c)$. Then, we seek the functions $U$, $g_1$, $g_2$, and the scale $y_c$ such that
	\be
	U (y)
	\, - \,
	g_2 (y_c)
	\; = \;
	g_1 (y/y_c). 
	\label{eq.log-req}
	\ee
It follows from~(\ref{eq.log-req}) that $g_2 (y) = U (y) - g_1 (1)$, $g_1 (y) = U (y) - g_2 (1)$, and $g_2 (1) = U (1) - g_1 (1)$. Therefore,~(\ref{eq.log-req}) can be rewritten as
	$
	U (y) 
	\, - \, 
	\big(
	U (y_c) - g_1 (1)
	\big) 
	\; = \;
	U (y/y_c) \, - \,
	\big(
	U (1) - g_1 (1)
	\big)
	$,
or $U (y) - U (y_c) = U (y/y_c) - U (1)$. The only functions that satisfy this constraint are the constant function and the logarithmic function and we have
	\be
	U (y) \; = \; d_1 \, + \, d_2 \log_{d_3} (y),
	~~~
	c \; = \; U (d_4 \,y_c),
	\label{eq.log-result}
	\ee
where $d_1$ to $d_4$ are constants. The wall-normal scale corresponds to the wall-normal location where $c = U (d_4 \, y_c)$. The constant $d_4$ is arbitrary since it enters as a coefficient in front of the scale $y_c$. We select $d_4 = 1$ such that $y_c$ is the critical wall-normal location corresponding to the wave speed $c$. Therefore, in the presence of a logarithmic mean velocity, the height of the resolvent modes scales with $y_c$.}
             	
\moa{
\subsection{Universal modes and self-similar modes}
\label{sec.summary-R-scale}
}

\begin{figure}
    \begin{center}
    \subfigure{\includegraphics[width=0.8\columnwidth]
    {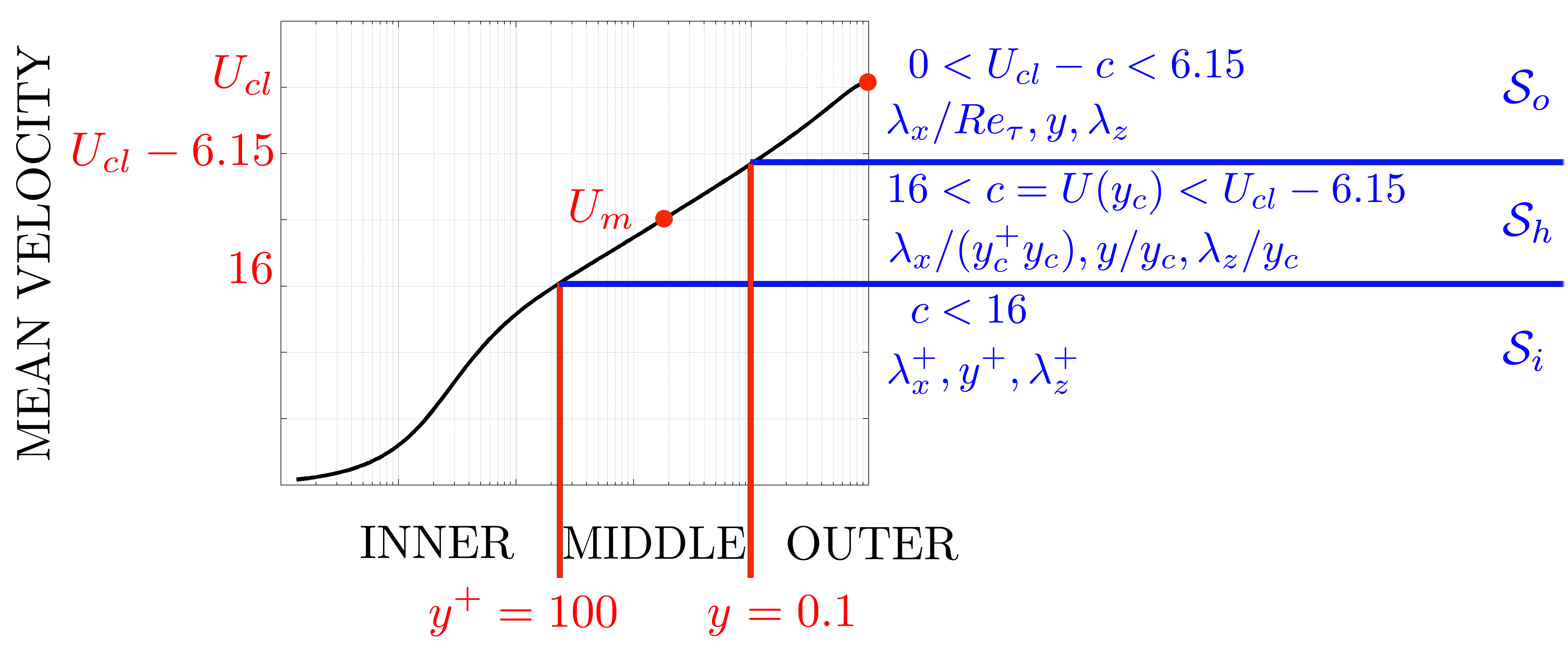}
    }
    \end{center}
    \caption{
    \moa{
Schematic of the different regions of the mean velocity and the associated classes of induced scales on the propagating waves. The mean velocity is denoted by $U_{cl}$ in the center plane and by $U_m$ in the geometric mean of the middle region. The inner, self-similar, and outer classes of modes are denoted by $\cS_i$, $\cS_h$, and $\cS_o$, respectively. See also table~\ref{table.subsets-R-scales} and figures~\ref{fig.U-scaling} and~\ref{fig.self-similar-sigma1-u1}.
    }
    }
    \label{fig.summary-U-scales}
    \end{figure}

\begin{figure}
    \begin{center}
    \begin{tabular}{cc}
    \subfigure{\includegraphics[width=0.45\columnwidth]
    {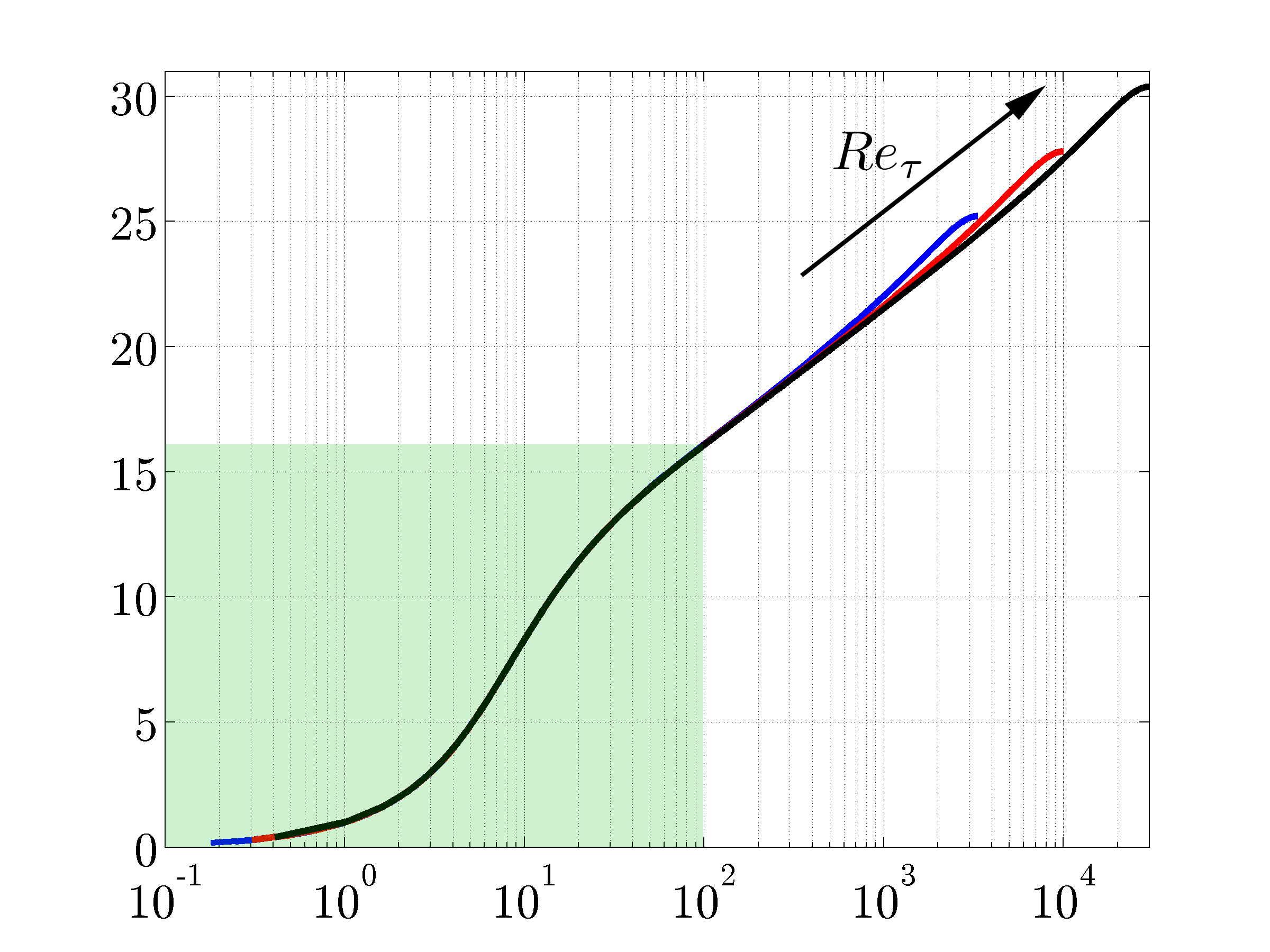}
    \label{fig.U-yp}}
    &
    \subfigure{\includegraphics[width=0.45\columnwidth]
    {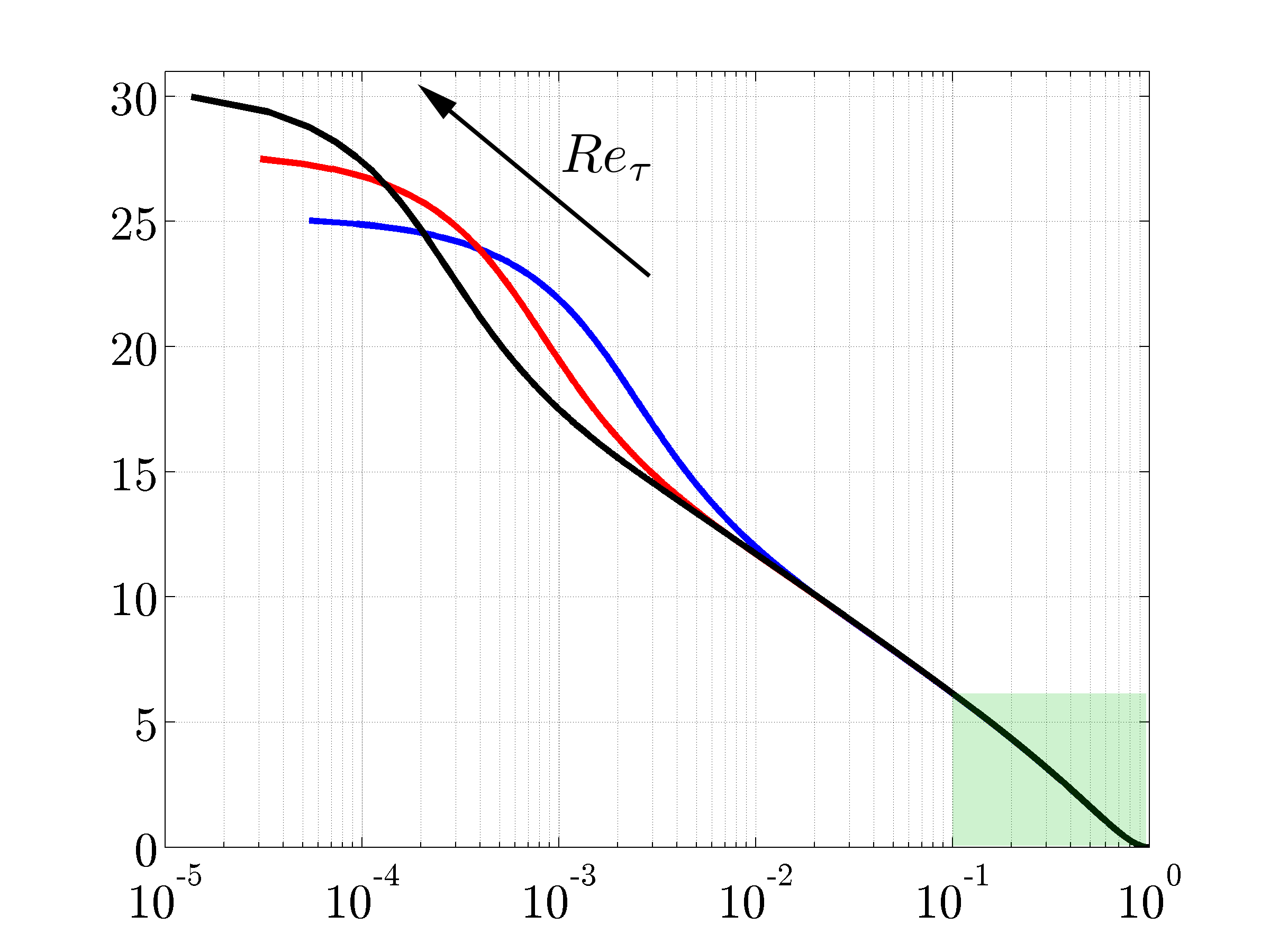}
    \label{fig.Udef-y}}
    \\[-3cm]
    \hskip-5.8cm
    \begin{turn}{90}
    $~~~~U$
    \end{turn}
    &
    \hskip-5.8cm
    \begin{turn}{90}
    $U_{cl} - U$
    \end{turn}
    \\[1.6cm]
    \tc{black}{$y^+$}
    &
    \tc{black}{$y$}
    \\[0.1cm]
    $(a)$
    &
    $(b)$
    \\[0.2cm]
    $\cS_{i}$
    &
    $\cS_{o}$
    \\[-0.18cm]
    \subfigure{\includegraphics[width=0.45\columnwidth]
    {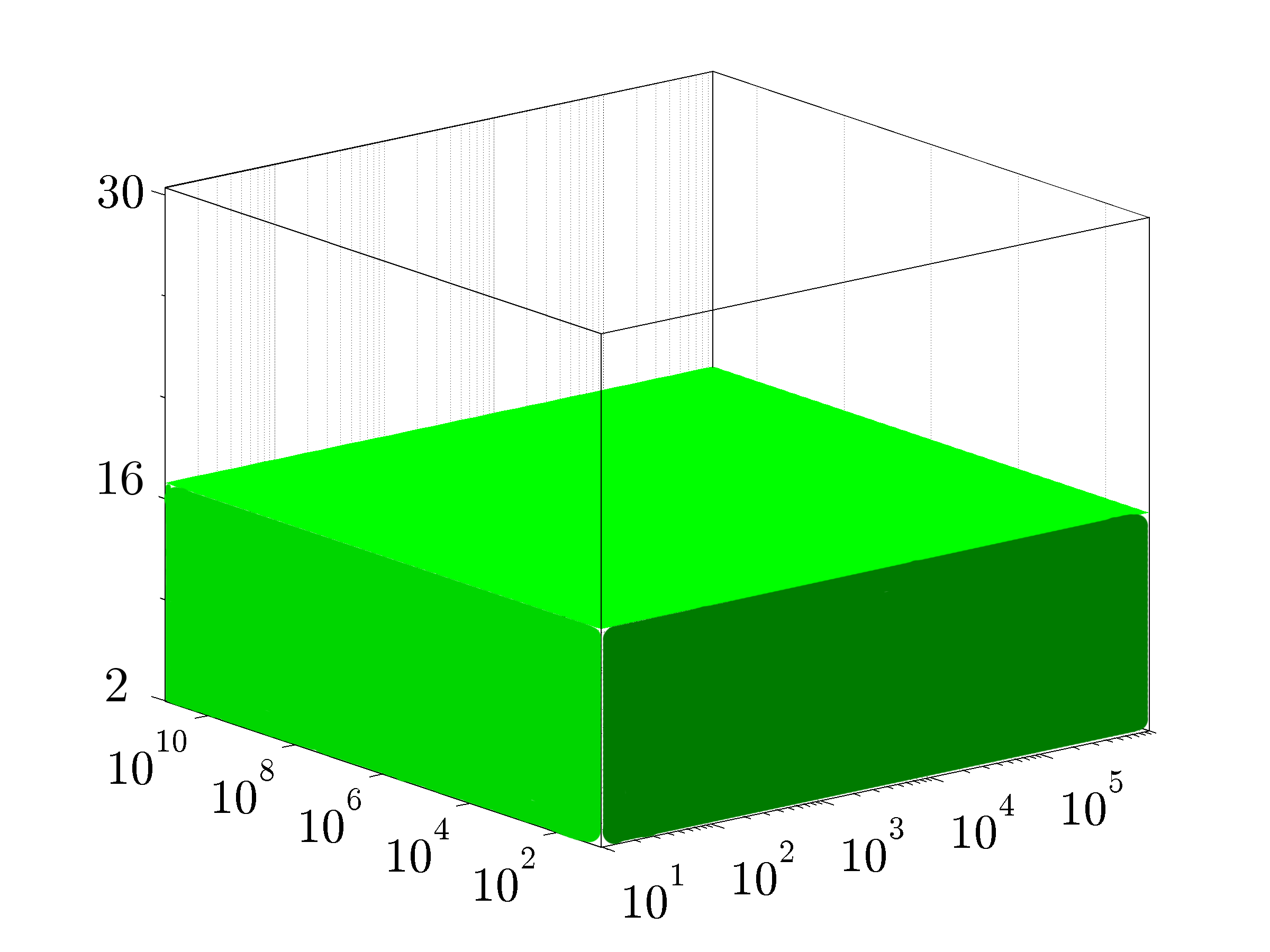}
    \label{fig.S-i}}
    &
    \subfigure{\includegraphics[width=0.45\columnwidth]
    {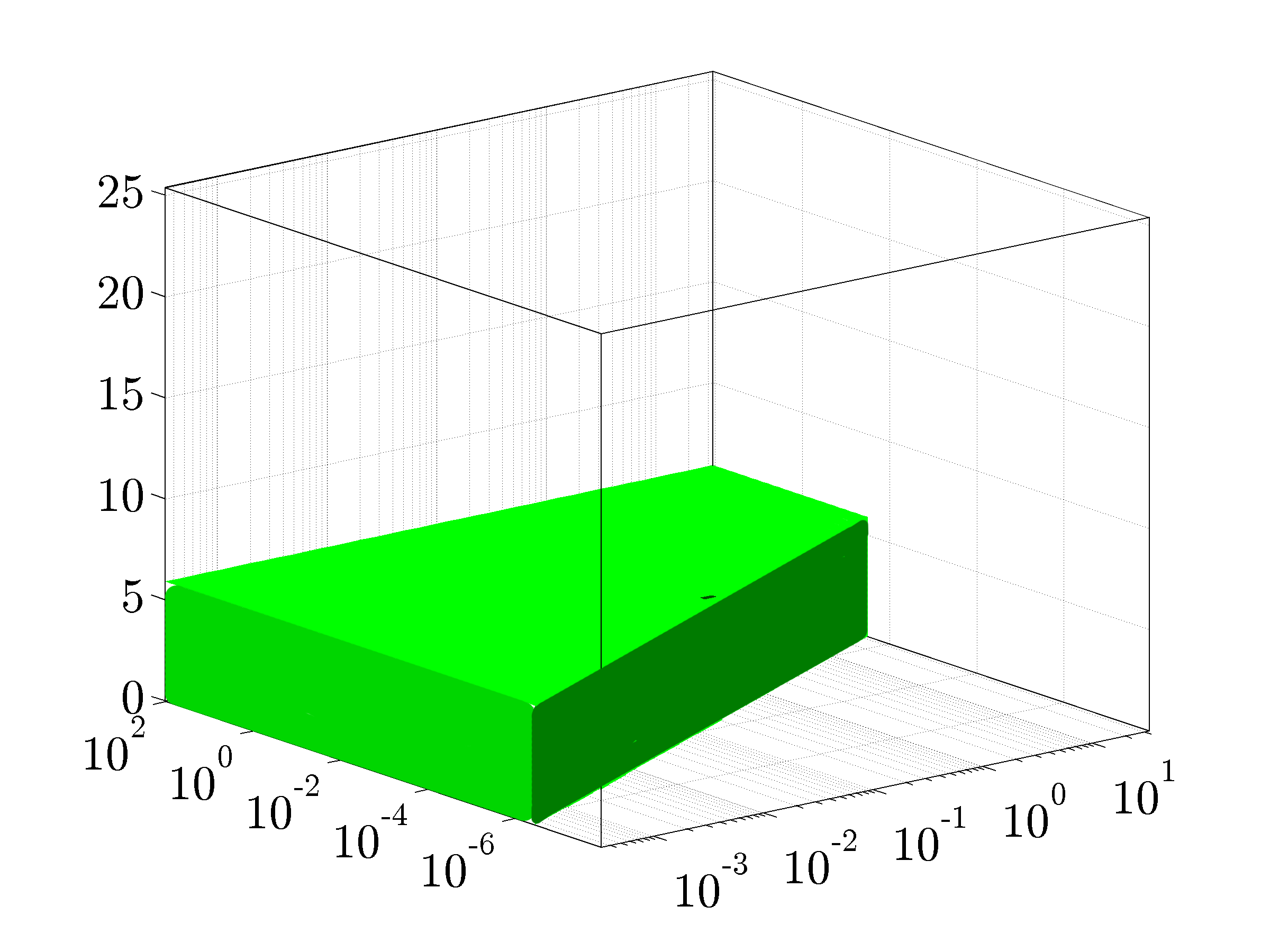}
    \label{fig.S-oy}}
    \\[-3cm]
    \hskip-5.8cm
    \begin{turn}{90}
    $~~~~c$
    \end{turn}
    &
    \hskip-5.8cm
    \begin{turn}{90}
    $U_{cl} - c$
    \end{turn}
    \\[1.6cm]
    \hskip-0.2cm
    \tc{black}{$\lambda_x^+$}
    \hskip2.4cm
    \tc{black}{$\lambda_z^+$}
    &
    \hskip-0.9cm
    \tc{black}{$\lambda_x/{Re}_\tau$}
    \hskip2cm
    \tc{black}{$\lambda_z$}
    \\[0.1cm]
    $(c)$
    &
    $(d)$
    \end{tabular}
    \end{center}
    \caption{
    \moa{
    (Color online)
    (a) The turbulent mean velocity $U (y^+)$ and
    (b) the defect velocity $U_{cl} - U (y)$ relative to the centerline,
    for ${Re}_\tau = 3333$ (blue), ${Re}_\tau = 10000$ (red), and ${Re}_\tau = 30000$ (black). The arrows indicate increase in the Reynolds number.
    Notice that the shaded regions are invariant with ${Re}_\tau$.
    (c) The inner class $\cS_{i}$ and
    (d) the outer class $\cS_{o}$ of wave parameters $(\lambda_x,\lambda_z,{c})$ that induce universal behavior with ${Re}_\tau$ on the low-rank approximation of $H$.
    $\cS_{o}$ is obtained for ${Re}_\tau > {Re}_{\tau,\mathrm{min}} = 3333$.
    }}
    \label{fig.U-scaling}
    \end{figure}

We start by reviewing the universal behavior of the turbulent mean velocity. This is a prerequisite to studying the universality of the principal propagating waves since the latter holds for critical modes only, as discussed in~\S~\ref{sec.self-sim-req}. In the commonly accepted picture~\citep{col56}, the mean velocity is divided into inner, logarithmic, and outer regions
	\be
	U
	\; = \;
	B (y^+)
	\, + \,
	(1/\kappa)
	\,
	\log (y^+)
	\, + \,
	(2 \Pi/\kappa)
	\,
	W (y),
	\label{eq.U-Coles}
	\ee
where $B$ is the inner-scaled wall function, $\Pi$ is the wake factor, $W$ is the outer-scaled wake function, and $\kappa$ is the von K{\'a}rm{\'a}n's constant also appearing in~(\ref{eq.U-nuT}). Consequently, $U - {c}$ is universal with ${Re}_\tau$ for certain intervals of wave speed and appropriate wall-normal scales; see figure~\ref{fig.summary-U-scales}. Figure~\ref{fig.U-yp} shows that $U (y^+) - c$ is universal for $y^+ \lesssim 100$ and fixed $c \lesssim 16$ (inner region). As shown in figure~\ref{fig.Udef-y}, the function $U (y) - c$ is universal for $y \gtrsim 0.1$ and fixed defect wave speeds relative to the centerline $0 \lesssim U_{cl} - c \lesssim 6.15$, with $U_{cl} = U (y = 1)$ (outer region). 

\moa{
The gap between the inner and outer regions \moar{of the turbulent mean velocity} is bridged by a middle region between $y^+ = 100$ and $y = 0.1$. There is an abundance of numerical and experimental evidence that support the presence of a logarithmic turbulent mean velocity in this region~\cite[see, for a recent summary,][]{smimckmar11}. In this study, we consider a logarithmic law throughout the middle region corresponding to $16 \leq U \leq U_{cl} - 6.15$, and note that recent experiments suggest that the lower bound on the logarithmic region depends on Reynolds number: $y^+ \sim Re_\tau^{1/2}$, see e.g.~\cite{marmonhulsmi13}. 
}

\moa{
The existence, at least approximately, of a logarithmic region in $U$ satisfies the necessary conditions in~\S~\ref{sec.Req-self-similar-log} for presence of self-similar resolvent modes. Owing to the locality of resolvent modes around the critical layer, the waves with speed $16 \leq c \leq U_{cl} - 6.15$ are at least one decade away from the walls and the center plane and the boundary effects are negligible. This eliminates the first two limitations for presence of self-similar modes, cf.~\S~\ref{sec.Req-self-similar-log}. The constants $d_1 = 5.28$, $d_2 = 1/\kappa$, and $d_3 = \mre$ in $U$ given by~(\ref{eq.log-result}) are obtained upon direct comparison with~(\ref{eq.U-Coles}).
}
        
\moa{
Associated with each region of the mean velocity, there is a class of wave parameters for which the low-rank approximation of $H$ exhibits either universal behavior with ${Re}_\tau$ or self-similar behavior with distance from the wall; see tables~\ref{table.subsets-R-scales} and~\ref{table.svd-R-scales} for a summary. As illustrated in figure~\ref{fig.summary-U-scales}, these classes are primarily distinguished by the wave speed. The identified scales represent inherent features of the linear mechanisms in the NSE and are not arbitrary: (i) The wall-normal length scale is inherited from the turbulent mean velocity at the critical layer, and (ii) the streamwise and spanwise length scales are determined from the balance between the viscous dissipation term, $(1/Re_\tau) \Delta$, and the mean advection terms, e.g. $\mri \kappa_x (U - c)$, in the resolvent in~(\ref{eq.RA}). In addition, the magnitude of the singular values and singular functions scale uniquely in each class of wave parameters, which induces unique scales on the premultiplied streamwise energy density $E_{uu} (y; \kappa_x, \kappa_z, c)$. Next, we separately discuss each class and refer the reader to Appendices~\ref{sec.inner-scales},~\ref{sec.outer-scales}, and~\ref{sec.self-similar-log-scales} for detailed derivation of the scales.
}

\subsubsection{The universal inner class $\cS_{i}$}
\label{sec.summary-R-scale-i}

For wave speeds in the inner region of the turbulent mean velocity, universality of $H$ requires constant $\lambda_x^+$, $y^+$, $\lambda_z^+$, and $c$; cf. Appendix~\ref{sec.inner-scales}, table~\ref{table.subsets-R-scales}, and figure~\ref{fig.S-i}. As a result, the time $T_c = \lambda_x/c$ over which the wave convects downstream for one wavelength relative to the wall reduces with ${Re}_\tau$ and the convective frequency $\omega_c = 2\pi/T_c$ increases with ${Re}_\tau$. In other words, a truly inner scale is induced on the length, height, width, and convective time of the waves that correspond to the principal resolvent modes. Therefore, the wall-normal support of the resolvent modes in outer units linearly decreases with ${Re}_\tau$, and the unit energy constraint on the resolvent modes requires that the magnitude of the resolvent modes increase with ${Re}_\tau^{1/2}$. The number of these waves per unit wall-parallel area and time increases with ${Re}_\tau^3$ as their length, width, and convective time decrease with ${Re}_\tau$. Since the singular values of $H$ linearly decrease with ${Re}_\tau$, the overall result is that $E_{uu} (y; \kappa_x, \kappa_z, c)$ increases with ${Re}_\tau^2$; cf. table~\ref{table.svd-R-scales}.

         \begin{table}
         \centering
         \begin{tabular}{lccccrl}
            Class
            \hspace{0.3cm}
            &
            \hspace{0.03cm}
            $x\/$-scale
            \hspace{0.03cm}
            &
            \hspace{0.03cm}
            $y\/$-scale
            \hspace{0.03cm}
            &
            \hspace{0.03cm}
            $z\/$-scale
            \hspace{0.03cm}
            &
            \hspace{0.03cm}
            $t\/$-scale
            \hspace{0.2cm}
            &
            &
            \hspace{0.03cm}
            Subset of wave parameters
            \\[0.2cm] 
            Inner
            &
            $\lambda_x^+$
            &
            $y^+$
            &
            $\lambda_z^+$
            &
            \hspace{0.03cm}
            $T_c^+$
            \hspace{0.2cm}
            &
            $\cS_i:$
            &
            $2 ~\lesssim~ c ~\lesssim~ 16$
            \\[0.3cm]
            Outer
            &
            $\dfrac{\lambda_x}{{Re}_\tau}$
            &
            $y$
            &
            $\lambda_z$
            &
            \hspace{0.03cm}
            ${Re}_\tau T_{cl}$
            \hspace{0.2cm}
            &
            $\cS_{o}:$
            &
            $\left\{
            \ba{l}
            0 ~\lesssim~ U_{cl} - {c} ~\lesssim~ 6.15
	      \\[0.2cm]
	      \lambda_x/\lambda_z ~\gtrsim~ \gamma {Re}_\tau/{Re}_{\tau,\mathrm{min}}
	      \ea
	      \right.$
	      \\[0.5cm]
            Self-similar
            &
            $\dfrac{\lambda_x}{y_c^+ y_c}$
            &
            $\dfrac{y}{y_c}$
            &
            $\dfrac{\lambda_z}{y_c}$
            &
            $-$
            &
            $\cS_{h}:$
            &
            $\left\{
            \ba{l}
            16 ~\lesssim~ {c} ~\lesssim~ U_{cl} - 6.15
            \\[0.2cm]
            c = U(y_c) = B + (1/\kappa) \log (y_c^+)
	      \\[0.2cm]
	      \lambda_x/\lambda_z ~\gtrsim~ \gamma
	      \ea
	      \right.$
	      \\[0.7cm]
            Middle
            &
            $\lambda_x$
            &
            $\sqrt{y y^+}$
            &
            $\sqrt{\lambda_z \lambda_z^+}$
            &
            \hspace{0.03cm}
            $T_m$
            \hspace{0.2cm}
            &
            $\cS_{m}:$
            &
            $\left\{
            \ba{l}
            \left|U_m - {c}\right| ~\lesssim~ d
	      \\[0.2cm]
	      \lambda_x/\lambda_z ~\gtrsim~ \gamma \sqrt{{Re}_\tau/{Re}_{\tau,\mathrm{min}}}
	      \ea
	      \right.$
         \end{tabular}
         \caption{
         \moa{
         Summary of the length scales and wave speeds for the universal modes of the transfer function $H$. See also figure~\ref{fig.summary-U-scales}.}}
         \label{table.subsets-R-scales}
         \end{table}
         
\subsubsection{The universal outer class $\cS_{o}$}
\label{sec.summary-R-scale-oy}

For wave speeds close to the centerline, universality of $H$ requires constant $\lambda_x/{Re}_\tau$, $y$, $\lambda_z$, and $U_{cl} - c$, such that an aspect ratio constraint $\lambda_x/\lambda_z \gtrsim \gamma {Re}_\tau/{Re}_{\tau,\mathrm{min}}$ is satisfied (a conservative value for $\gamma$ is $\sqrt{10}$); cf. Appendix~\ref{sec.outer-scales}, table~\ref{table.subsets-R-scales}, and figure~\ref{fig.S-oy}. As a result, the time $T_{cl} = \lambda_x/(U_{cl} - c)$ over which the wave convects upstream for one wavelength relative to an observer with speed $U_{cl}$ increases with ${Re}_\tau$ and the convective frequency $\omega_{cl} = 2\pi/T_{cl}$ decreases with ${Re}_\tau$. In addition, the aspect ratio $\lambda_x/\lambda_z$ of the universal waves increases as ${Re}_\tau$. This explains why universality for this class holds for the waves with aspect ratios larger than a threshold: As the aspect ratio of the resolvent modes increases with Reynolds number, the Laplacian operator in the resolvent becomes independent of $\kappa_x$. Therefore, the necessary condition for the Laplacian to be universal with ${Re}_\tau$ is that $\kappa_z$ dominates $\kappa_x$ even for the smallest Reynolds number ${Re}_{\tau,\mathrm{min}}$ that is considered. This poses the above-mentioned aspect ratio constraint on the universal waves. The magnitude of resolvent modes is independent of ${Re}_\tau$ since the resolvent modes scale with outer units in the wall-normal direction. The number of waves per unit area and time decreases with ${Re}_\tau^2$ since their length and convective time increase with ${Re}_\tau$. The singular values increase with ${Re}_\tau^2$ and the overall result is that the energy density $E_{uu} (y; \kappa_x, \kappa_z, c)$ increases with ${Re}_\tau^2$; cf. table~\ref{table.svd-R-scales}.

The waves in the outer class asymptotically approach the streamwise constant fluctuations, i.e. $\kappa_x = 0$, as ${Re}_\tau$ increases. These infinitely long fluctuations exhibit the largest linear transient growth in response to initial perturbations in laminar~\citep{gus91,butfar92,redhen93} and turbulent~\citep{butfar93,alajim06,pujgarcosdep09} flows. In addition, they are the most highly amplified by the linear dynamics in laminar~\citep{farioa93,bamdah01,jovbamJFM05} and turbulent~\citep{hwacos10} flows subject to stochastic disturbances.

The effect of Reynolds number on the streamwise constant fluctuations has been studied in laminar flows. For example,~\cite{gus91} showed that the peak of linear transient growth scales with the square of centerline Reynolds number $Re_{cl} = U_{cl} h/\nu$. For the flow subject to harmonic disturbances,~\cite{jovbamJFM05} showed that the singular values of $H$ increase as $Re_{cl}^2$ when the temporal frequency $\omega$ linearly decreases with $Re_{cl}$. No other scales for the singular values were found since the laminar mean velocity $U/U_{cl} = 2y-y^2$ is universal with Reynolds number throughout the channel. Our study shows that the singular values in the turbulent flow increase quadratically with ${Re}_\tau$ for the waves with defect speeds, $U_{cl} - {c} \lesssim 6.15$, and streamwise wavelengths that linearly increase with ${Re}_\tau$, i.e. $\lambda_x \gtrsim \gamma\, \lambda_z {Re}_\tau/{Re}_{\tau,\mathrm{min}}$; cf. table~\ref{table.svd-R-scales}.

         \begin{table}
         \centering
         \begin{tabular}{lcccccccccc}
            Class
            \hspace{0.6cm}
            &
            \hspace{0cm}
            Subset
            \hspace{0cm}
            &
            \hspace{0.1cm}
            &
            \hspace{0.15cm}
            $\kappa_x$
            \hspace{0.15cm}
            &
            \hspace{0.15cm}
            $(\mrd/\mrd y)$
            \hspace{0.15cm}
            &
            \hspace{0.15cm}
            $\kappa_z$
            \hspace{0.15cm}
            &
            \hspace{0.3cm}
            $\omega_{i,o,m}$
            \hspace{0.3cm}
            &
            \hspace{0cm}
            &
            \hspace{0.15cm}
            $\sigma_1$
            \hspace{0.15cm}
            &
            \hspace{0.15cm}
            $u_1$
            \hspace{0.15cm}
            &
            \hspace{0.15cm}
            $E_{uu}$
            \hspace{0.15cm}
            \\[0.2cm] 
            Inner
            &
            $\cS_i$
            &
            &
            ${Re}_\tau$
            &
            ${Re}_\tau$
            &
            ${Re}_\tau$
            &
            ${Re}_\tau$
            &
            &
            ${Re}_\tau^{-1}$
            &
            ${Re}_\tau^{1/2}$
            &
            ${Re}_\tau^{2}$
            \\[0.2cm]
            Outer
            &
            $\cS_{o}$
            &
            &
            ${Re}_\tau^{-1}$
            &
            $1$
            &
            $1$
            &
            ${Re}_\tau^{-1}$
            &
            &
            ${Re}_\tau^{2}$
            &
            $1$
            &
            ${Re}_\tau^{2}$
            \\[0.2cm]
            Self-similar
            &
            $\cS_{h}$
            &
            &
            $(y_c^+ y_c)^{-1}$
            &
            $y_c^{-1}$
            &
            $y_c^{-1}$
            &
            $-$
            &
            &
            $(y_c^+)^2 y_c$
            &
            $y_c^{-1/2}$
            &
            ${Re}_\tau^{2}$
            \\[0.2cm]
            Middle
            &
            $\cS_{m}$
            &
            &
            $1$
            &
            ${Re}_\tau^{1/2}$
            &
            ${Re}_\tau^{1/2}$
            &
            $1$
            &
            &
            ${Re}_\tau^{1/2}$
            &
            ${Re}_\tau^{1/4}$
            &
            ${Re}_\tau^{2}$
         \end{tabular}
         \caption{
         \moa{
         Summary of the growth/decay rates (with respect to $Re_\tau$ or $y_c$) of the wall-parallel wavenumbers, the wall-normal derivative, the convective frequency, the principal singular value and the principal streamwise singular function of $H$, and the premultiplied three-dimensional streamwise energy density for the classes of universal waves outlined in table~\ref{table.subsets-R-scales}.}}
         \label{table.svd-R-scales}
         \end{table}
          
\moa{
\subsubsection{The geometrically self-similar class $\cS_{h}$}
\label{sec.summary-R-scale-i}
}

\moa{
The logarithmic region of the turbulent mean velocity yields a hierarchy of geometrically self-similar resolvent modes that are uniquely parameterized by the critical wall-normal distance $y_c$, i.e. $c = U(y_c)$; see Appendix~\ref{sec.self-similar-log-scales} for derivation. As summarized in table~\ref{table.subsets-R-scales}, the height and width of the self-similar modes scale with $y_c$ and their length with $y_c^+ y_c$. In addition, the self-similar modes satisfy an aspect ratio constraint, $\lambda_x/\lambda_z \gtrsim \gamma$, where a conservative value for $\gamma$ is $\sqrt{10}$. This agrees with the observation of~\cite{hwacos10} that the optimal responses were approximately similar for $\kappa_x \ll \kappa_z$. Notice that the difference between the streamwise scaling of the self-similar resolvent modes $\lambda_x \sim y_c^+ y_c$ and the scaling $\lambda_x \sim y$ chosen in original developments of the attached-eddy hypothesis~\citep{tow76,percho82} does not contradict the philosophy of self-similar attached eddies, i.e. the resolvent modes are still self-similar. 

Any hierarchy is a subset of $\cS$ and can be described by a representative mode with $\lambda_{x,r}$, $\lambda_{z,r}$, and $c_r = U (y_{c_r})$ 
	\be
	\cS_h (\lambda_{x,r}, \lambda_{z,r}, c_r)
	\; = \;
	\Bigg\{
	(\lambda_x,\lambda_z,{c})
	~|~
	\left\{
	\ba{l}
	\lambda_x = \lambda_{x,r} \Big(\dfrac{y_c^+ y_c}{y_{c_r}^+ y_{c_r}}\Big),
	\\[0.4cm]
	\lambda_z = \lambda_{z,r} \Big(\dfrac{y_c}{y_{c_r}}\Big),
	\\[0.4cm]
	\,c~\,= B + (1/\kappa) \log y_{c}^+,
	\ea
	\right.
	\dfrac{100}{{Re}_\tau} \leq y_{c_1} \leq y_c \leq 0.1
	\Bigg\}.
	\label{eq.S_h}
	\ee
Here, $y_{c_1}$ is the critical wall-normal location associated with the smallest wave speed $c_1$ above which the aspect ratio constraint is satisfied, see Appendix~\ref{sec.self-similar-log-scales}. 

}

    \begin{figure}
    \begin{center}
    \begin{tabular}{cc}
    \subfigure{\includegraphics[width=0.46\columnwidth]
    {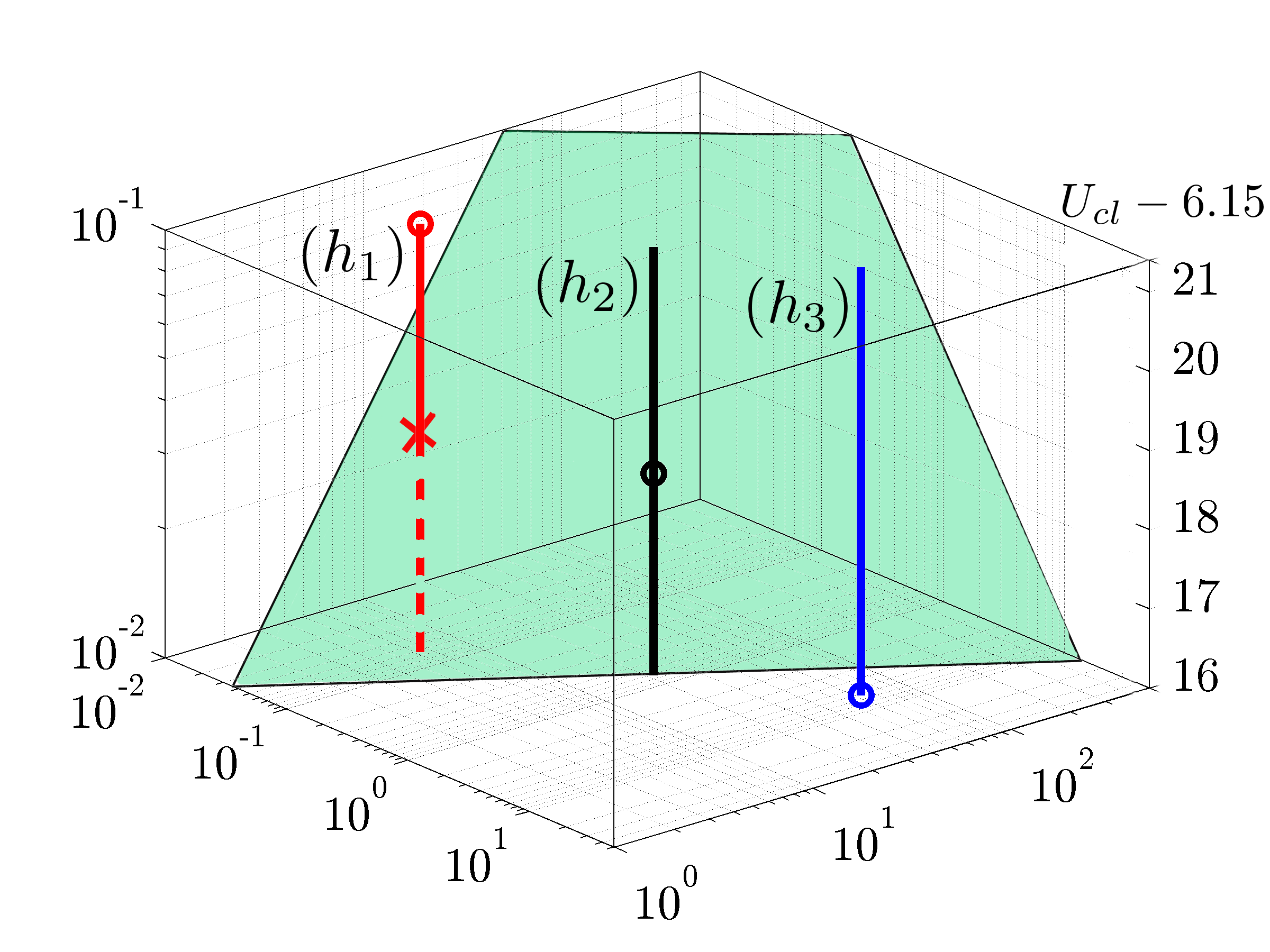}
    \label{fig.log-region-hierarchy-lxlzUpyp-R1e4-kx1-kz10}}
    &
    \hskip0cm
    \subfigure{\includegraphics[width=0.46\columnwidth]
    {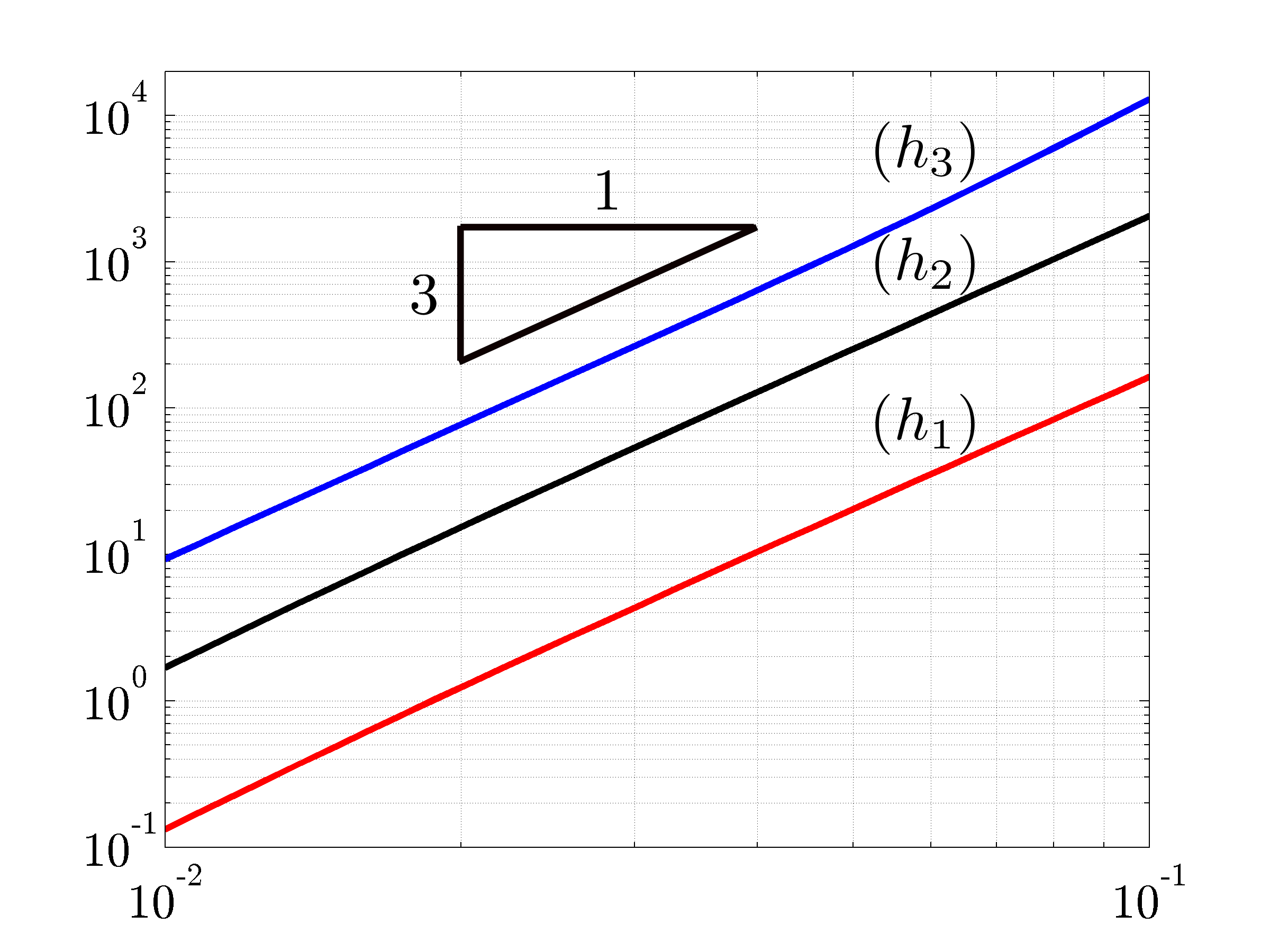}
    \label{fig.log-region-hierarchy-sigma1-vs-yp-ypstar0p1r-Up2_3Uck-yp100_Rb-kx1kz10-R1e4}}
    \\[0.6cm]
    $(a)$
    &
    $(b)$
    \end{tabular}
    \begin{tabular}{c}
    \\[-5cm]
    \begin{tabular}{c}
    \hskip-5.8cm
    \begin{turn}{90}
    \tc{black}{$~~~~~y_c$}
    \end{turn}
    \hskip5.6cm
    \begin{turn}{90}
    \tc{black}{$c = U(y_c)$}
    \end{turn}
    \hskip0.1cm
    \begin{turn}{90}
    \tc{black}{$~~~~~\sigma_1$}
    \end{turn}
    \end{tabular}
    \\[2.1cm]
    \begin{tabular}{c}
    \hskip-2.2cm
    \tc{black}{$\lambda_x/(y_c^+ y_c)$}
    \hskip2.2cm
    \tc{black}{$\lambda_z/y_c$}
    \hskip4.1cm
    \tc{black}{$y_c$}
    \end{tabular}
    \end{tabular}
    \\[-0.2cm]
    \begin{tabular}{cc}
    \subfigure{\includegraphics[width=0.46\columnwidth]
    {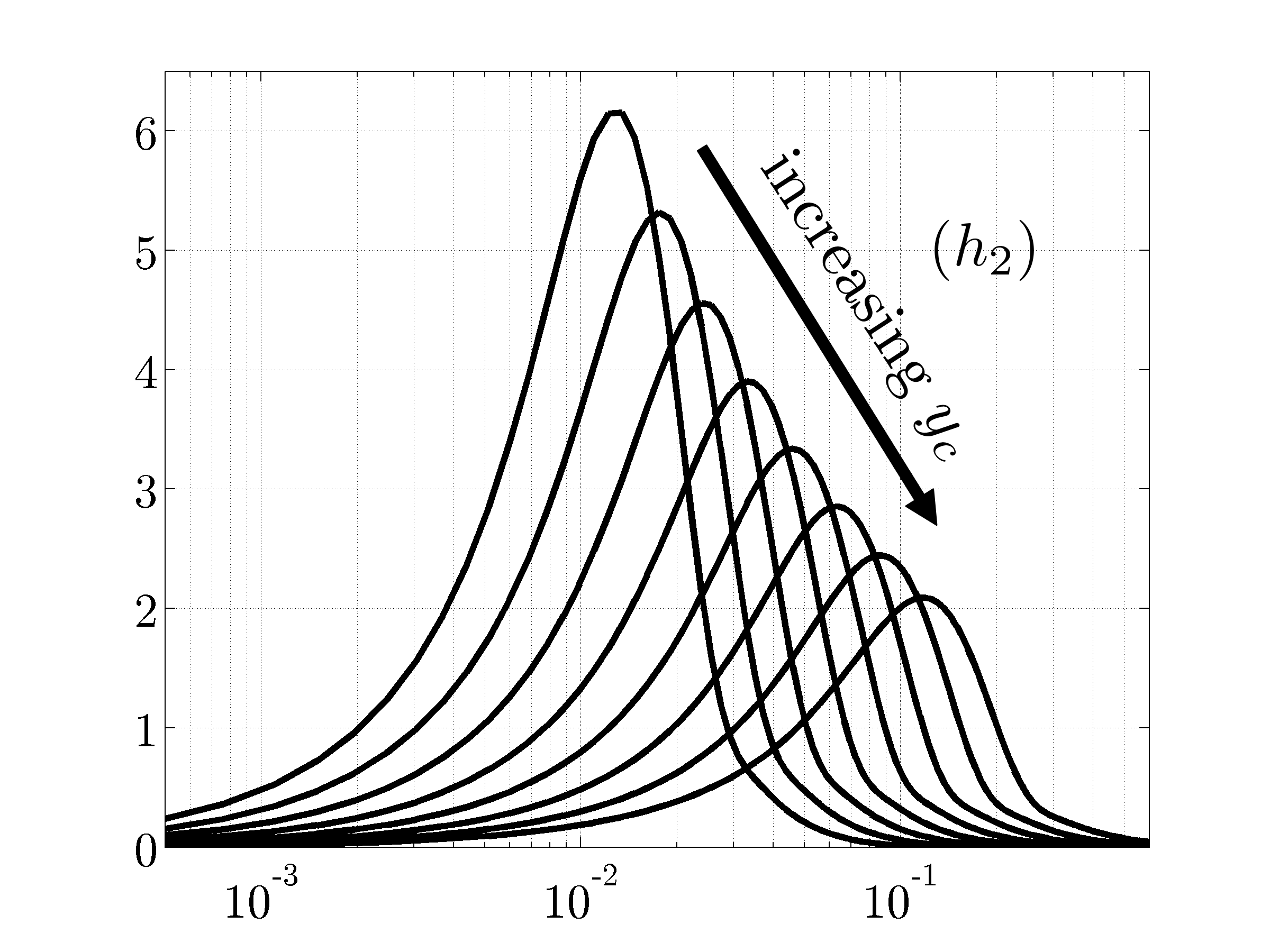}
    \label{fig.log-region-hierarchy-u1-vs-y-Upstar2_3Uc-kx1kz10-R1e4}}
    &
    \hskip0cm
    \subfigure{\includegraphics[width=0.46\columnwidth]
    {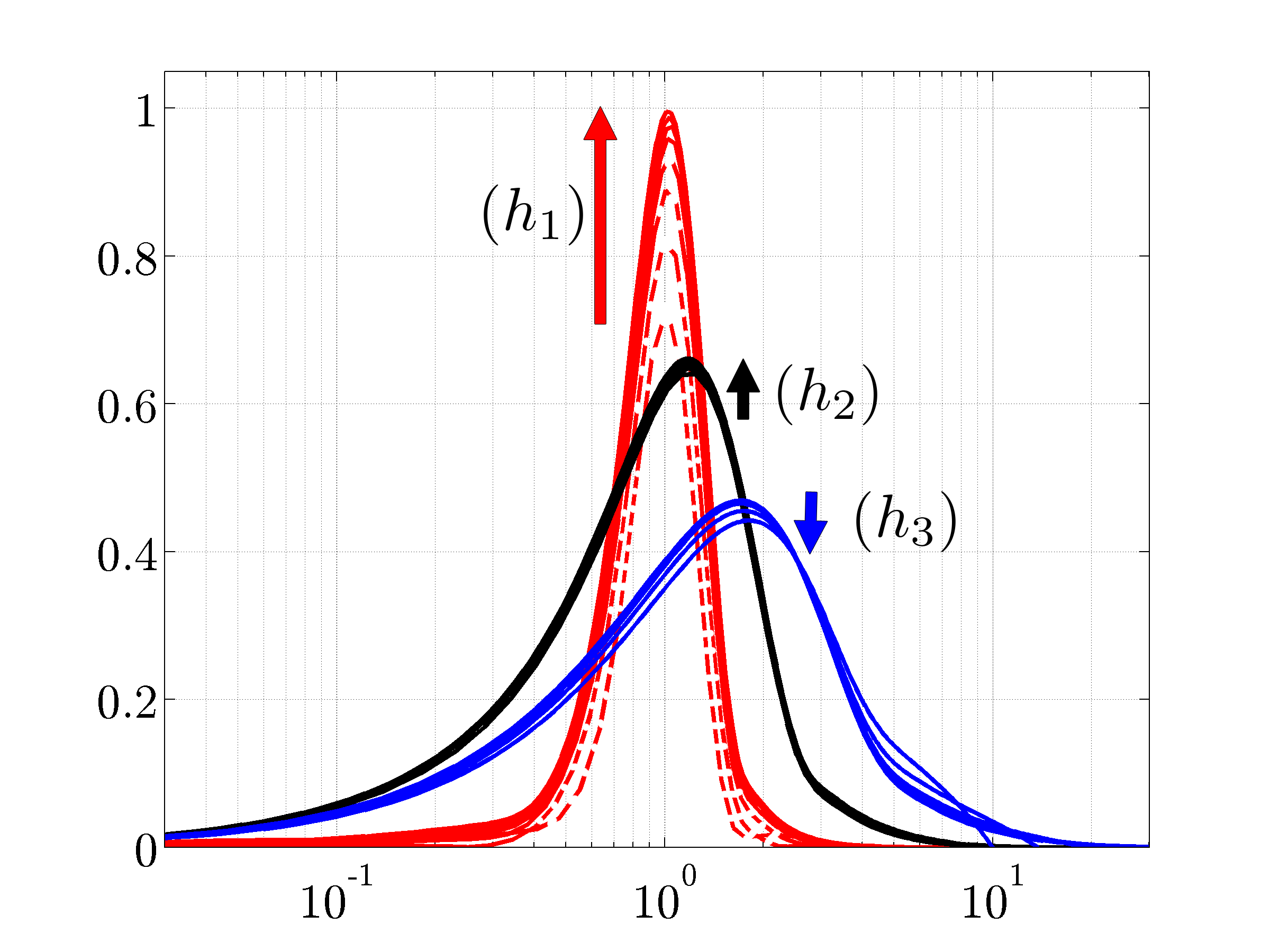}
    \label{fig.log-region-hierarchy-u-similarity-vs-y-yp0p1r-Up2_3Uck-yp100_Rb-kx1kz10-R1e4}}
    \\[0.6cm]
    $(c)$
    &
    $(d)$
    \end{tabular}
    \begin{tabular}{c}
    \\[-5.1cm]
    \begin{tabular}{c}
    \hskip-6.1cm
    \begin{turn}{90}
    \tc{black}{$~~~~~~\abs{u_1}$}
    \end{turn}
    \hskip5.7cm
    \begin{turn}{90}
    \tc{black}{$~~~~\sqrt{y_c} \; \abs{u_1}$}
    \end{turn}
    \end{tabular}
    \\[2cm]
    \begin{tabular}{c}
    \hskip0.2cm
    \tc{black}{$y_c$}
    \hskip6cm
    \tc{black}{$y/y_c$}
    \end{tabular}
    \end{tabular}
    \end{center}
    \caption{
    \moa{
    (Color online)
    (a) The vertical lines are the loci of wave parameters that belong to the hierarchies with representative modes (open circles) $\kappa_{x,r} = 1$, $\kappa_{z,r} = 10$, and $c_r = U_{cl} - 6.15$ ($h_1$, red), $(2/3) U_{cl}$ ($h_2$, black), and $16$ ($h_3$, blue) for ${Re}_\tau = 10000$. The shaded threshold plane corresponds to the wavenumbers with aspect ratio $\lambda_x/\lambda_z = \sqrt{10}$. The modes below this plane do not belong to any hierarchy.
    (b) The principal singular values along the hierarchies in figure~\ref{fig.log-region-hierarchy-lxlzUpyp-R1e4-kx1-kz10}. 
    (c) The principal streamwise resolvent modes that belong to the hierarchy with $c_r = (2/3) U_{cl}$ ($h_2$, black) in figure~\ref{fig.log-region-hierarchy-lxlzUpyp-R1e4-kx1-kz10}.
    (d) The normalized and scaled (according to table~\ref{table.subsets-R-scales}) principal resolvent modes along the hierarchies in figure~\ref{fig.log-region-hierarchy-lxlzUpyp-R1e4-kx1-kz10}.
    The arrows show the direction of increasing $y_c$ with $100/{Re}_\tau \leq y_c \leq 0.1$.
    }}
    \label{fig.self-similar-sigma1-u1}
    \end{figure}  
    
\moa{
The concept of hierarchies is illustrated in figure~\ref{fig.log-region-hierarchy-lxlzUpyp-R1e4-kx1-kz10} where the loci of wave parameters that belong to three demonstrative hierarchies with representative modes marked by open circles are shown. The mode with $\kappa_{x,r} = 1$, $\kappa_{z,r} = 10$, and $c_r = (2/3) U_{cl}$ ($h_2$, black) is representative of the very large-scale motions~\citep{mcksha10}. The representative modes for the other hierarchies have the same wavenumbers but different speeds, i.e. $c_r = 16$ ($h_3$, blue) and $U_{cl} - 6.15$ ($h_1$, red), corresponding to the mean velocity at the upper limit of the inner region and the lower limit of the outer region. Each locus constitutes a vertical line after normalizing the length, width, and height of the modes according to the scales in~(\ref{eq.S_h}) obtained from the resolvent. In fact, the resolvent modes are self-similar along any vertical line as long as $\lambda_x/\lambda_z > \gamma$. The aspect ratio constraint requires that the wave parameters lie above the shaded threshold plane $\lambda_x/\lambda_z = \gamma$ in figure~\ref{fig.log-region-hierarchy-lxlzUpyp-R1e4-kx1-kz10}. For example, the waves corresponding to the dashed segment of the hierarchy with $c_r = U_{cl} - 6.15$ do not belong to any hierarchy.
}

\moa{
Owing to the self-similar behavior, the principal singular values and singular functions of $H$ for all the modes in a given hierarchy can be determined from its representative mode. The principal singular value $\sigma_1$ corresponding to the waves that belong to the hierarchies in figure~\ref{fig.log-region-hierarchy-lxlzUpyp-R1e4-kx1-kz10} are shown in figure~\ref{fig.log-region-hierarchy-sigma1-vs-yp-ypstar0p1r-Up2_3Uck-yp100_Rb-kx1kz10-R1e4}. The singular values grow with $(y_c^+)^2 (y_c)$ as theoretically predicted, cf. table~\ref{table.svd-R-scales}. Figure~\ref{fig.log-region-hierarchy-u1-vs-y-Upstar2_3Uc-kx1kz10-R1e4} shows the principal streamwise resolvent mode $u_1$ corresponding to the hierarchy with $\kappa_{x,r} = 1$, $\kappa_{z,r} = 10$, and $c_r =  (2/3) U_{cl}$ for $100/{Re}_\tau \leq y_c \leq 0.1$. The arrow shows the direction of increasing $y_c$. Normalizing and scaling the resolvent modes according to table~\ref{table.subsets-R-scales} collapses the resolvent modes for different wave speeds; see black curves marked as $h_2$ in figure~\ref{fig.log-region-hierarchy-u-similarity-vs-y-yp0p1r-Up2_3Uck-yp100_Rb-kx1kz10-R1e4}. This figure also shows the scaled resolvent modes corresponding to the hierarchies with $\kappa_{x,r} = 1$, $\kappa_{z,r} = 10$, and $c_r = 16$ ($h_3$, blue) and $U_{cl} - 6.15$ ($h_1$, red). We see that the normalized and scaled resolvent modes lie on the top of each other for the hierarchy with $c_r = 16$. For the hierarchy with $c_r = U_{cl} - 6.15$, the resolvent modes for large $y_c$ collapse on each other while the resolvent modes for small $y_c$ are considerably different. This is expected since the aspect ratios of the modes fall below $\gamma$ as $y_c$ decreases. Notice that for this hierarchy, the modes corresponding to small $y_c$ lie below the threshold plane in figure~\ref{fig.log-region-hierarchy-lxlzUpyp-R1e4-kx1-kz10}. 
}

\vskip0.3cm

\moa{
\subsubsection{The universal middle class $\cS_{m}$}
\label{sec.summary-R-scale-m}
}

\moa{The Reynolds number scaling of the self-similar class depends on the wave speed and is consistent with the inner and outer classes of resolvent modes, cf. tables~\ref{table.subsets-R-scales} and~\ref{table.svd-R-scales}. For example, when the wave speed is fixed as $Re_\tau$ changes, $y_c^+$ remains constant and the inner scale is recovered. When the defect wave speed $U_{cl} - c$ is fixed, $y_c$ remains constant and the outer scale is recovered.} \moar{Consequently, the energy density corresponding to the complete range of wave speeds in the self-similar region, $16 < c < U_{cl} - 6.15$, is centered around the geometric mean of the middle region of the turbulent mean velocity, i.e. $y_m = \sqrt{10/Re_\tau}$. The self-similar class is primarily concerned with geometric self-similarity of the resolvent modes. We next construct a middle class of modes $\cS_m$, a subset of the self-similar class $\cS_h$, with unique Reynolds number scalings.}

The wave speeds in the middle class are confined to $|U_m - c| < d$, with $d$ denoting a radius around $U_m = U (y_m)$. For the resolvent modes in the middle class, universality of $H$ requires constant $\lambda_x$, $\sqrt{y^+ y}$, $\sqrt{\lambda_z^+ \lambda_z}$, and $U_m - c$ such that the aspect ratio constraint $\lambda_x/\lambda_z \gtrsim \gamma \sqrt{{Re}_\tau/{Re}_{\tau,\mathrm{min}}}$ is satisfied; cf. table~\ref{table.subsets-R-scales}. These scales are equal to the geometric mean of the scales in the inner and outer classes, and can also be recovered from the scales of the self-similar class for fixed $\sqrt{y_c^+ y_c}$ as $Re_\tau$ changes. \moar{When only one Reynolds number is considered, we have $Re_{\tau,\mathrm{min}} = Re_\tau$ and the aspect ratio constraints in the self-similar and middle classes are equivalent, i.e. the constraint $\lambda_x/\lambda_z \gtrsim \gamma \sqrt{Re_\tau/Re_{\tau,\mathrm{min}}} = \gamma$ in the middle class is the same as $\lambda_x/\lambda_z \gtrsim \gamma$ in the self-similar class. When the self-similar modes are compared across more than one Reynolds number, the aspect ratio constraint on the middle class is more restrictive. This is because, the modes remain self-similar as the Reynolds number increases. However, they do not remain independent of $Re_\tau$ unless their aspect ratio is larger than $\gamma$ even for the smallest Reynolds number $Re_{\tau,\mathrm{min}}$ that is considered, resulting in the modified aspect ratio constraint $\lambda_x/\lambda_z \gtrsim \gamma \sqrt{Re_\tau/Re_{\tau,\mathrm{min}}}$ for the middle class. This constraint can be obtained similarly to the constraint for the outer class, cf. \S~\ref{sec.summary-R-scale-oy} and Appendix~\ref{sec.outer-scales}.}

The time $T_m = \lambda_x/\abs{U_m - c}$ over which a wave in the middle class convects away for one wavelength relative to an observer with speed $U_m$ remains independent of ${Re}_\tau$ and same does the convective frequency $\omega_m = 2\pi/T_m$. These waves have the same scales as the structures in the meso-layer; see, for example,~\cite{lonche81,afz84,sresah97,weififklemcm05}. The aspect ratio constraint follows from similar arguments to those discussed for the outer class. The magnitude of the corresponding resolvent modes increases with ${Re}_\tau^{1/4}$ because of the unit energy constraint. The number of waves per unit area and time increases with ${Re}_\tau^{1/2}$. In addition, the principal singular values of the waves in $\cS_{m}$ increase with ${Re}_\tau^{1/2}$. The above scales result in growth of $E_{uu} (y; \kappa_x, \kappa_z, c)$ with ${Re}_\tau^2$; cf. table~\ref{table.svd-R-scales}.

\moar{There is a direct relationship between $\cS_m$ and $\cS_h$: The union of the middle class of modes equals the union of the geometrically self-similar modes with speeds $|U_m - c| < d$ and aspect ratio constraint $\lambda_x/\lambda_z \gtrsim \gamma \sqrt{{Re}_\tau/{Re}_{\tau,\mathrm{min}}}$. Since the difference between the middle class and the self-similar class becomes larger as $Re_\tau$ increases, our ongoing research is focused on analytical developments using the scalings of the self-similar class that bridges the gap between the inner and outer classes.}

\subsection{Universality of the streamwise energy density}
\label{sec.R-scale-energy-spectra}

\begin{figure}
    \begin{center}
    \begin{tabular}{ccc}
    $\cS_{i}$
    &
    \hskip-0.6cm
    $\cS_{m}$
    &
    \hskip-0.6cm
    $\cS_{o}$
    \\[-0.1cm]
    \subfigure{\includegraphics[width=0.34\columnwidth]
    {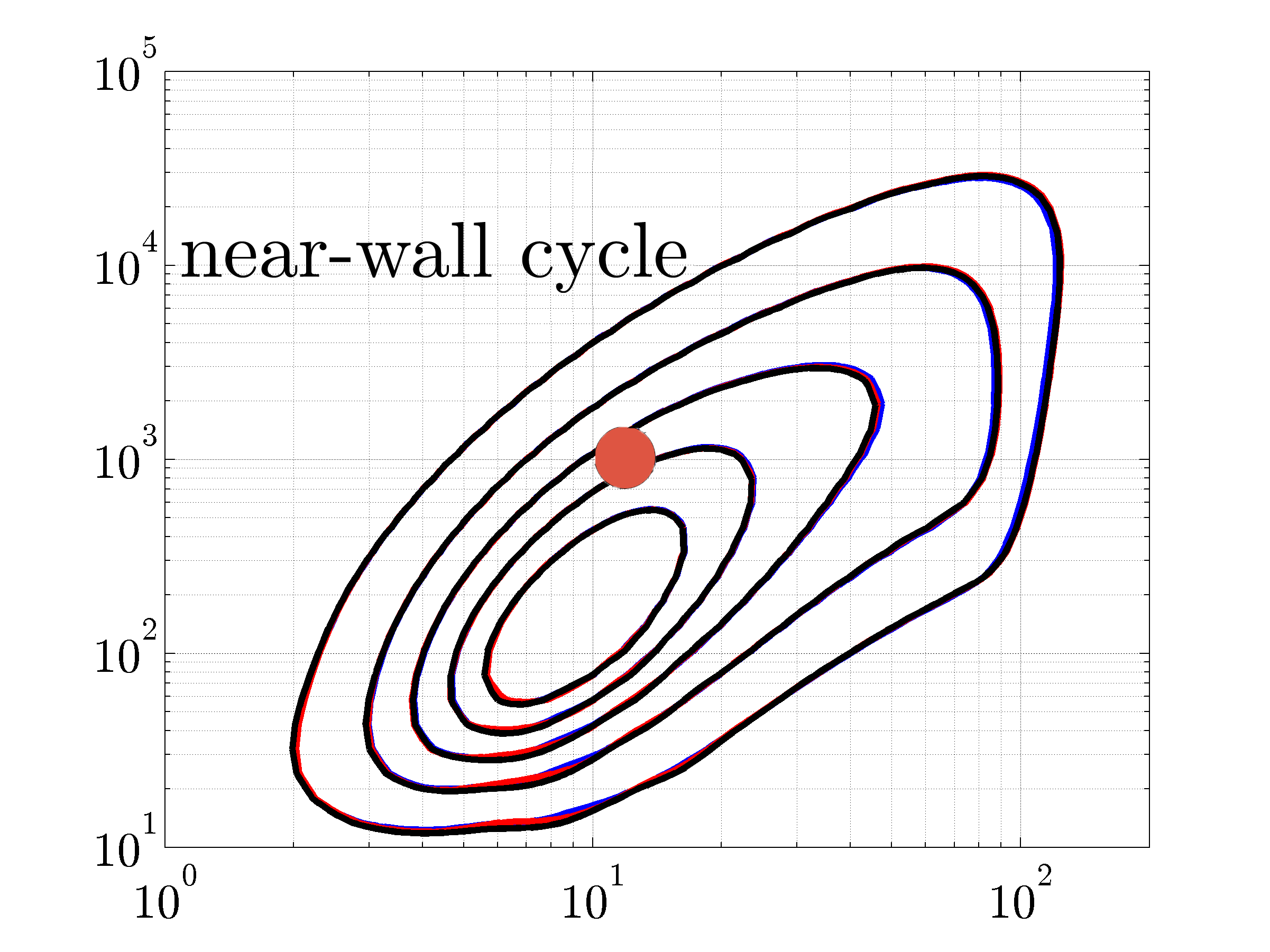}
    \label{fig.Euu-Rm2-kzkx2-vs-yp-lxp-R3333b-10000r-30000k-sigma1-inner-levels-0p05-0p05-0p3}}
    &
    \hskip-0.6cm
    \subfigure{\includegraphics[width=0.34\columnwidth]
    {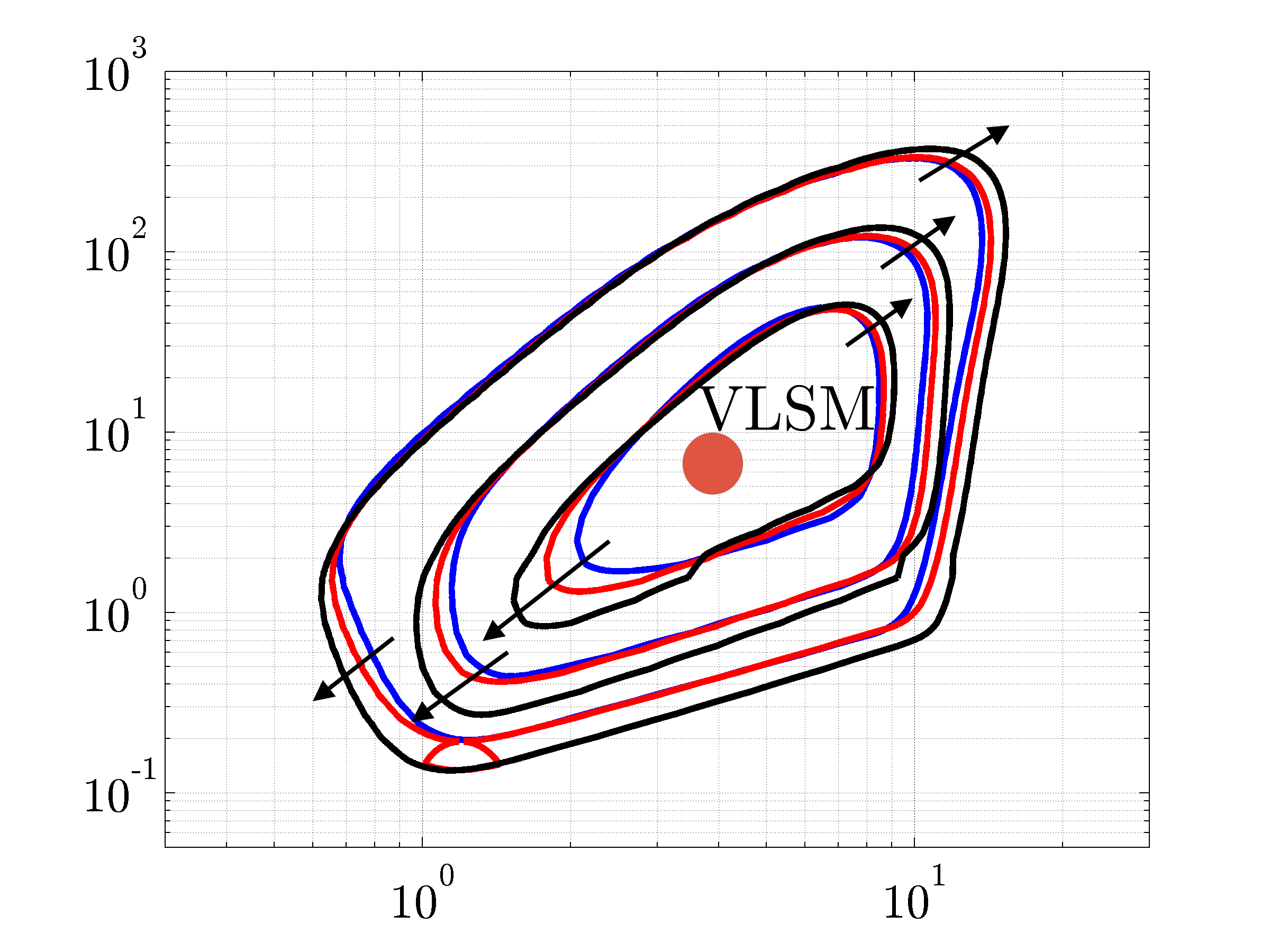}
    \label{fig.Euu-Rm2-kzkx2-vs-yh-lx-R3333b-10000r-30000k-sigma1-outerx-levels-0p05-0p05-0p3}}
    &
    \hskip-0.6cm
    \subfigure{\includegraphics[width=0.34\columnwidth]
    {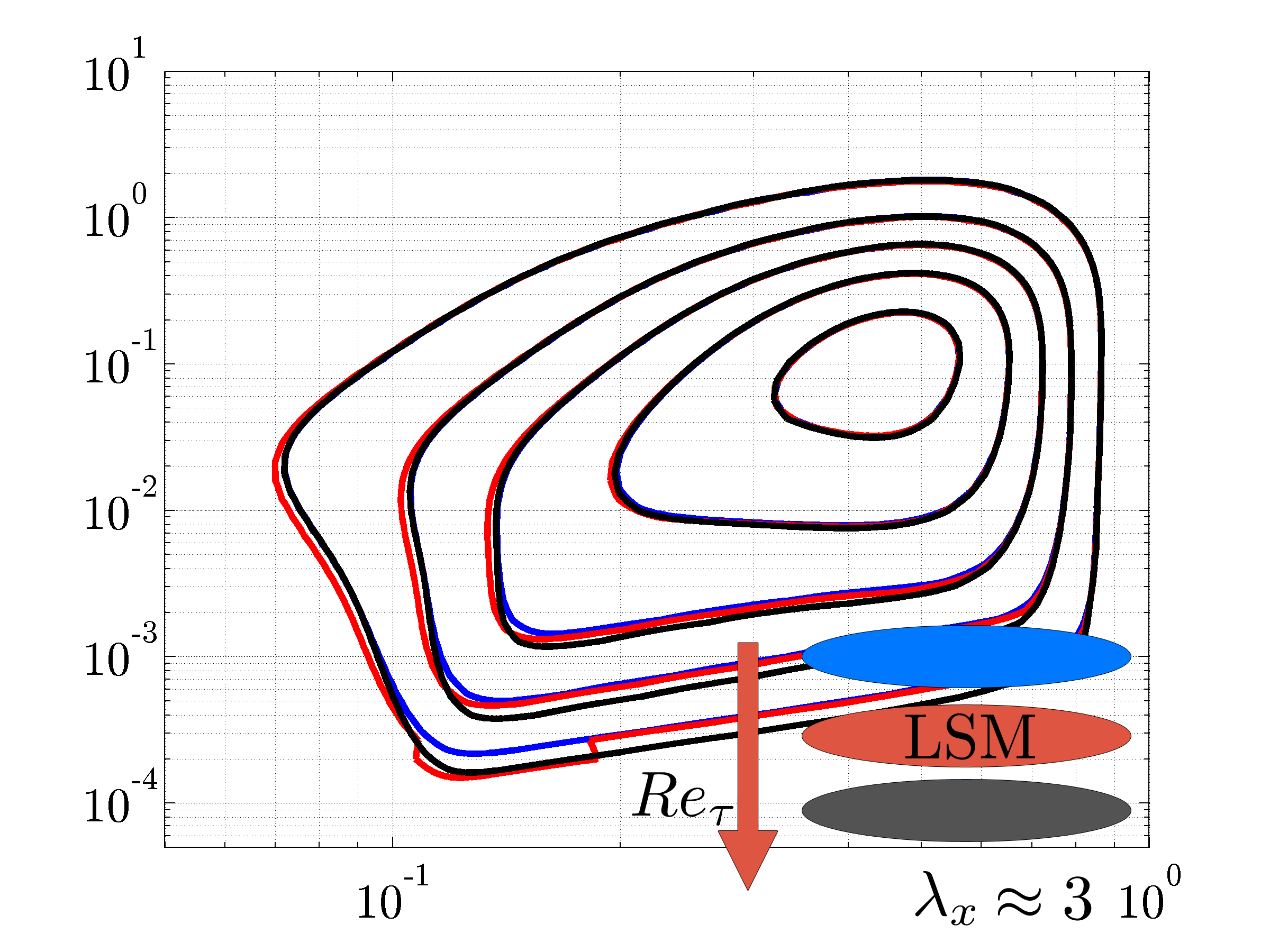}
    \label{fig.Euu-Rm2-kzkx2-vs-y-lxm-R3333b-10000r-30000k-sigma1-outery-levels-0p05-0p05-0p3}}
    \\[-2.5cm]
    \hskip-4.5cm
    \begin{turn}{90}
    $~~\,\lambda_x^+$
    \end{turn}
    &
    \hskip-4.5cm
    \hskip-0.6cm
    \begin{turn}{90}
    $~~~\lambda_x$
    \end{turn}
    &
    \hskip-4.5cm
    \hskip-0.6cm
    \begin{turn}{90}
    $\lambda_x/{Re}_\tau$
    \end{turn}
    \\[1.3cm]
    \tc{black}{$y^+$}
    &
    \hskip-0.6cm
    \tc{black}{$\sqrt{y y^+}$}
    &
    \hskip-0.6cm
    \tc{black}{$y$}
    \\[0.1cm]
    $(a)$
    &
    \hskip-0.6cm
    $(b)$
    &
    \hskip-0.6cm
    $(c)$
    \\[-0.1cm]
    \subfigure{\includegraphics[width=0.34\columnwidth]
    {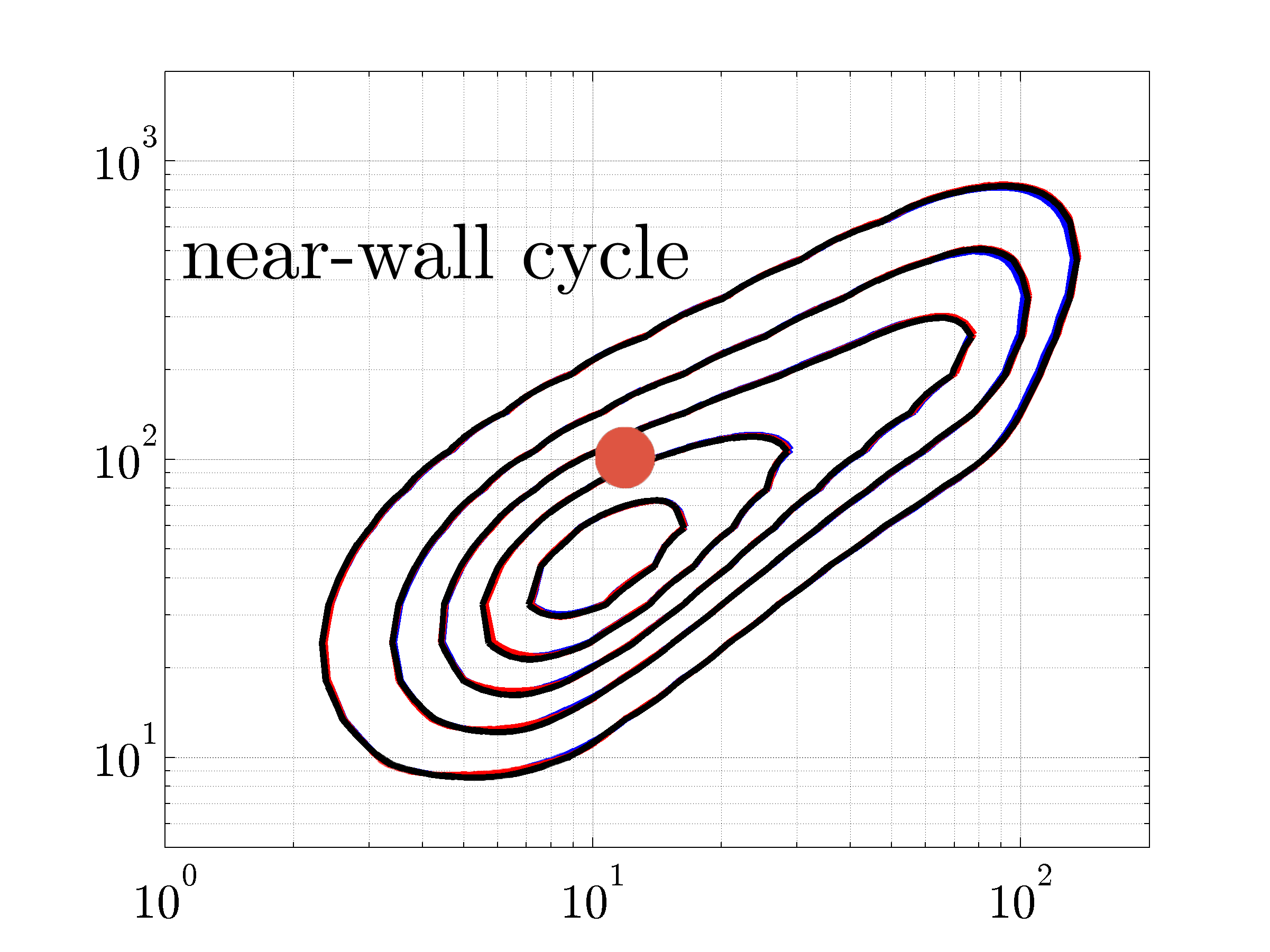}
    \label{fig.Euu-Rm2-kzkx2-vs-yp-lzp-R3333b-10000r-30000k-sigma1-inner-levels-0p1-0p1-1}}
    &
    \hskip-0.6cm
    \subfigure{\includegraphics[width=0.34\columnwidth]
    {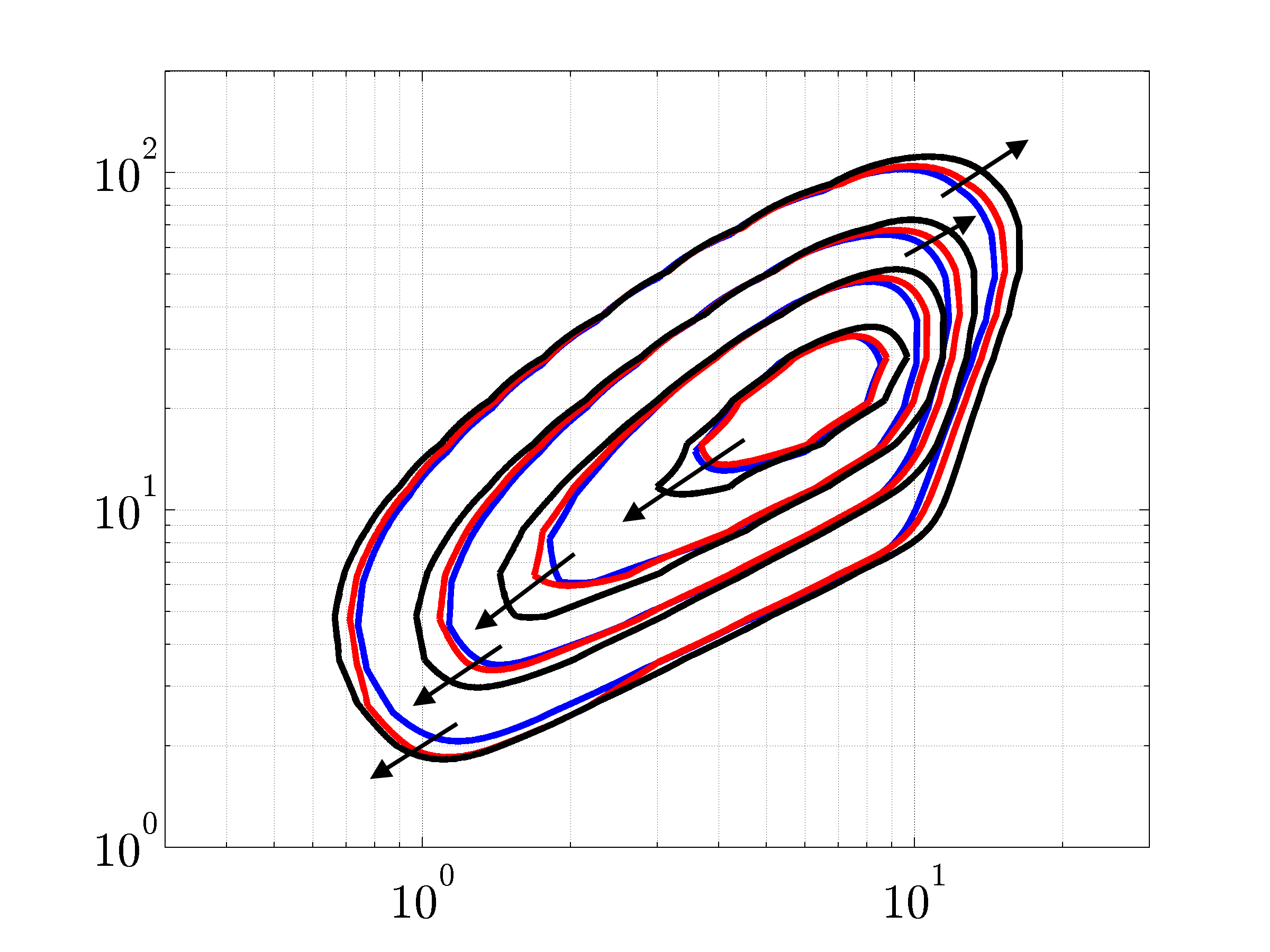}
    \label{fig.Euu-Rm2-kzkx2-vs-yh-lzh-R3333b-10000r-30000k-sigma1-outerx-levels-0p1-0p1-1}}
    &
    \hskip-0.6cm
    \subfigure{\includegraphics[width=0.34\columnwidth]
    {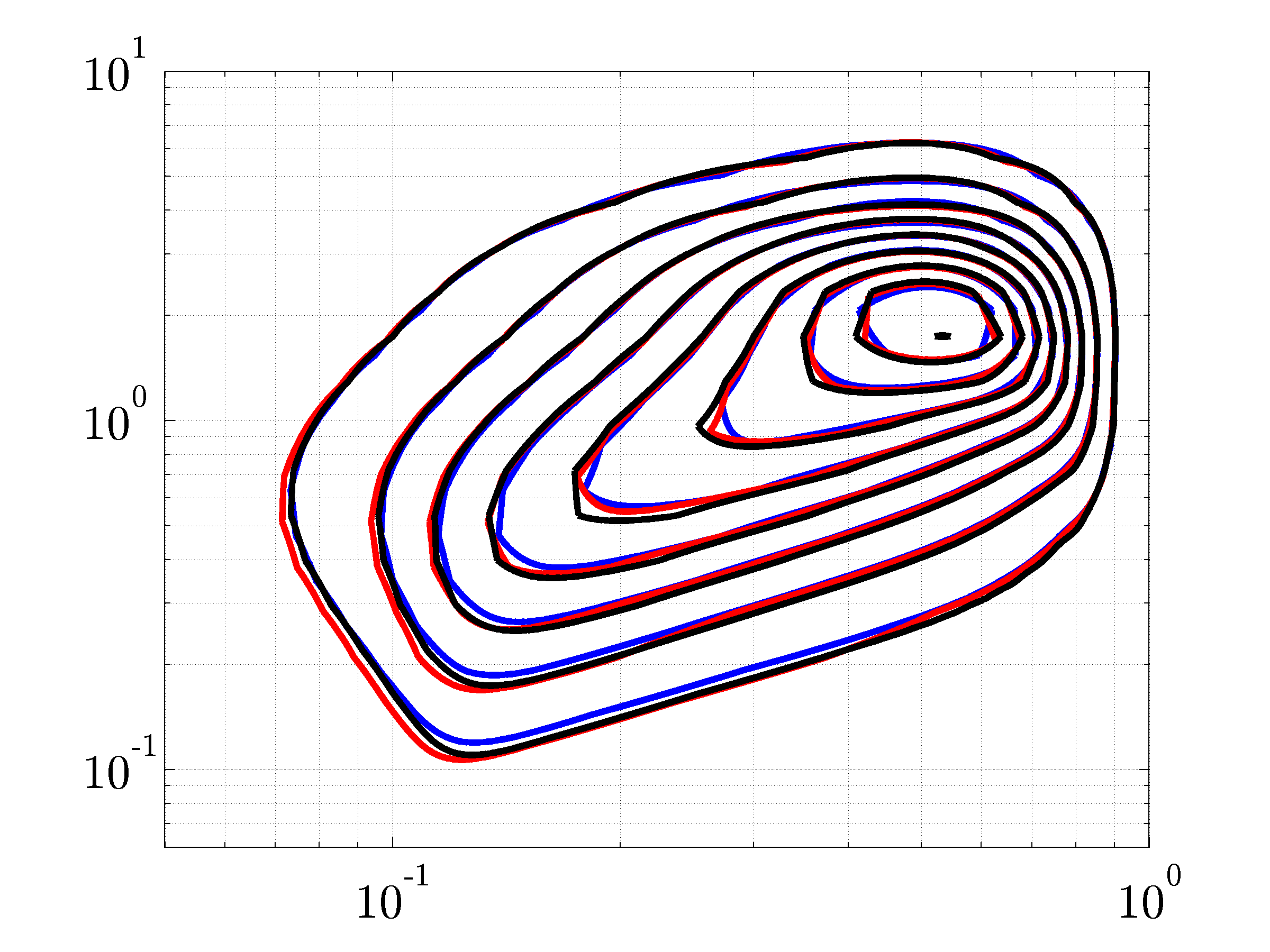}
    \label{fig.Euu-Rm2-kzkx2-vs-y-lz-R3333b-10000r-30000k-sigma1-outery-levels-0p1-0p1-1}}
    \\[-2.5cm]
    \hskip-4.5cm
    \begin{turn}{90}
    $~~\,\lambda_z^+$
    \end{turn}
    &
    \hskip-4.5cm
    \hskip-0.6cm
    \begin{turn}{90}
    $\sqrt{\lambda_z \lambda_z^+}$
    \end{turn}
    &
    \hskip-4.5cm
    \hskip-0.6cm
    \begin{turn}{90}
    $~~~\lambda_z$
    \end{turn}
    \\[1.3cm]
    \tc{black}{$y^+$}
    &
    \hskip-0.6cm
    \tc{black}{$\sqrt{y y^+}$}
    &
    \hskip-0.6cm
    \tc{black}{$y$}
    \\[0.1cm]
    $(d)$
    &
    \hskip-0.6cm
    $(e)$
    &
    \hskip-0.6cm
    $(f)$
    \\[-0.1cm]
    \subfigure{\includegraphics[width=0.34\columnwidth]
    {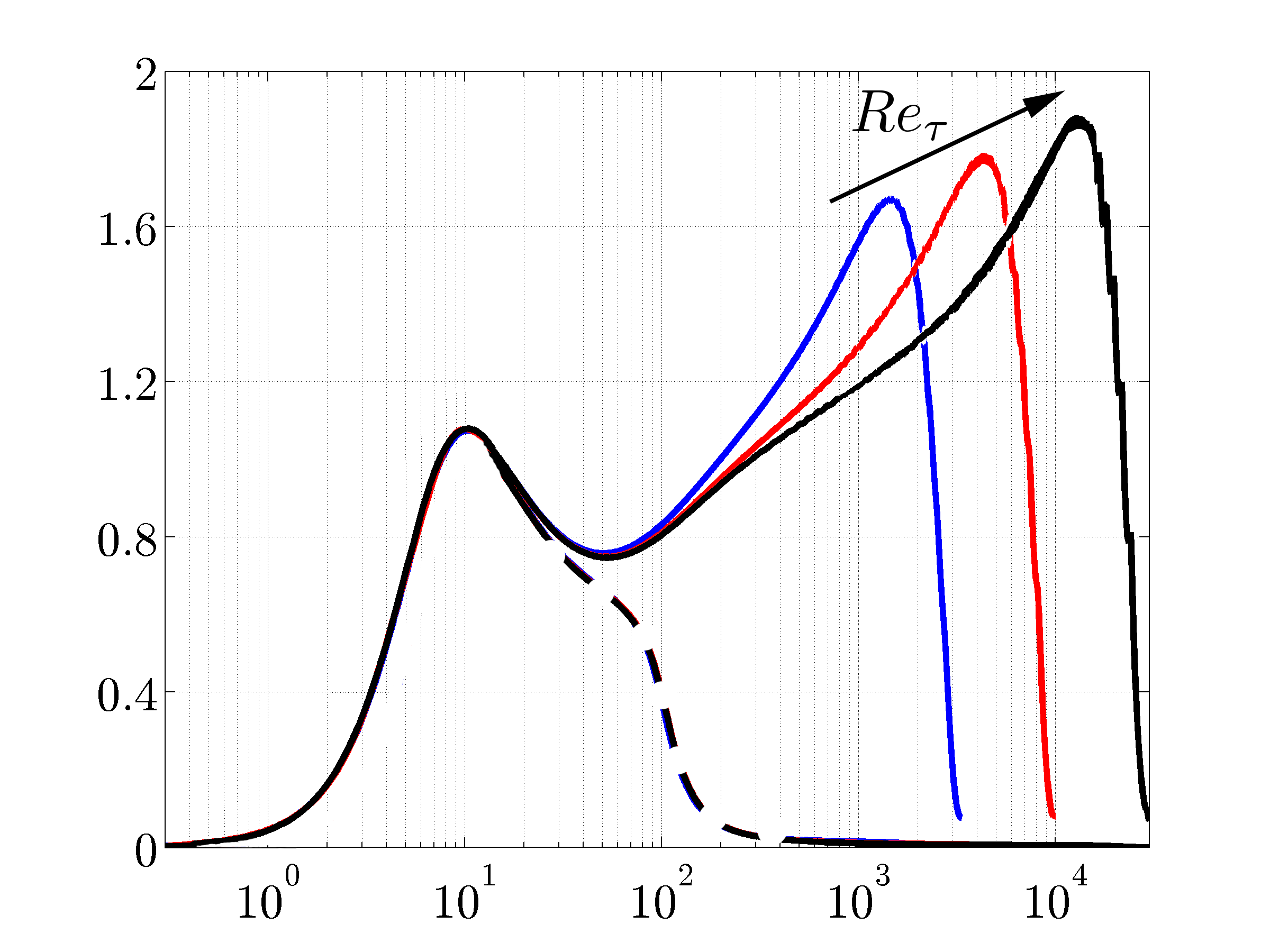}
    \label{fig.Euu-Rm2-kzkx2-vs-yp-R3333b-10000r-30000k-sigma1}}
    &
    \hskip-0.6cm
    \subfigure{\includegraphics[width=0.34\columnwidth]
    {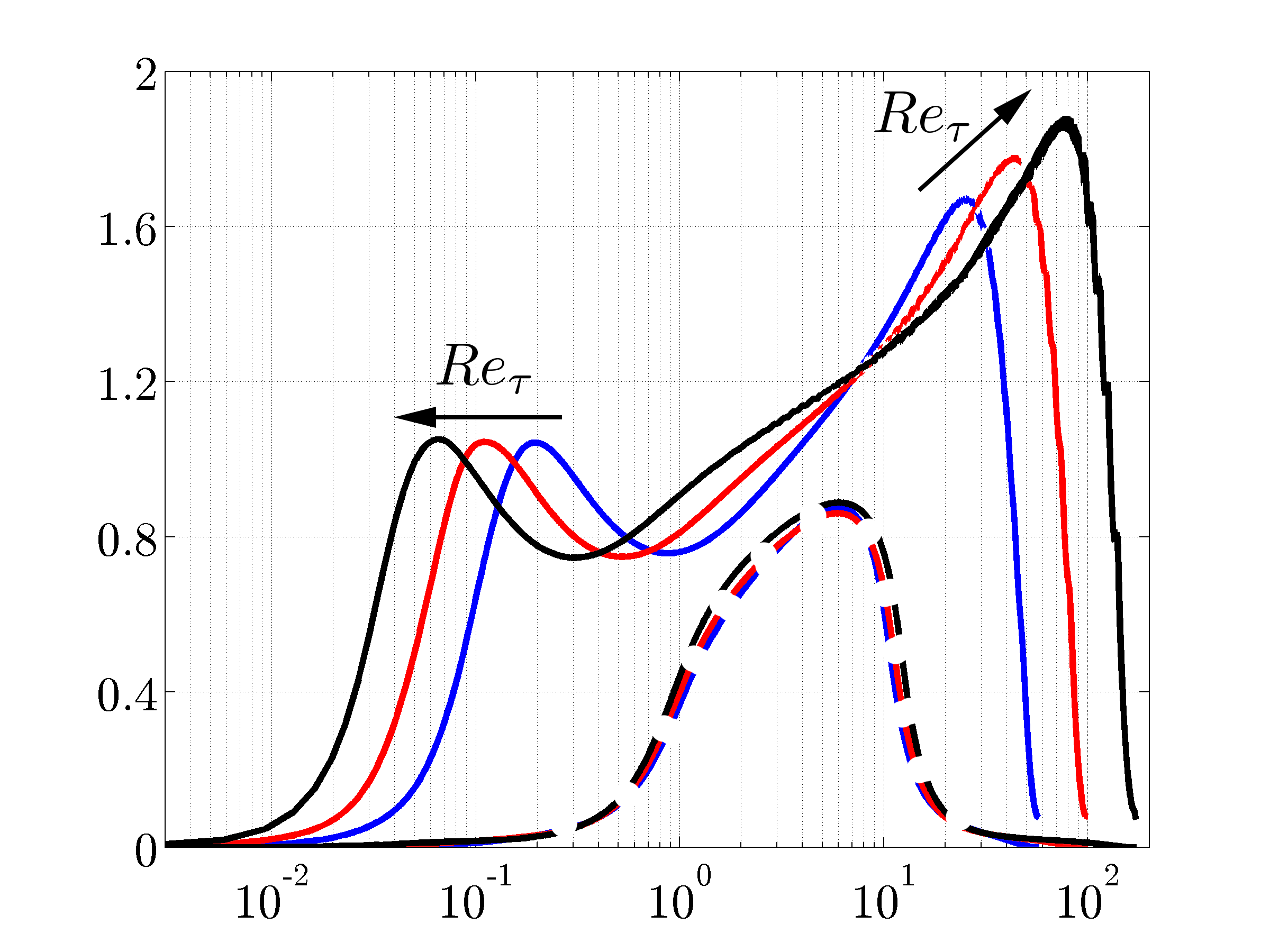}
    \label{fig.Euu-Rm2-kzkx2-vs-yh-R3333b-10000r-30000k-sigma1}}
    &
    \hskip-0.6cm
    \subfigure{\includegraphics[width=0.34\columnwidth]
    {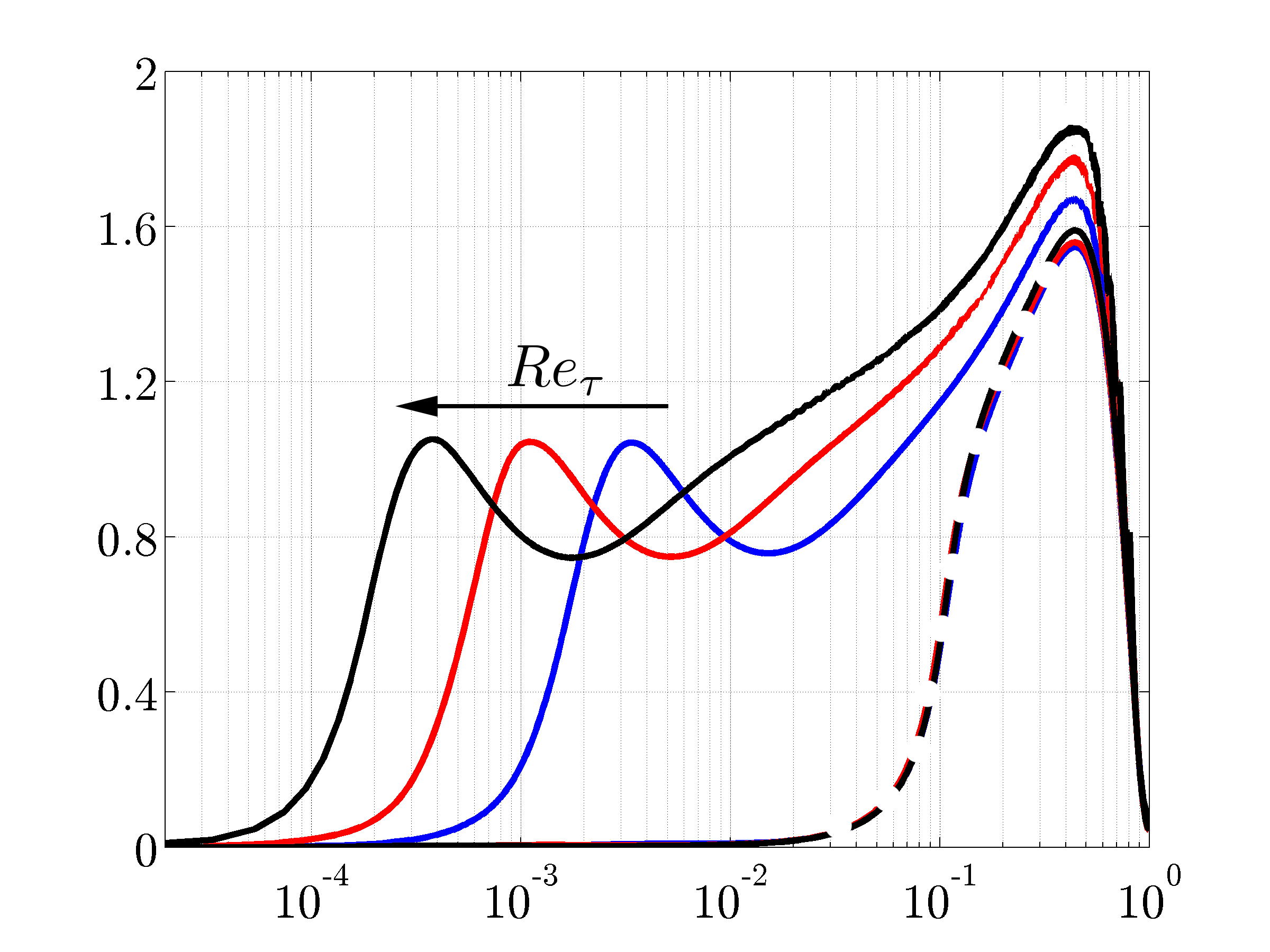}
    \label{fig.Euu-Rm2-kzkx2-vs-y-R3333b-10000r-30000k-sigma1}}
    \\[-0.05cm]
    \tc{black}{$y^+$}
    &
    \hskip-0.6cm
    \tc{black}{$\sqrt{y y^+}$}
    &
    \hskip-0.6cm
    \tc{black}{$y$}
    \\[0.1cm]
    $(g)$
    &
    \hskip-0.6cm
    $(h)$
    &
    \hskip-0.6cm
    $(i)$
    \end{tabular}
    \end{center}
    \caption{(Color online) The premultiplied one-dimensional streamwise energy density ${Re}_\tau^{-2} \, E_{uu} (y,\lambda_x)$ in (a)-(c); and ${Re}_\tau^{-2} \, E_{uu} (y,\lambda_z)$ in (d)-(f); and the streamwise energy intensity ${Re}_\tau^{-2} \, E_{uu} (y)$, dashed curves in (g)-(i) for the rank-1 model with broadband forcing.
    The wave parameters are confined to
    $\cS = \cS_{i}$ in (a), (d), (g);
    $\cS = \cS_{m}$ in (b), (e), (h); and
    $\cS = \cS_{o}$ in (c), (f), (i).
    The Reynolds numbers are
    ${Re}_\tau = 3333$ (blue), ${Re}_\tau = 10000$ (red), and ${Re}_\tau = 30000$ (black).
    The contour levels decrease by $0.05$ from their maximum value of $0.25$ (a); $0.15$ (b); and $0.25$ (c),
    and by $0.1$ from their maximum value of $0.5$ (d); $0.4$ (e); and $0.8$ (f).
    The solid curves in (g)-(i) are obtained by confining the wave parameters to $\cS = \cS_{e}$. The arrows indicate increase in the Reynolds number.}
    \label{fig.Euu-y-lx-lz-scales}
    \end{figure}

We compute the streamwise energy density of the rank-1 model with broadband forcing and illustrate its universal behavior with ${Re}_\tau$. These computations build the basis for prediction of the streamwise energy intensity at the technologically relevant values of ${Re}_\tau$ in~\S~\ref{sec.energy-density-weighted}. Because of the unique scales in the inner, middle, and outer classes of wave parameters, we distinguish the corresponding intervals of wave speeds by expanding the premultiplied energy density into the following three integrals:
\be
\ba{rcl}
E_{uu} (y, \kappa_x, \kappa_z)
&\!\!=\!\!&
\ds{
\int_{2}^{16}
}
\;
E_{uu} (y, \kappa_x, \kappa_z, c)
\,
\mrd c
\, + \,
\\[0.2cm]
&\!\! \!\!&
\ds{
\int_{16}^{U_{cl}-6.15}
}
\;
E_{uu} (y, \kappa_x, \kappa_z, c)
\,
\mrd c
\, + \,
\ds{
\int_{U_{cl}-6.15}^{U_{cl}}
}
\;
E_{uu} (y, \kappa_x, \kappa_z, c)
\,
\mrd c.
\ea
\label{eq.Euu-expand}
\ee
Similar expansions can be written for $E_{uu} (y,\kappa_x)$, $E_{uu} (y,\kappa_z)$, and $E_{uu} (y)$. In spite of the different behavior of singular values and singular functions, the energy density increases with ${Re}_\tau^2$ in all three classes of wave parameters. Figure~\ref{fig.Euu-y-lx-lz-scales} shows the premultiplied one-dimensional energy densities and the energy intensity confined to each class of wave parameters and normalized by ${Re}_\tau^{2}$. The same contour levels are used for all Reynolds numbers, ${Re}_\tau = 3333$ (blue), $10000$ (red), and $30000$ (black), in figures~\ref{fig.Euu-Rm2-kzkx2-vs-yp-lxp-R3333b-10000r-30000k-sigma1-inner-levels-0p05-0p05-0p3}-\ref{fig.Euu-Rm2-kzkx2-vs-y-lz-R3333b-10000r-30000k-sigma1-outery-levels-0p1-0p1-1}. Notice that confining the wavenumbers to $\cS_{i}$, $\cS_{m}$, and $\cS_{o}$ yields a universal energy density as summarized in tables~\ref{table.subsets-R-scales} and~\ref{table.svd-R-scales}.
	
The inner peak of streamwise energy density in the rank-1 model with broadband forcing occurs at $y^+ \approx 11$, $\lambda_x^+ \approx 184$, and $\lambda_z^+ \approx 44$; see figures~\ref{fig.Euu-Rm2-kzkx2-vs-yp-lxp-R3333b-10000r-30000k-sigma1-inner-levels-0p05-0p05-0p3} and~\ref{fig.Euu-Rm2-kzkx2-vs-yp-lzp-R3333b-10000r-30000k-sigma1-inner-levels-0p1-0p1-1}. The location of the above wall-normal peak represents an integral effect over all wave parameters in $\cS_{i}$ and corresponds to the critical speed ${c} \approx 8.5$. The inner peak is comparable with the location, length, and spacing of the most energetic structures associated with the near-wall cycle, i.e. $y^+ \approx 15$, $\lambda_x^+ \approx 700-1000$, and $\lambda_z^+ \approx 100$~\cite[see, for example,][]{hoyjim06}. Figures~\ref{fig.Euu-Rm2-kzkx2-vs-y-lxm-R3333b-10000r-30000k-sigma1-outery-levels-0p05-0p05-0p3} and~\ref{fig.Euu-Rm2-kzkx2-vs-y-lz-R3333b-10000r-30000k-sigma1-outery-levels-0p1-0p1-1} show that the outer peak takes place at $y \approx 0.45$ (corresponding to critical defect speed $U_{cl} - {c} \approx 2$) for $\lambda_x/{Re}_\tau \approx 0.1$, and $\lambda_z \approx 2$. This peak points to much longer structures relative to the LSM structures observed in experiments, i.e. $\lambda_x \approx 3$~\cite[see, for example in boundary layers,][]{adrmeitom00}. The same length scales are reported for channels and pipes~\cite[see, for example,][]{guahomadr06,monhutngmarcho09}. 

The middle peak takes place at $\sqrt{y y^+} \approx 5.2-5.8$ (corresponding to critical speed $c \approx U_m$), $\lambda_x \approx 12-16$, and $\sqrt{\lambda_z \lambda_z^+} \approx 20$; see figures~\ref{fig.Euu-Rm2-kzkx2-vs-yh-lx-R3333b-10000r-30000k-sigma1-outerx-levels-0p05-0p05-0p3} and~\ref{fig.Euu-Rm2-kzkx2-vs-yh-lzh-R3333b-10000r-30000k-sigma1-outerx-levels-0p1-0p1-1}. It has the same streamwise and wall-normal scalings as the VLSMs and its location is comparable with the most energetic VLSMs located at $\sqrt{y y^+} \approx 3.9$ and $\lambda_x \approx 6$ in boundary layers~\cite[see, for example,][]{marmathut10} and at $\lambda_x \approx 12-20$ in pipes and channels~\cite[see, for example,][]{monhutngmarcho09}. \moa{Spanwise correlations of experimental data show that the spanwise wavelength of modes in the middle region increases approximately linearly with $y$, see e.g.~\cite{monstewilcho07,baihulsmi08}. The self-similar scales of the resolvent modes in the middle region are consistent with these results. The Reynolds number scaling of the spanwise wavelength appears to still be under investigation. Since the spanwise peak of the one-dimensional spectrum is obtained by including a range of modes with speeds $|U_m - c| < d$ (instead of focusing on one mode), the Reynolds number scaling of the spanwise peak is similar to the wall-normal scaling of the modes.} \moar{The organization of the self-similar coherent motions in the logarithmic layer of real turbulent flows has been studied by many authors, e.g. see~\cite{tomadr03,deljimzanmos06,flojim10}. In addition,~\cite{hwacos11} addressed the self-sustaining mechanisms of these coherent motions. Studying the implications of the identified scalings of the resolvent modes on these structures is a topic of future research.} 

\moa{In making the above comparisons, it is important to note the distinction between the resolvent modes and the real turbulent flow that can be represented by a weighted superposition of the resolvent modes. The agreement between the admitted scales of the principal resolvent modes and the scalings observed in real flows is striking considering the simplicity of the rank-1 model subject to broadband forcing. This agreement emphasizes the role of linear mechanisms and critical layers in determining the scaling of turbulent flows. In addition, the differences between the scalings highlights the role of nonlinearity in shaping the weights of the resolvent modes. We also note that the experimentally obtained outer peak in the two-dimensional spectrum and the wavelengths associated with VLSMs and LSMs may be contaminated by use of the Taylor's hypothesis and lack of sufficient scale separation at relatively low $Re_\tau$.}

The one-dimensional energy densities can be integrated in the remaining wall-parallel direction to obtain the streamwise energy intensity $E_{uu} (y)$ for the rank-1 model with broadband forcing. The dashed curves in figures~\ref{fig.Euu-Rm2-kzkx2-vs-yp-R3333b-10000r-30000k-sigma1}-\ref{fig.Euu-Rm2-kzkx2-vs-y-R3333b-10000r-30000k-sigma1} are the energy intensities normalized by ${Re}_\tau^2$ obtained by confining the wave parameters to $\cS_{i}$, $\cS_{m}$, and $\cS_{o}$, respectively. As expected, the energy intensities are independent of Reynolds number when confined to the universal classes of wave parameters. The solid curves are obtained by integrating the energy density over all wavenumbers and wave speeds $2 \leq {c} \leq U_{cl}$, i.e. by confining the wave parameters to $\cS_{e}$. These figures highlight the selection of two local peaks by the linear amplification mechanism where the inner and outer peaks dominate the middle peak. The inner peak takes place close to the inner peak of the streamwise intensity in real turbulent flows. While the energy intensity of real flows exhibits outer scales near the center of the channel, there is no strong evidence for presence of an outer peak even for high ${Re}_\tau$.

As evident from figure~\ref{fig.Euu-Rm2-kzkx2-vs-yp-R3333b-10000r-30000k-sigma1}, the universal inner waves contribute to more than $96 \%$ of the total energy intensity for $y^+ < 20$ for all Reynolds numbers. On the other hand, figure~\ref{fig.Euu-Rm2-kzkx2-vs-y-R3333b-10000r-30000k-sigma1} shows that the universal outer waves capture a smaller amount of the total intensity for $y = 0.45$ as ${Re}_\tau$ increases; $95 \%$ for ${Re}_\tau = 3333$ vs. $86 \%$ for ${Re}_\tau = 30000$. This is because the aspect ratio constraint in $\cS_{o}$ excludes more wavenumbers from $\cS_{e}$ as ${Re}_\tau$ increases. The excluded waves are not universal with ${Re}_\tau$ and their contribution to the energy intensity is not completely negligible. A similar reasoning explains why the universal middle scale captures $82 \%$ of the total energy intensity at $\sqrt{y y^+} = \sqrt{10}$ for ${Re}_\tau = 3333$ vs. $72 \%$ for ${Re}_\tau = 30000$; cf. figure~\ref{fig.Euu-Rm2-kzkx2-vs-yh-R3333b-10000r-30000k-sigma1}.

At the end of this section, we recall that the streamwise energy densities and intensities thus far were obtained for the model with broadband forcing in $\lambda_x$, $\lambda_z$, and $c$. In~\S~\ref{sec.energy-density-weighted}, we consider a non-broadband forcing by introducing an optimally shaped energy density.

\section{Predicting the streamwise energy intensity}
\label{sec.energy-density-weighted}

In this section, we introduce a model for predicting the energy intensity of real turbulent flows by considering a non-broadband forcing in wave speed. This is done by incorporating a positive weight function $W ({c})$ that amplifies or attenuates the energy density $E_{uu} (y, {c})$ of the rank-1 model with broadband forcing. Even though $W (c)$ differs from a true forcing spectrum (that also depends on the wall-parallel wavelengths), it provides the model with sufficient degrees of freedom for predicting the energy intensity. In addition, since each wave speed is associated with a certain class of wavelengths, $W (c)$ affects different classes of wavelengths as the wave speed changes.

First, we show that $W (c)$ can be optimally shaped such that the model-based streamwise energy intensity,
	\be
	E_{uu,W} (y)
	\; = \;
	\ds{
	\int_{2}^{U_{cl}}
	}
	W ({c}) \, E_{uu} (y, {c}) \, \mrd {c},
	\label{eq.Euu-W}
	\ee
matches the intensity of real flows at low Reynolds numbers. Then, we estimate similarity laws to approximate the optimal weight functions at high values of ${Re}_\tau$. These weight functions in conjunction with the energy density of the rank-1 model with broadband forcing enable prediction of the streamwise energy intensity at technologically relevant Reynolds numbers.

\subsection{Optimal weights for small Reynolds numbers}
\label{sec.weights}

The weight function $W ({c})$ is determined by minimizing the deviation between $E_{uu,W} (y)$ in~(\ref{eq.Euu-W}) and the streamwise energy intensity obtained from DNS, $E_{uu,\mathrm{DNS}} (y)$, in the interval $y^+ \geq 1$ and $y \leq 0.8$. We do not enforce matching for $y \geq 0.8$ since it requires significantly large values of $W ({c})$ for wave speeds close to $U_{cl}$. This is because $E_{uu} (y, {c})$ is considerably smaller and more localized near the centerline compared to other locations and results in sensitivity of $W ({c})$; see, for example, figure~\ref{fig.Euu-Rm2-kzkx2-vs-yp-Ux-R30000-sigma1-log10} for $y^+ > 24000$ corresponding to $y > 0.8$ for ${Re}_\tau = 30000$. Note that the main amplification mechanisms for waves with speeds close to $U_{cl}$ is critical behavior of the resolvent modes since the non-normality effect is small as the mean shear approaches zero. This results in small gains and resolvent modes that are localized in the wall-normal direction.

Since $E_{uu} (y, {c})$ scales with ${Re}_\tau^{2}$ (cf. table~\ref{table.svd-R-scales}) while $E_{uu,\mathrm{DNS}} (y)$ does not, we find the normalized weight function $\overline{W} ({c}) = {Re}_\tau^{2} \, W ({c})$ that solves the following optimization problem
      \be
	\ba{ll}
	\mbox{minimize:}
       &
       \dfrac{
	\norm{
       E_{uu,\mathrm{DNS}} (y)
       \, - \,
       E_{uu,W} (y)
	}_e^2
	}{
	\norm{E_{uu,\mathrm{DNS}} (y)}_e^2
	}
       \, + \,
       \gamma_w \;
       \norm{\overline{W} ({c})}_w^2,
       \\[0.4cm]
       \mbox{subject to:}
       &
       \overline{W} ({c}) > 0,~
       2 \leq {c} \leq U_{cl}.
       \ea
	\label{eq.opt-F-Up}
	\ee
Here, $\norm{g (y)}_e$ is defined as (note integration in $\log y^+$)
	\be
       \norm{g (y)}_e^2
       \; = \;
       \ds{
	\int_{0}^{\log (0.8 {Re}_\tau)}
	}
       g^2 (\log y^+)
       \,
       \mrd \log y^+,
       \non
       \ee
to equally penalize the deviation of energy intensities near the inner peak as well as in the channel core. The second term in the objective function,
	\be
       \norm{\overline{W} (c)}_w^2
       \; = \;
       \dfrac{1}{U_{cl} - 2}
       \ds{
	\int_{2}^{U_{cl}}
	}
       \overline{W}^2 (c)
       \,
       \mrd c,
       \non
       \ee
provides the weight function with smoothness by penalizing the magnitude of $\overline{W}$, and $\gamma_w \geq 0$ controls the importance of smoothness relative to matching the model-based and DNS-based energy intensities.

The optimization problem~(\ref{eq.opt-F-Up}) is solved using CVX, a package for specifying and solving convex programs in Matlab~\citep{cvx,graboy-cvx-08}. We find the optimal weights for the largest Reynolds numbers that have been simulated to date using DNS, i.e. ${Re}_\tau = 934$~\citep{deljimzanmos04} and ${Re}_\tau = 2003$~\citep{hoyjim06}. Even though these are orders of magnitude smaller than the Reynolds numbers for which we predict the energy intensity in~\S~\ref{sec.predict}, they are free of measurement errors and useful for finding the optimal weights. We choose $\gamma_w = 0.2$ to strike a balance between matching error $\norm{E_{uu,\mathrm{DNS}} (y) - E_{uu,W} (y)}_e^2$ and smoothness of $\overline{W} (c)$. The optimization problem is robust with respect to the choice of $\gamma_w$. For example, changing $\gamma_w$ by a factor of two has negligible effect on the matching error and the optimal weights for $7 \lesssim c \lesssim U_{cl} - 2$ while slightly modifying $\overline{W} (c)$ elsewhere.

    \begin{figure}
    \begin{center}
    \begin{tabular}{cc}
    \hskip0.1cm
    \subfigure{\includegraphics[width=0.5\columnwidth]
    {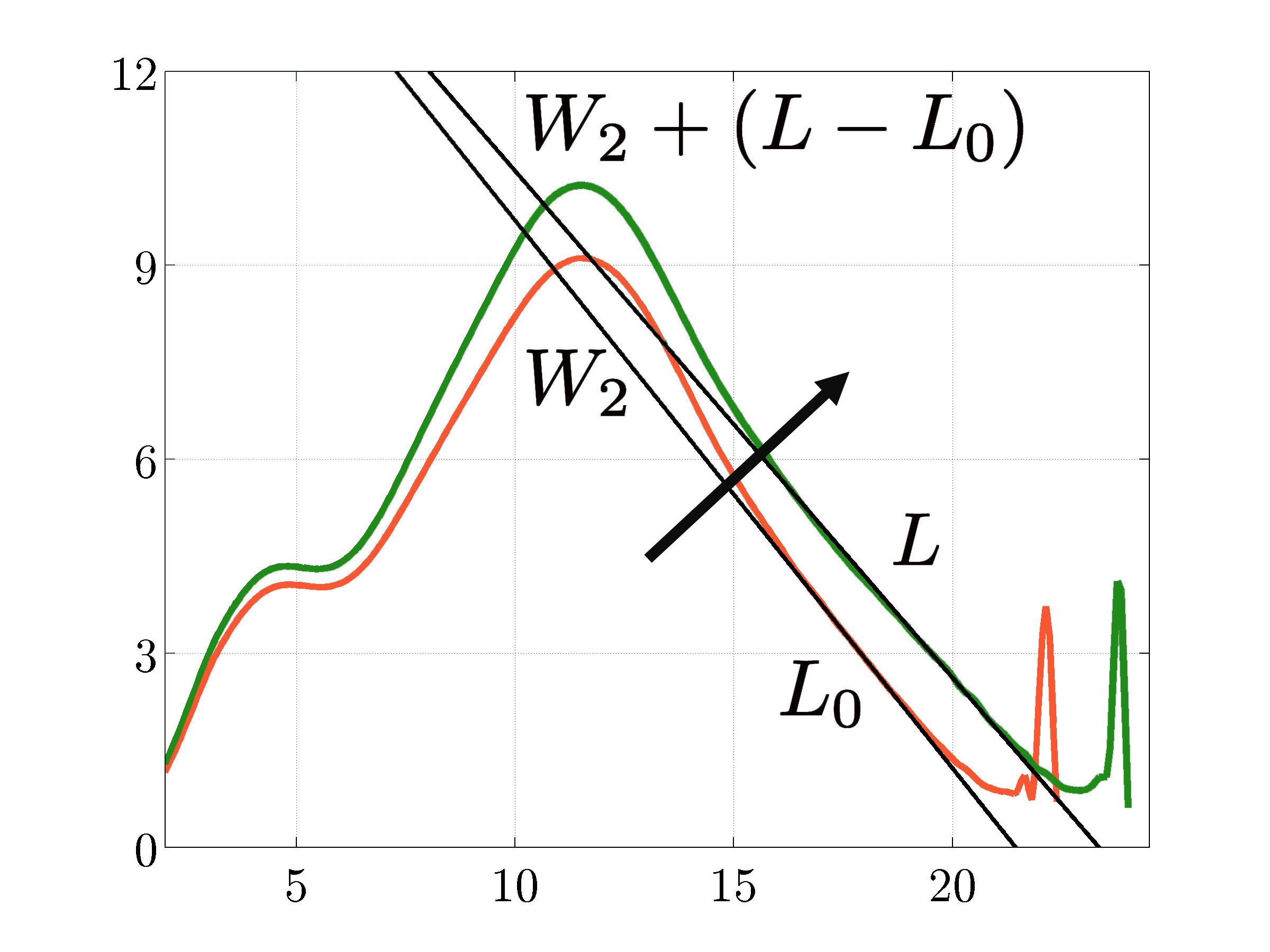}
    \label{fig.F-vs-Up-R934-2003-lines-area-F2-110912}}
    &
    \hskip-0.3cm
    \subfigure{\includegraphics[width=0.5\columnwidth]
    {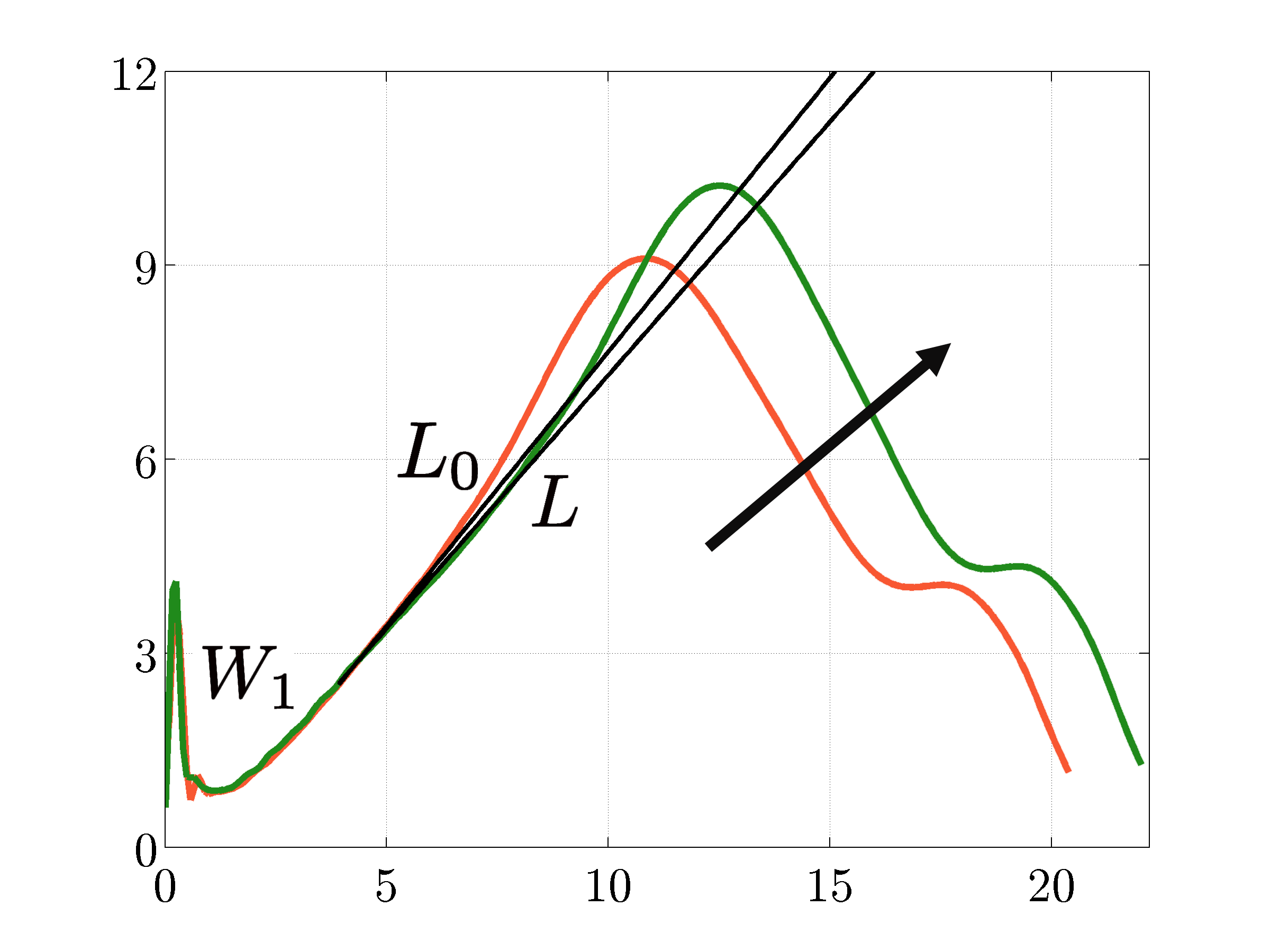}
    \label{fig.F-vs-UcmUp-R934-2003-lines-F1-110912}}
    \\[-0.2cm]
    \hskip0.1cm
    \tc{black}{$c$}
    &
    \hskip-0.3cm
    \tc{black}{$U_{cl} - c$}
    \\[0.1cm]
    \hskip0.1cm
    $(a)$
    &
    \hskip-0.3cm
    $(b)$
    \\[-3.6cm]
    \hskip-6.1cm
    \hskip0.1cm
    \begin{turn}{90}
    $\overline{W}$
    \end{turn}
    &
    \hskip-6.1cm
    \hskip-0.3cm
    \begin{turn}{90}
    $\overline{W}$
    \end{turn}
    \\[3.1cm]
    \hskip0.1cm
    \subfigure{\includegraphics[width=0.5\columnwidth]
    {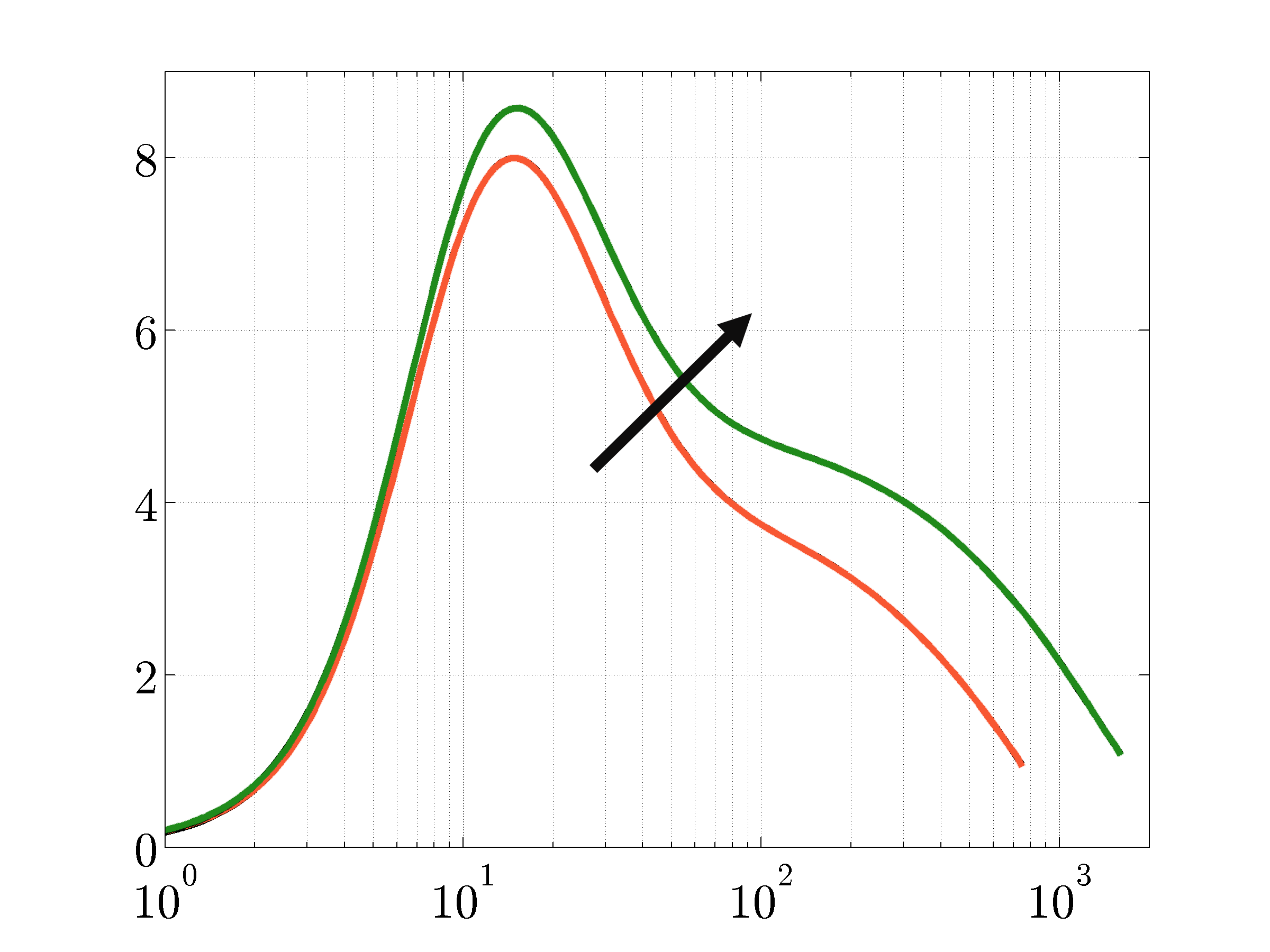}
    \label{fig.Euu-vs-yp-match-F-Up-R934-2003}}
    &
    \hskip-0.3cm
    \subfigure{\includegraphics[width=0.5\columnwidth]
    {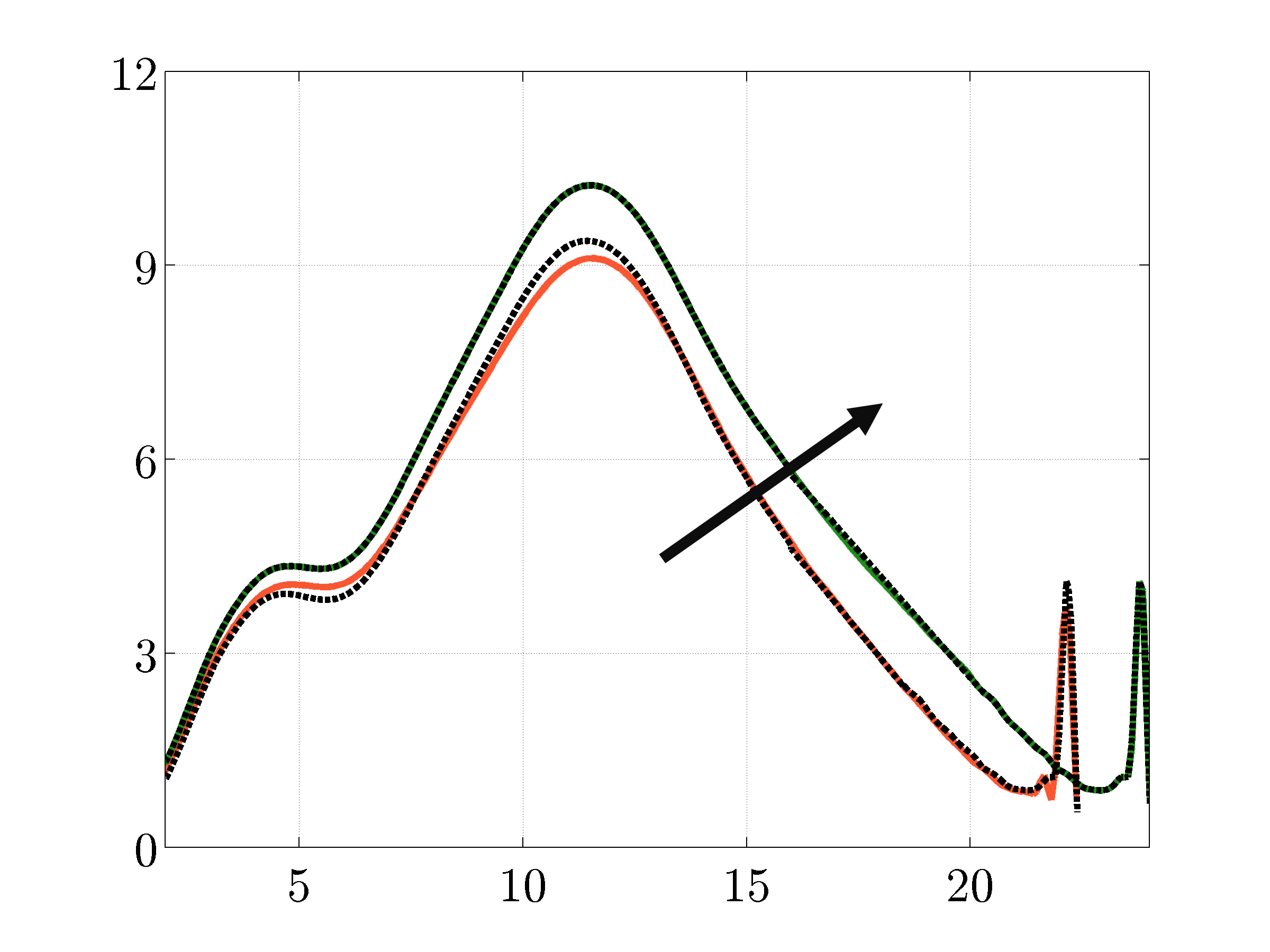}
    \label{fig.F-vs-UcmUp-R934-2003-lines-area-F1-110912-compare-model}}
    \\[-0.2cm]
    \hskip0.1cm
    \tc{black}{$y^+$}
    &
    \hskip-0.3cm
    \tc{black}{$c$}
    \\[0.1cm]
    \hskip0.1cm
    $(c)$
    &
    \hskip-0.3cm
    $(d)$
    \\[-4.3cm]
    \hskip-6.1cm
    \hskip0.1cm
    \begin{turn}{90}
    $E_{uu,W}; E_{uu,\mathrm{DNS}}$
    \end{turn}
    &
    \hskip-6.1cm
    \hskip-0.3cm
    \begin{turn}{90}
    $~~~~~~~~~~~\overline{W}$
    \end{turn}
    \\[2.1cm]
    \end{tabular}
    \end{center}
    \caption{(Color online) (a)-(b): The optimal weight functions $\overline{W}$ as a function of ${c}$ in (a) and $U_{cl} - {c}$ in (b) for ${Re}_\tau = 934$ (orange) and $2003$ (green). The tangent lines, $L$, to $\overline{W} (U_{cl} - {c})$ at $U_{cl} - {c} = 3.9$ for each ${Re}_\tau$ are shown in black.
    (c) The model-based streamwise energy intensity $E_{uu,\mathrm{W}} (y^+)$ for ${Re}_\tau = 934$ (orange) and $2003$ (green) and the DNS-based intensity $E_{uu,\mathrm{DNS}} (y^+)$ (black) are optimally matched by solving~(\ref{eq.opt-F-Up}) for each ${Re}_\tau$. The respective curves lie on the top of each other.
    (d) The optimal weight functions $\overline{W}$ for ${Re}_\tau = 934$ (orange) and $2003$ (green) are compared with the weight functions obtained using the similarity law~(\ref{eq.F-sim}) (black dots).
    The arrows indicate increase in the Reynolds number.}
    \label{fig.F-R934-2003}
    \end{figure}

Figures~\ref{fig.F-vs-Up-R934-2003-lines-area-F2-110912} and~\ref{fig.F-vs-UcmUp-R934-2003-lines-F1-110912} show the optimal weights as a function of $c$ and $U_{cl} - c$ for ${Re}_\tau = 934$ and $2003$. These weight functions match $E_{uu,\mathrm{DNS}}$ and $E_{uu,W}$ with a relative error of approximately $0.2 \%$; see~figure~\ref{fig.Euu-vs-yp-match-F-Up-R934-2003}. As expected, $\overline{W}$ is qualitatively similar for $c \lesssim 16$ and $U_{cl} - c \lesssim 6.15$ since both the model-based and DNS-based intensities exhibit inner and outer scaling in the respective regions. Figure~\ref{fig.F-vs-UcmUp-R934-2003-lines-F1-110912} shows that $\overline{W} (U_{cl} - {c})$ approximately coincides for ${Re}_\tau = 934$ and $2003$ for $U_{cl} - {c} \leq 3.9$. We denote this universal function by $W_1 (U_{cl} - {c})$. For simplicity, the weights are approximated by linear functions in the \moar{self-similar} region for $16 \leq c \leq U_{cl} - 3.9$. These lines are denoted by $L$ (black) and intersect for $L (c = -2) = 19.88$ and $L (U_{cl} - {c} = 3.9) = 2.54$. This gives an analytical expression for $L$ as a function of wave speed and Reynolds number since $U_{cl}$ varies with ${Re}_\tau$
	\be
	L ({c}; {Re}_\tau)
	\; = \;
	2.54
	\, + \,
	17.34
	\left(
	\dfrac{U_{cl} - {c} - 3.9}{U_{cl} - 1.89}
	\right).
	\label{eq.L}
	\ee
As the Reynolds number increases, $\overline{W} ({c})$ is shifted upward for $c \leq 16$ by the kick that it receives from the \moar{self-similar} region. This is expected since the DNS-based energy intensity increases with ${Re}_\tau$ in the inner region while $E_{uu} (y^+,c)$ remains constant. More discussion about the relationship between the weights in the \moar{self-similar} and inner regions are provided in~\S~\ref{sec.predict}. Motivated by these observations, we formulate a similarity law for the weight function:
	\be
	\overline{W} ({c}; {Re}_\tau)
	\; = \;
	\left\{
	\ba{ll}
	W_2 ({c})
	\, + \,
	\big( L ({c}; {Re}_\tau) - L_0 (c) \big),
	&~~~~~~~~~~~~~~~
	2 \leq {c} \leq 16,
	\\[0.15cm]
	L ({c}; {Re}_\tau),
	&~~~~~~~~~~~~~~
	16 < {c} < U_{cl} - 3.9,
	\\[0.15cm]
	W_1 (U_{cl} - {c}),
	&~~~~~\,
	U_{cl} - 3.9 \leq {c} \leq U_{cl},
	\ea
	\right.
	\label{eq.F-sim}
	\ee
that consists of three segments: A universal outer segment represented by $W_1 (U_{cl} - {c})$ for $U_{cl} - 3.9 \leq {c} \leq U_{cl}$; a Reynolds number dependent linear segment $L ({c}; {Re}_\tau)$ that is analytically determined by~(\ref{eq.L}); and an inner segment composed of a universal function $W_2 (c)$ for $2 \leq {c} \leq 16$ superposed by a linear function $L ({c}; {Re}_\tau) - L_0 (c)$ where $L_0 (c) = L(c;{Re}_\tau = 934)$. Figure~\ref{fig.F-vs-UcmUp-R934-2003-lines-area-F1-110912-compare-model} shows that the optimal weights computed by solving~(\ref{eq.opt-F-Up}) are well-captured by the weights formulated using the similarity law~(\ref{eq.F-sim}). \moa{We note that more complex approximations could be used if more than two DNS datasets were available. Efforts to determine these weights analytically are ongoing.}

    \begin{figure}
    \begin{center}
    \begin{tabular}{cc}
    \subfigure{\includegraphics[width=0.5\columnwidth]
    {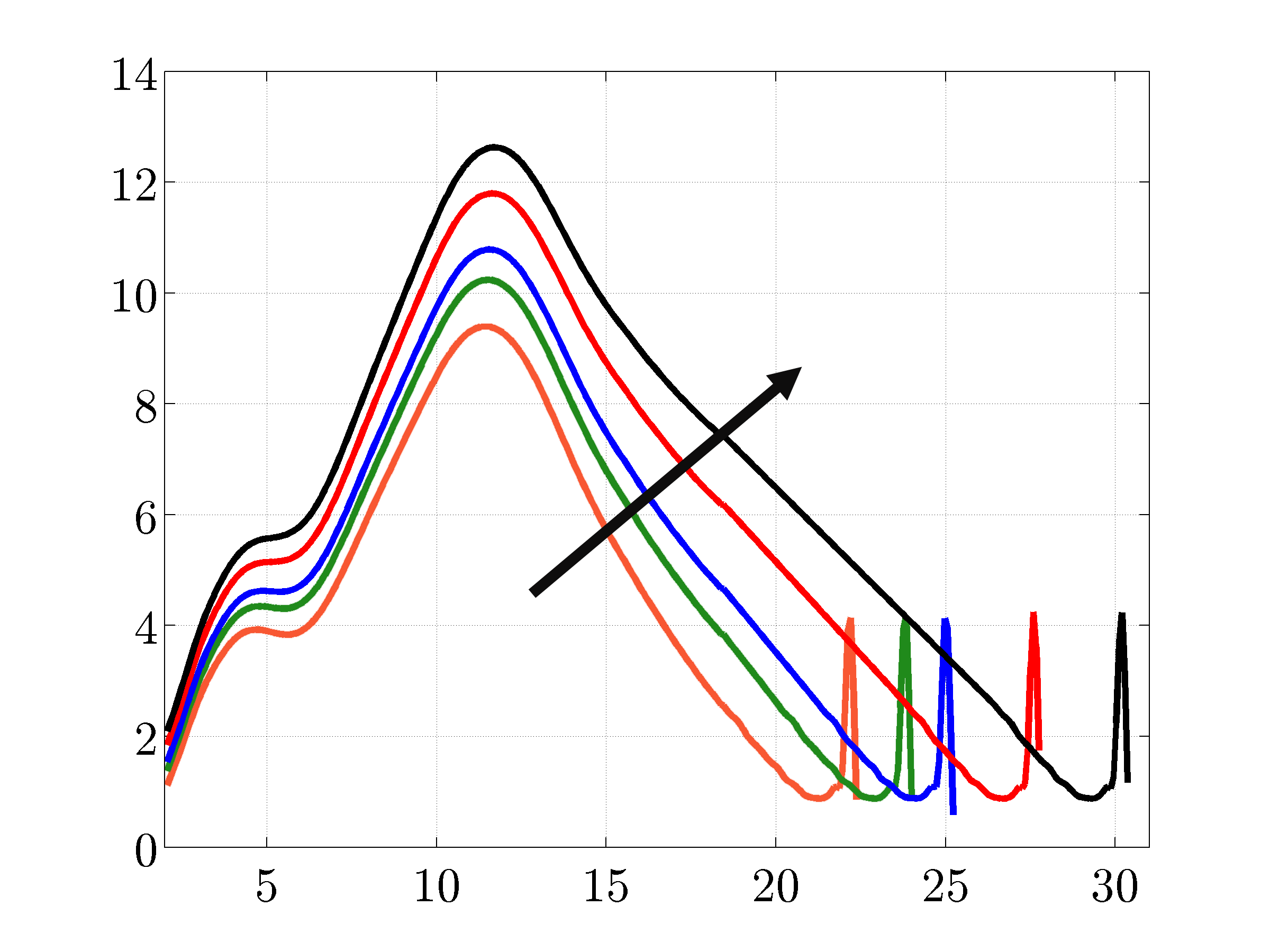}
    \label{fig.F-vs-Up-R934c-2003g-3333b-10000r-30000k}}
    &
    \hskip-0.4cm
    \subfigure{\includegraphics[width=0.5\columnwidth]
    {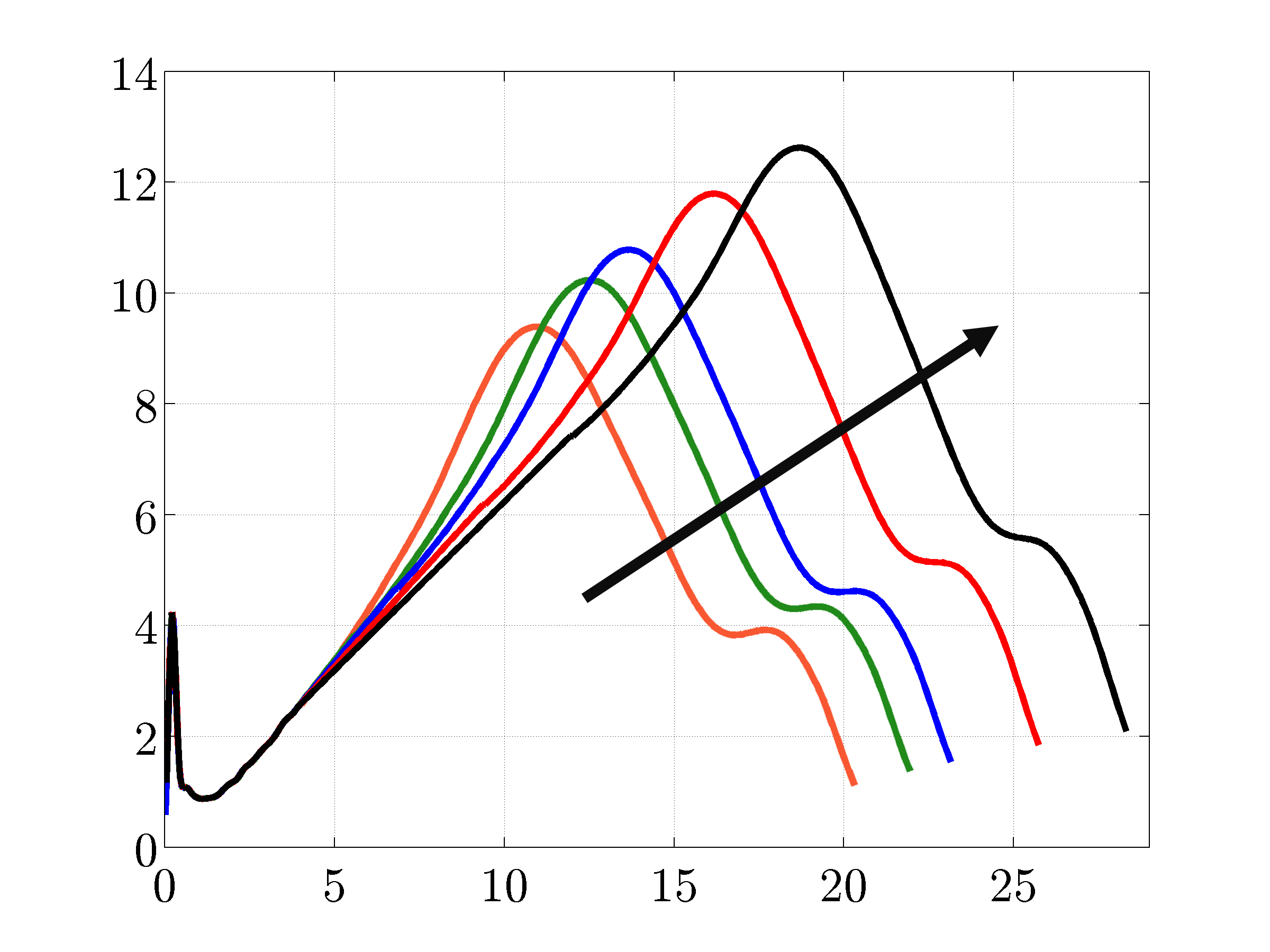}
    \label{fig.F-vs-UcmUp-R934c-2003g-3333b-10000r-30000k}}
    \\[-0.2cm]
    \hskip0.1cm
    \tc{black}{$c$}
    &
    \hskip-0.3cm
    \tc{black}{$U_{cl} - c$}
    \\[0.1cm]
    \hskip0.1cm
    $(a)$
    &
    \hskip-0.3cm
    $(b)$
    \\[-3.6cm]
    \hskip-6.1cm
    \hskip0.1cm
    \begin{turn}{90}
    $\overline{W}$
    \end{turn}
    &
    \hskip-6.1cm
    \hskip-0.3cm
    \begin{turn}{90}
    $\overline{W}$
    \end{turn}
    \\[3cm]
    \subfigure{\includegraphics[width=0.5\columnwidth]
    {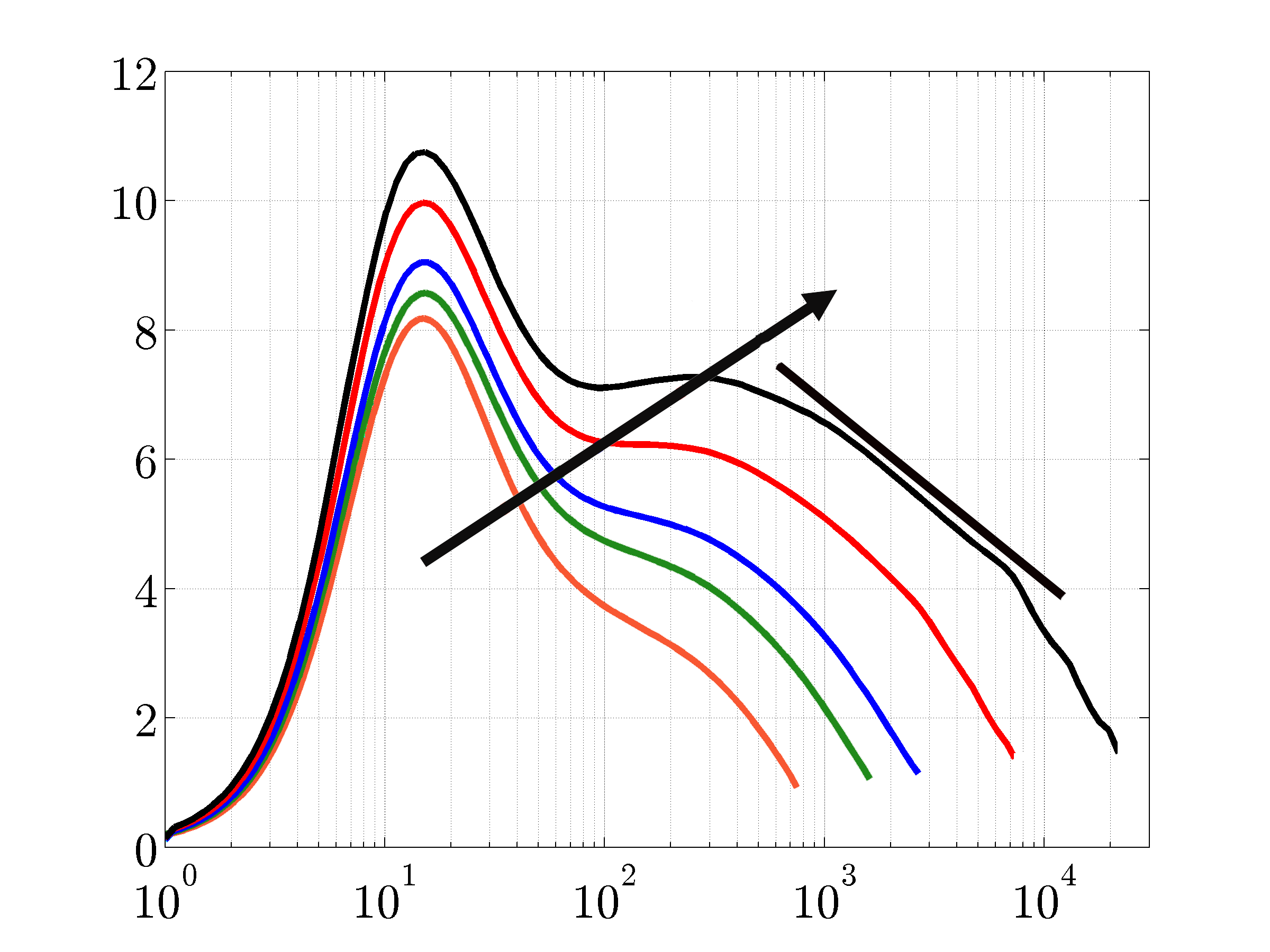}
    \label{fig.Euu-vs-yp-match-F-Up-R934c-2003g-3333b-10000r-30000k-110912-line}}
    &
    \hskip-0.4cm
    \subfigure{\includegraphics[width=0.5\columnwidth]
    {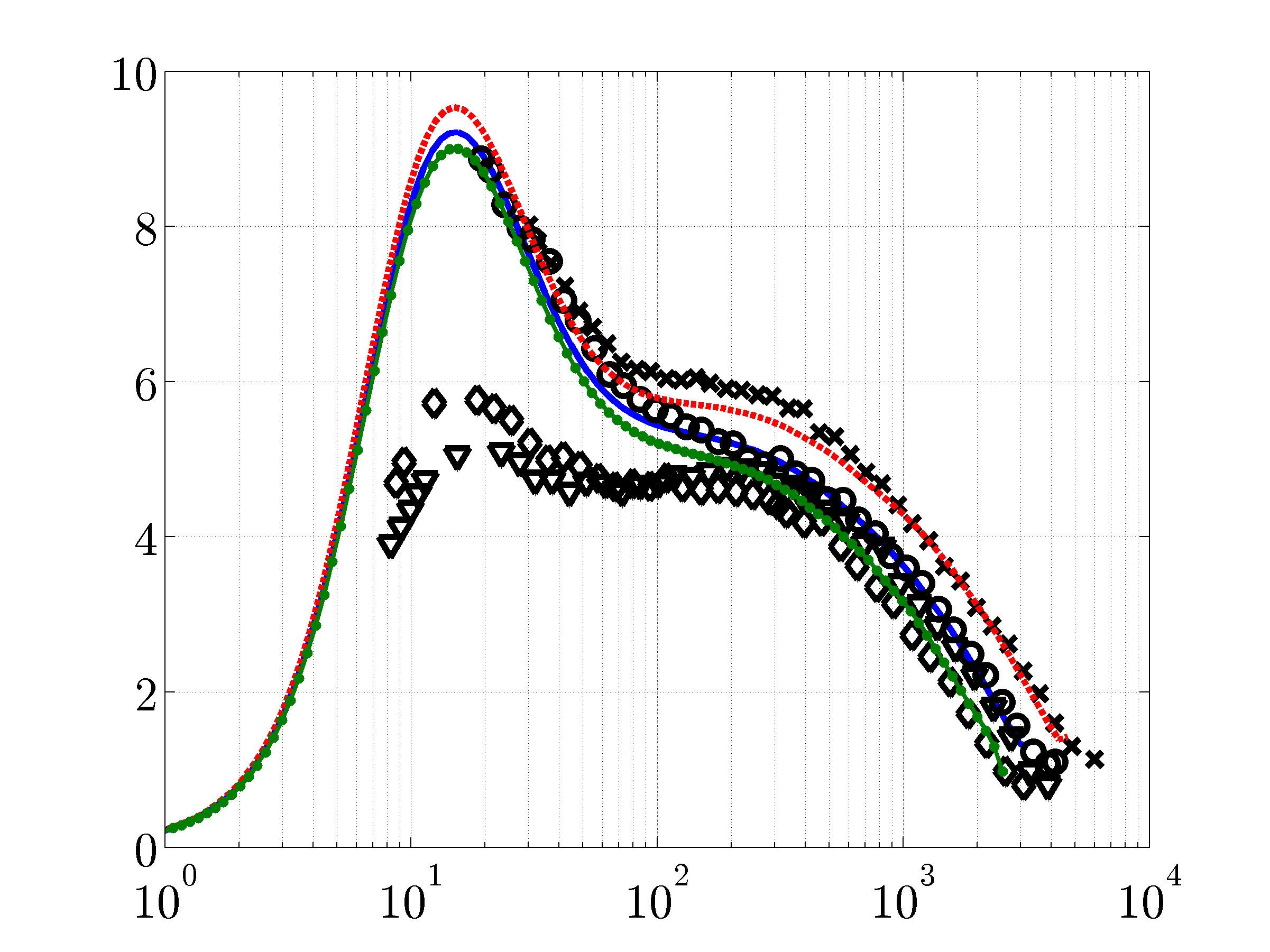}
    \label{fig.Euu-vs-yp-match-F-Up-R3165-4000-5813-3165d-4000v-4000o-6000x-Exp-data-110912-color}}
    \\[-0.2cm]
    \hskip0.1cm
    \tc{black}{$y^+$}
    &
    \hskip-0.3cm
    \tc{black}{$y^+$}
    \\[0.1cm]
    \hskip0.1cm
    $(c)$
    &
    \hskip-0.3cm
    $(d)$
    \\[-3.9cm]
    \hskip-6.1cm
    \hskip0.1cm
    \begin{turn}{90}
    $E_{uu,W}$
    \end{turn}
    &
    \hskip-6.1cm
    \hskip-0.3cm
    \begin{turn}{90}
    $E_{uu,W}$
    \end{turn}
    \\[2.7cm]
    \subfigure{\includegraphics[width=0.5\columnwidth]
    {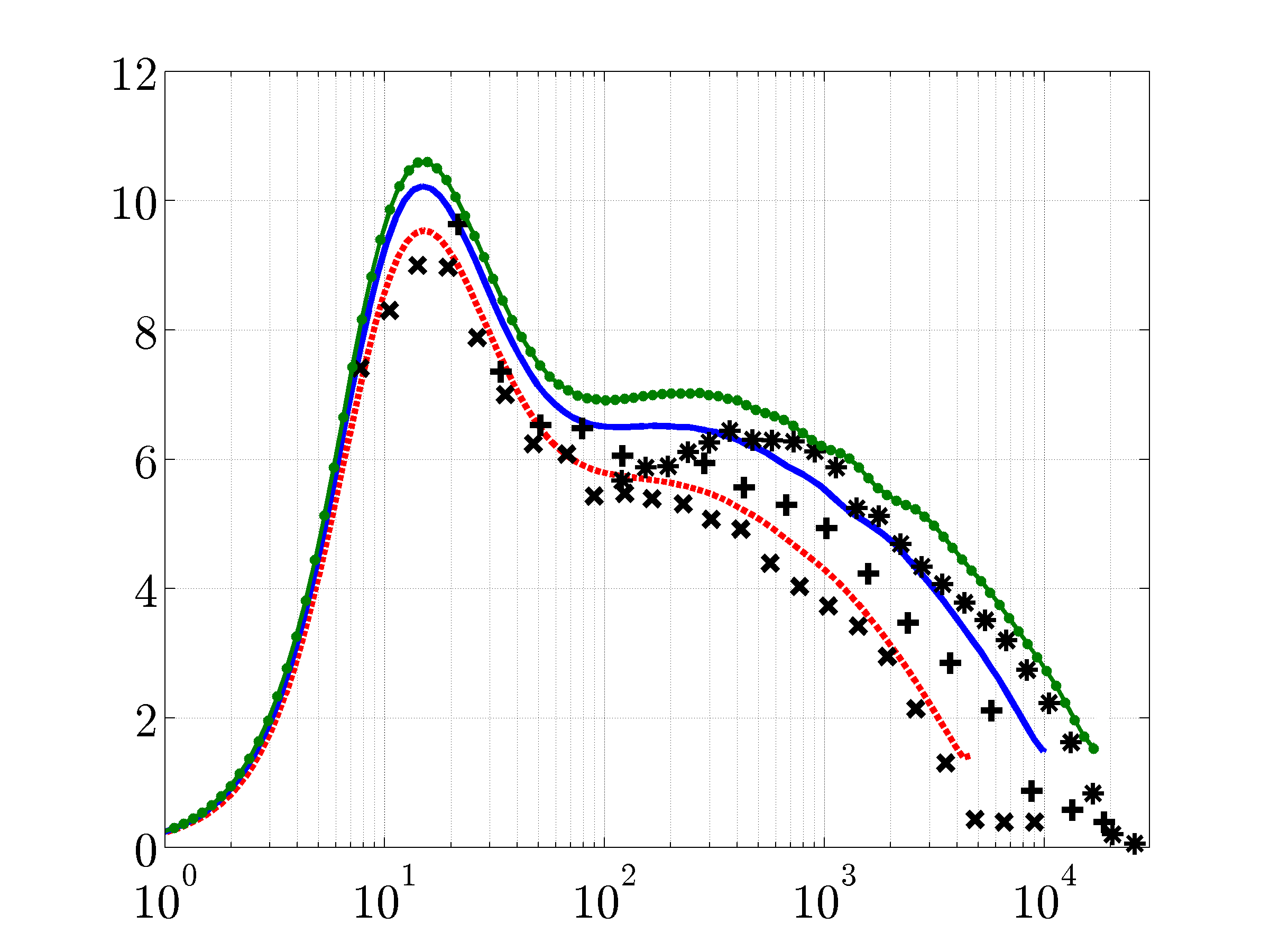}
    \label{fig.Euu-vs-yp-match-F-Up-R5813x-13490+-23013star-Exp-data-110912-color}}
    &
    \hskip-0.4cm
    \subfigure{\includegraphics[width=0.5\columnwidth]
    {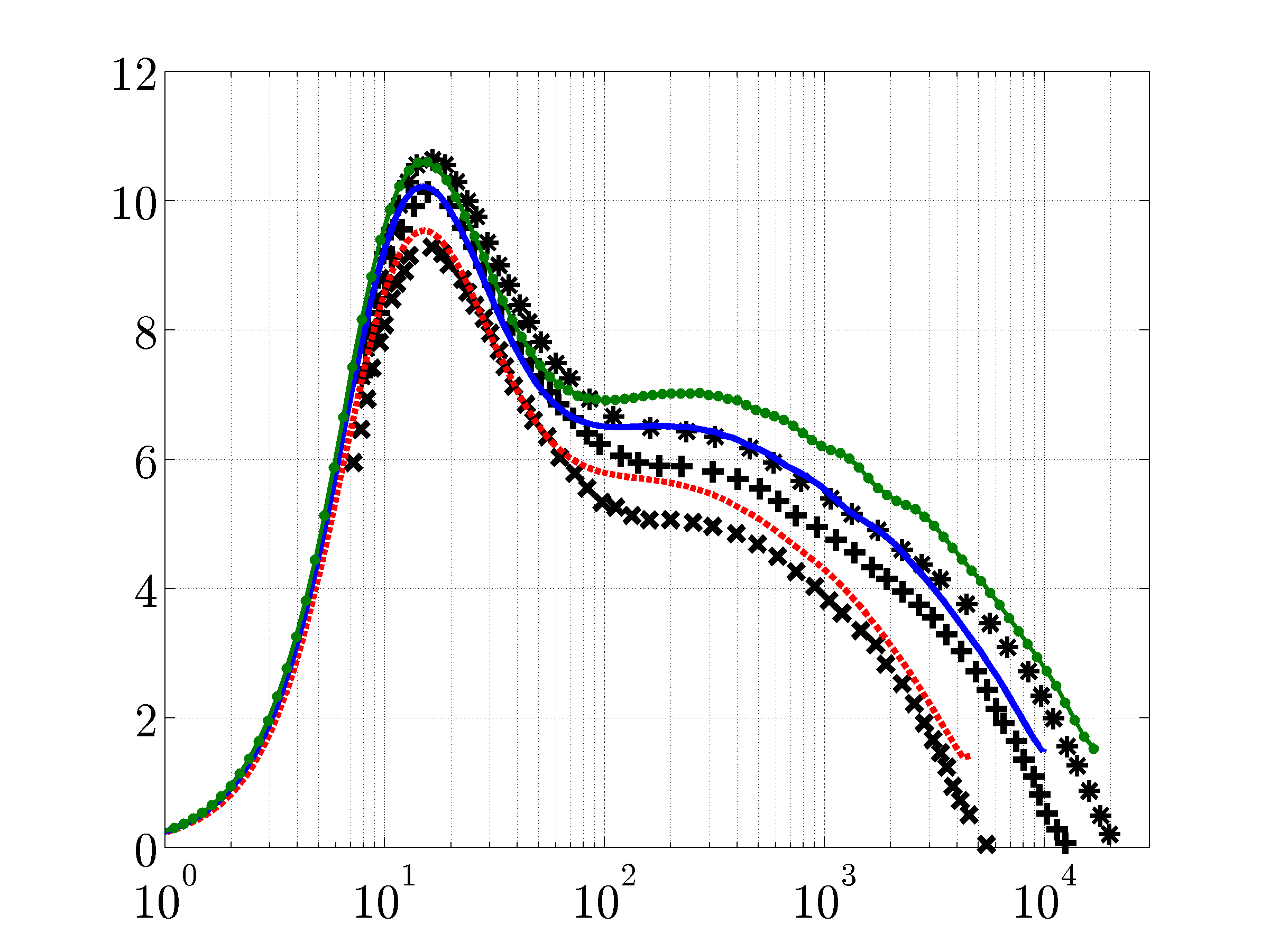}
    \label{fig.Euu-vs-yp-match-F-Up-R5813x-13490+-23013star-Exp-cor-110912-color}}
    \\[-0.2cm]
    \hskip0.1cm
    \tc{black}{$y^+$}
    &
    \hskip-0.3cm
    \tc{black}{$y^+$}
    \\[0.1cm]
    \hskip0.1cm
    $(e)$
    &
    \hskip-0.3cm
    $(f)$
    \\[-3.9cm]
    \hskip-6.1cm
    \hskip0.1cm
    \begin{turn}{90}
    $E_{uu,W}$
    \end{turn}
    &
    \hskip-6.1cm
    \hskip-0.3cm
    \begin{turn}{90}
    $E_{uu,W}$
    \end{turn}
    \\[2.7cm]
    \end{tabular}
    \end{center}
    \caption{\moa{
    (Color online) (a)-(b): The optimal weight functions $\overline{W}$ obtained using the similarity law~(\ref{eq.F-sim}) as a function of ${c}$ in (a) and $U_{cl} - {c}$ in (b).
    (c)-(f): The model-based streamwise energy intensity $E_{uu,\mathrm{W}} (y^+)$ for ${Re}_\tau = 934$ (orange), $2003$ (green), $3333$ (blue), $10000$ (red), and $30000$ (black) in (c); ${Re}_\tau = 3165$ (green dotted), $4000$ (blue solid), and $5813$ (red dashed) in (d); and ${Re}_\tau = 5813$ (red dashed), $13490$ (blue solid), and $23013$ (green dotted) in (e)-(f).
    The arrows in (a)-(c) indicate increase in the Reynolds number and the line in (c) shows logarithmic scaling.
    The symbols in (d) are experimental data from channel flows for ${Re}_\tau = 3165$; $\diamond$ and $4000$; $\triangledown$~\citep{monty-thesis-05}, and for ${Re}_\tau = 4000$; $\circ$ and $6000$; $\times$~\citep{schfla13}.
    The symbols in (e) are experimental data from boundary layers for ${Re}_\tau = 5813$; $\times$, $13490$; $+$~\citep{degeat00}, and $23013$; $\star$~\citep{ferkranocsch95}. The symbols in (f) are the corrected~\citep{kunmar06} data in the inner region of (e) using attached eddy hypothesis.}
    }
    \label{fig.F-R934-2003-joint}
    \end{figure}

\subsection{Predictions at high Reynolds numbers}
\label{sec.predict}

The similarity law in~(\ref{eq.L})-(\ref{eq.F-sim}) is used to predict the weight functions, and consequently, the streamwise energy intensity at high ${Re}_\tau$ using~(\ref{eq.Euu-W}). Figures~\ref{fig.F-vs-Up-R934c-2003g-3333b-10000r-30000k}-\ref{fig.Euu-vs-yp-match-F-Up-R934c-2003g-3333b-10000r-30000k-110912-line} show the predicted weights and energy intensities for ${Re}_\tau = 934$, $2003$, $3333$, $10000$, and $30000$. An approximately logarithmic dependence of the energy intensity on the distance from the wall is predicted at high Reynolds numbers in the logarithmic region of the mean velocity which is consistent with recent experiments~\citep{hulvalbaismi12,marmonhulsmi13} and predictions of the attached eddy hypothesis~\cite[see, for example,][]{percho82}. \moa{As evident from figure~\ref{fig.Euu-vs-yp-match-F-Up-R3165-4000-5813-3165d-4000v-4000o-6000x-Exp-data-110912-color}, the model-based predictions are consistent with the experiments of channel flows for ${Re}_\tau = 3165$, $4000$, and $6000$; especially note the comparison at $Re_\tau = 4000$ with the data of~\cite{schfla13} that maintains a sufficient spatial resolution down to the wall-normal location of the inner peak. Note that the data of~\cite{monty-thesis-05} and the data of~\cite{schfla13} at $Re_\tau = 6000$ are not fully spatially resolved near the wall. In the absence of channel flow data at higher Reynolds numbers, figure~\ref{fig.Euu-vs-yp-match-F-Up-R5813x-13490+-23013star-Exp-data-110912-color} compares the model-based streamwise intensities with the data from boundary layer experiments for ${Re}_\tau = 5813$ and $13490$~\citep{degeat00} and $Re_\tau = 23013$~\citep{ferkranocsch95}.~\cite{monhutngmarcho09} showed that the behavior of boundary layers, pipes, and channels is similar in the near-wall region in spite of the differences between channels/pipes and boundary layers further away from the wall.} The experimental measurements are not accurate near the wall as they suffer from spatial resolution issues~\cite[see, for example,][]{hutnicmarcho09}. Figure~\ref{fig.Euu-vs-yp-match-F-Up-R5813x-13490+-23013star-Exp-cor-110912-color} shows that the predicted values of the inner peak are consistent with the boundary layer measurements that are corrected~\citep{kunmar06} based on the attached eddy hypothesis. On the other hand, our predictions of the energy intensity in channel flow are larger than the data in boundary layers close to the outer peak in the middle region. This is expected since recent experiments have shown that the large structures are more energetic in internal flows such as channels and pipes compared to boundary layers~\cite[see, for example,][]{monhutngmarcho09}. This difference was attributed to the observations suggesting that the VLSMs are longer in internal flows than in boundary layers.

Obtaining the results of figure~\ref{fig.F-R934-2003-joint} requires computation of the streamwise energy density of the rank-1 model with broadband forcing at the respective values of ${Re}_\tau$. Alternatively, the universal behavior of $E_{uu} (y; \kappa_x, \kappa_z, c)$ can be used to avoid these computations. In the present study, we employ the universality (invariance with ${Re}_\tau$) of $E_{uu} (y; \kappa_x, \kappa_z, c)$ for $\cS_{i}$ to predict the inner peak of the streamwise intensity at arbitrary high ${Re}_\tau$. Expanding the weighted energy density according to the wave speed and substituting for the weight function using the similarity law~(\ref{eq.F-sim}) yields
	  \be
	  \begin{tabular}{rcl}
        \hskip-0.2cm
        $
        E_{uu,W} (y)
        $
        &
        $
        \!\! = \!\!
        $
        &
        $
        \ds{\int_{2}^{16}}
        ~
        \underset{\mbox{{\footnotesize
        \begin{tabular}{c}
        universal
        \end{tabular}
        }}}{\underbrace{W_2 (c) \big( {Re}_\tau^{-2} E_{uu} (y, c) \big)}}
        \,
        \bigg(
        1
        \, + \,
        \dfrac{L (c; {Re}_\tau) - L_0 (c)}{W_2 (c)}
        \bigg)
        \,
        \mrd c
        \, + \,
        $
        \\[0.2cm]
        &&
        $
        \ds{\int_{16}^{U_{cl} - 3.9}}
        L(c;{Re}_\tau) \, \big( {Re}_\tau^{-2} E_{uu} (y, c) \big)
        \, \mrd c
        \, + \,
        \ds{\int_{U_{cl} - 3.9}^{U_{cl}}}
        W_1(c) \, \big( {Re}_\tau^{-2} E_{uu} (y, c) \big)
        \, \mrd c.
        $
        \end{tabular}
        \label{eq.W-expand}
        \ee
The first integral, corresponding to the inner class of wave parameters $\cS_{i}$, contains a universal function multiplied by a coefficient $L(c; {Re}_\tau)$ that also appears in the second integral for the faster and larger waves in the \moar{self-similar} region. It represents the contribution from the inner class of wave parameters that are coupled with and amplified by the large scales in the \moar{self-similar} region. This is similar to the model that~\cite{marmathut10-science} proposed to capture the influence of the large scales $u_{L}$ (close to the geometric mean of the middle region of $U$) on the small scales $u_{S}$ close to the inner peak of the energy intensity
        \be
        u_{S}
        \, = \,
        u^{*} (1+\beta \, u_{L})
        \, + \,
        \alpha \, u_{L}.
        \label{eq.marusic}
        \ee
For the purpose of the present study,~(\ref{eq.marusic}) implies that the small structures are determined by a universal inner-scaled function $u^*$ multiplied by a coefficient $1+\beta \, u_{L}$ that increases with the energy of the large structures. Physically, the first term in~(\ref{eq.marusic}) describes the amplitude modulation of small scales by the large scales and the second term represents the direct superimposition of the large scales on the inner-scaled near-wall peak~\citep{marmathut10-science}.

    \begin{center}
    \begin{figure}
    \begin{tabular}{c}
    \begin{tabular}{cc}
    \hskip0.1cm
    \subfigure{\includegraphics[width=0.5\columnwidth]
    {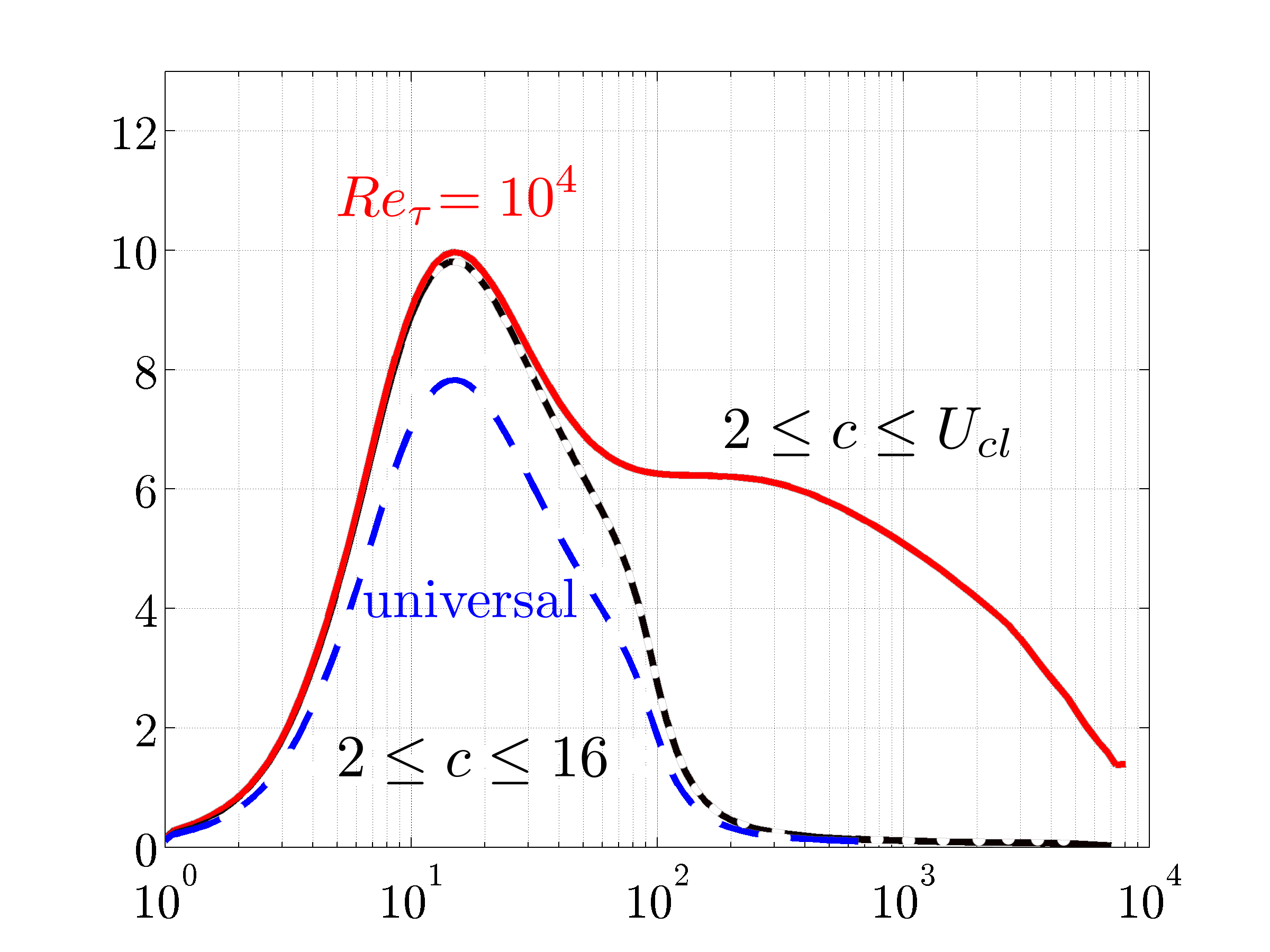}
    \label{fig.Euu-vs-yp-match-F-Up-R10000r-UcLess16-universal2k-110912}}
    &
    \hskip-0.3cm
    \subfigure{\includegraphics[width=0.5\columnwidth]
    {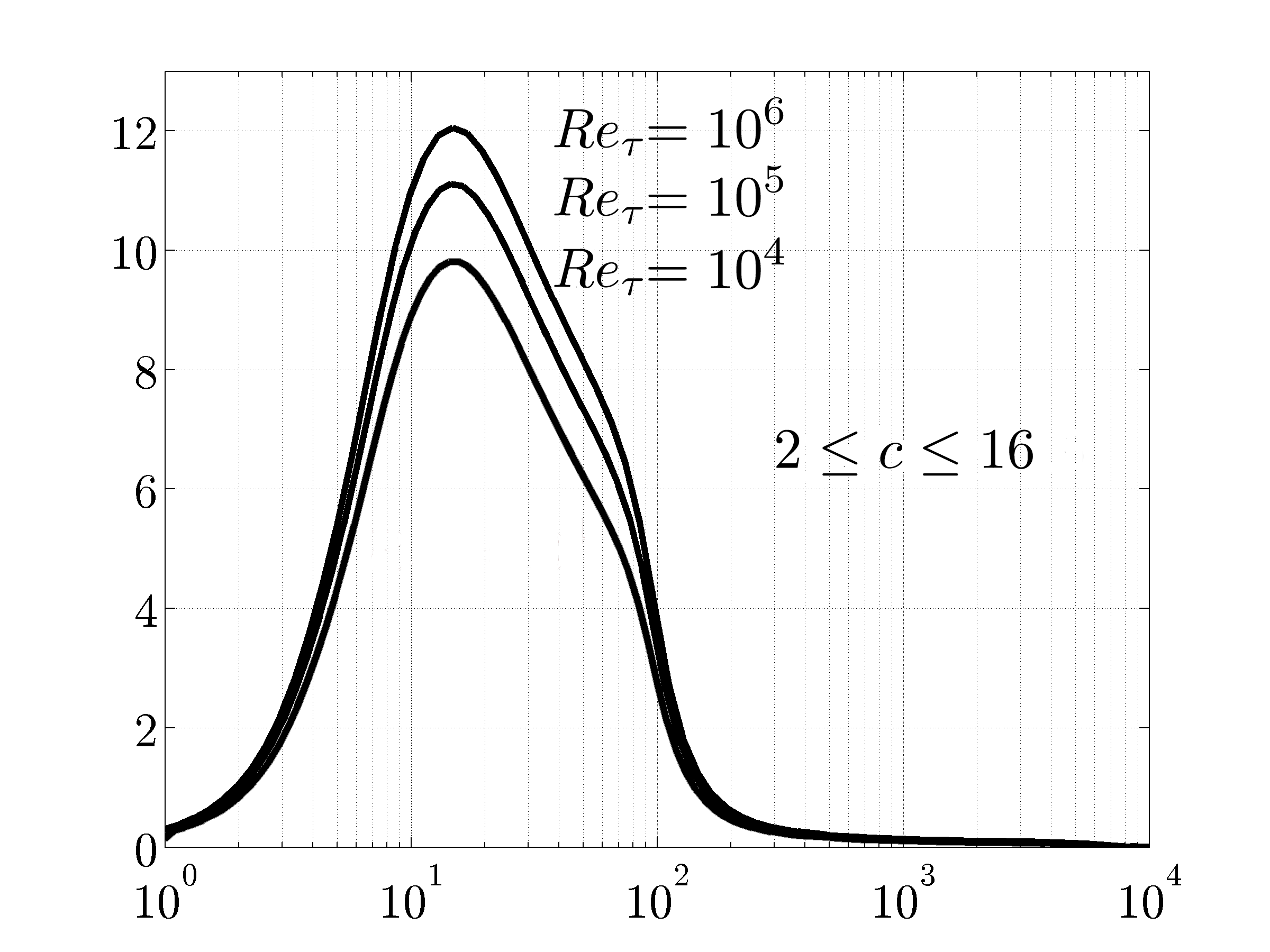}
    \label{fig.Euu-vs-yp-match-F-Up-R10000r-R1e5k-1e6k-UcLess16-110912}}
    \\[-0.2cm]
    \hskip0.1cm
    $y^+$
    &
    \hskip-0.3cm
    $y^+$
    \\[0.1cm]
    \hskip0.1cm
    (a)
    &
    \hskip-0.3cm
    (b)
    \end{tabular}
    \end{tabular}
    \caption{(Color online) (a) The red solid curve is the model-based energy intensity $E_{uu,W} (y)$ for ${Re}_\tau = 10^4$. The blue dashed curve is the contribution from the universal function in~(\ref{eq.W-expand}), and the black dotted curve is the contribution from the inner class of wave parameters $\cS_{i}$. (b) The contribution to the energy intensity from the inner class of wave parameters for ${Re}_\tau = 10^4$, $10^5$, and $10^6$.}
    \label{fig.inner-peak-predict}
    \end{figure}
    \end{center}

The blue dashed curve in figure~\ref{fig.Euu-vs-yp-match-F-Up-R10000r-UcLess16-universal2k-110912} shows the contribution of the universal function in~(\ref{eq.W-expand}) to the energy intensity. This is equal to the contribution of the inner class of wave parameters to the energy intensity for ${Re}_\tau = 934$, i.e. for $L = L_0$. In other words, the inner class of wave parameters is not influenced by the large scales in the middle region for ${Re}_\tau = 934$. \moa{This is expected since at $Re_\tau \approx 1000$, the inner and outer scales are separated, in the (temporal) frequency domain, by the wave speed $c = 16$: i.e. inner scales for $c < 16$ and outer scales for $c > U_{cl} - 6.15 \approx 16$. Therefore, ${Re}_\tau \approx 1000$ is the smallest Reynolds number where the purely inner and outer scales are separated in the wavenumber-frequency domain. Notice that the above-mentioned scale separation in the frequency domain does not contradict the weak scale separation in the premultiplied spectra at $Re_\tau \approx 1000$. The latter is a consequence of time-averaging that overlays the separated scales in the frequency domain such that the distinction of different scales in the spatial spectra becomes difficult.}

The black dotted curve shows the contribution of the first integral in~(\ref{eq.W-expand}) to the streamwise intensity for ${Re}_\tau = 10^4$. Notice that the large scales from the \moar{self-similar} region increase the inner peak by amplifying the universal function through the coefficient $1 + (L - L_0)/W_2$. The red solid curve is the total intensity obtained by integrating the contribution of all wave parameters $\cS_{e}$. The inner peak is captured by the first integral and the direct superimposition of the large scales on the inner peak is negligible. Therefore, the first integral readily yields the behavior of the streamwise intensity near the inner peak. For example, figure~\ref{fig.Euu-vs-yp-match-F-Up-R10000r-R1e5k-1e6k-UcLess16-110912} illustrates how the more energetic large scales at ${Re}_\tau = 10^5$ and $10^6$ further increase the inner peak relative to ${Re}_\tau = 10^4$ by amplifying the universal function $W_2$.

    \begin{figure}
    \begin{center}
    \includegraphics[width=0.95\columnwidth]
    {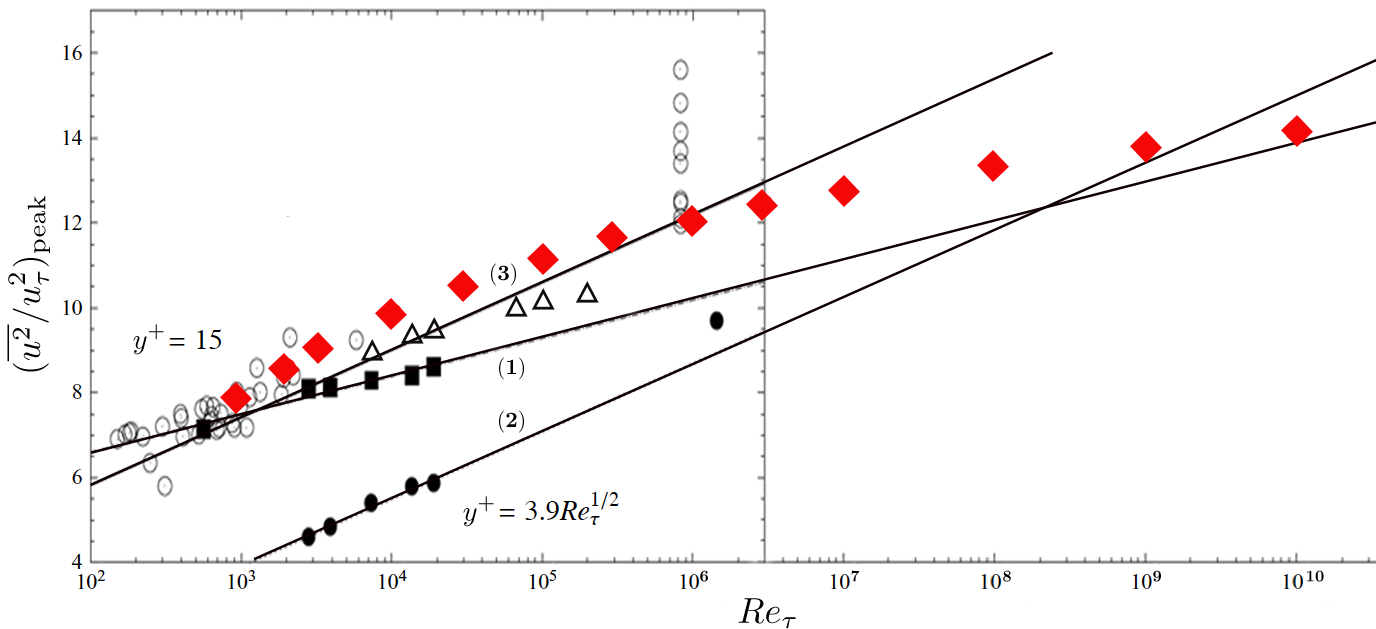}
    \end{center}
    \caption{(Color online) Variation of the inner ($y^+ = 15$) and outer ($y^+ = 3.9 {Re}_\tau^{1/2}$) peaks of the streamwise energy intensity with Reynolds number. The figure is adapted from~\cite{marmathut10}. The black open symbols are experimental and simulation data from channels and boundary layers, see~\cite{hutmar07-JFM} for a full list of references. The open triangles are from large-eddy simulations of boundary layers~\citep{inomatmarpul12}. The diamonds are the predicted inner peak intensities obtained from the present model for turbulent channels.}
    \label{fig.inner-outer-peak-vs-Rtau}
    \end{figure}

Figure~\ref{fig.inner-outer-peak-vs-Rtau} is adapted from figure~8 in~\cite{marmathut10} where the DNS and experimental data from channels and boundary layers are summarized (open and filled black symbols). The black filled squares and circles, respectively, show the magnitude of the inner ($y^+ = 15$) and outer ($y^+ = 3.9 {Re}_\tau^{1/2}$) peaks in recent boundary layer experiments~\citep{marmathut10}. Using these data, the authors proposed two possibilities for the behavior of the inner peak at high Reynolds numbers. The first possibility is to extrapolate following the trend suggested by the filled black squares (line 1). The second possibility, motivated by the fact that the large scales increase the energy of the small scales, is to extrapolate following line 3 which is parallel to line 2 that captures the variation of the outer peak with ${Re}_\tau$. The data (open triangles) from large-eddy simulations of boundary layers~\citep{inomatmarpul12} combined with the wall-model of~\cite{marmathut10-science} are shown for comparison. \moa{The current understanding, at least for relatively small intervals of Reynolds numbers, suggests logarithmic growth of the inner peak. However, due to lack of sufficient spatial resolution close to the wall, the available experimental data conducted for different ranges of $Re_\tau$, predict different rates for the logarithmic growth, e.g. see~\cite{marmathut10}. Therefore, the available data is not sufficient for predicting the exact behavior of the inner peak as ${Re}_\tau$ increases.}

The diamonds in figure~\ref{fig.inner-outer-peak-vs-Rtau} show the model-based prediction of the inner peak of the streamwise intensity up to ${Re}_\tau = 10^{10}$. These predictions are made at no additional cost using the universal energy density for the inner class of wave parameters and the similarity law for the weight functions. These results are obtained for channels and are potentially different than boundary layers. \moa{In spite of an approximately logarithmic growth of the predicted inner peak up to $Re_\tau \sim 10^6$, a sub-logarithmic behavior becomes evident when seven decades of $Re_\tau$ are considered. As shown in equation~(\ref{eq.W-expand}), the linear part of the weight function, modeling the influence of large outer-scaled modes on the small inner-scaled modes, affects the growth of the inner peak with $Re_\tau$. The sub-logarithmic growth of the predicted inner peak can be attributed to the decrease in the slope of $L(c)$ as $Re_\tau$ increases, cf. equation~(\ref{eq.L}) and figure~\ref{fig.F-vs-Up-R934-2003-lines-area-F2-110912}. Understanding the Reynolds number dependence of $L(c)$ is an essential part of our ongoing research which is focused on analysis of the self-similar modes in the logarithmic region.}

\section{Concluding remarks}
\label{sec.conclusion}

Starting from the NSE, we highlighted the low-rank nature of the resolvent, formulated for individual wall-parallel wavenumbers and frequencies, and illustrated its power by showing that the most energetic motions of real turbulent flows correspond to wavenumbers and frequencies whose resolvent is approximately rank-1 (in the wall-normal direction). Motivated by this observation, we studied the streamwise energy density of the rank-1 model subject to forcings in the wall-parallel directions and time that were broadband and optimized, or ``trained'', with respect to the available DNS data.

\bjm{Our analysis consists of two steps: firstly identifying the modes that are highly amplified by the NSE and their scaling (essentially an analysis of the resolvent operator), and then calculating weighting functions (by matching to DNS results) which determine which of these modes will be sustained in the real flow (connecting the linear system of resolvent operators back to the full NSE).}

\moa{
\subsection{Scaling of the most amplified resolvent modes}
\label{sec.modes}
}

It was shown that the resolvent admits three classes of wavenumbers and wave speeds where the corresponding principal singular values and singular functions exhibit universal behavior with Reynolds number. These classes are directly related to the universal regions of the turbulent mean velocity (which is assumed known \emph{a priori}) and thus are primarily distinguished by the wave speed: (i) a truly inner-scaled class of waves with constant speeds in the inner region of the turbulent mean velocity; (ii) a class of waves with outer-scaled height and width and constant defect speeds relative to the centerline; and (iii) a class of waves with outer-scaled length and constant defect speeds relative to geometric mean of the middle region of the turbulent mean velocity. 
\moa{In addition, we showed that hierarchies of geometrically self-similar modes whose length and width respectively scale quadratically and linearly with their height are admitted by the resolvent in the presence of a logarithmic mean velocity.}

\moa{The integral role of wave speed and critical layers in characterizing the classes of universal modes with Reynolds number and the geometrically self-similar modes with the wall-normal distance is understood and emphasized for the first time. The conventional understanding about the scales of turbulent flows comes from the time-averaged velocity spectra in DNS and experiments. Upon integration in time, the separated scales in the (temporal) frequency domain are overlaid, and distinction of different scales in the spatial spectra becomes difficult. Therefore, the identified scales have significant implications for understanding the scaling of wall turbulence. They are inherent features of the linear mechanisms in the NSE and, consequently, the energy extraction mechanisms from the mean velocity. In both the universal and self-similar classes, the wall-normal length scale is inherited from the turbulent mean velocity, and the wall-parallel length scales are determined from the balance between the viscous dissipation term, $(1/Re_\tau) \Delta$, and the mean advection terms in the resolvent, e.g. $\mri \kappa_x (U - c)$.}

\moa{The main results of the present paper, i.e. the identified scalings in~\S~\ref{sec.Re-scaling-theory} rely on the accepted scales of the turbulent mean velocity and, otherwise, do not depend on the exact shape of $U$. Therefore, the choice of eddy viscosity or the von K{\'a}rm{\'a}n's constant $\kappa$ does not change our main results. On the other hand, the debate on the universality and/or exact value of $\kappa$ is ongoing, e.g. see~\cite{nagcha08} and using the turbulent viscosity given in~(\ref{eq.U-nuT}) can result in inaccuracies in the considered mean velocity. This can affect the quantitative results of~\S~\ref{sec.Re-scaling-theory} and~\S~\ref{sec.energy-density-weighted}, e.g. the shape of the resolvent modes and the predicted growth rate of the inner peak. Characterizing these effects is a topic of future work\bjm{, and the sensitivity is known to be highest in the region of highest shear, close to the wall.} Since closing the feedback loop in figure~\ref{fig.block-diag} eventually generates the exact turbulent mean velocity, we do not over-emphasize the quantitative results of the present study.}

\moa{We highlight the uniqueness of the identified scales meaning that there are no other scales that result in universal or geometrically self-similar principal resolvent modes. In addition, the difference between the scalings of the resolvent modes and real turbulent flows implies the need for distinguishing the resolvent modes from the weighted modes that represent the real turbulent flow. For example, the scaling admitted by the self-similar resolvent modes yields $\lambda_z^3 \sim \lambda_x y_c$ which is different from the trend $\lambda_z^2 \sim \lambda_x y$ observed in the DNS-based two-dimensional streamwise spectrum~\citep{jimhoyJFM08}. Understanding the scaling differences between the resolvent modes and the weighted modes requires detailed scrutiny of the weights and the nonlinear effects, a topic of ongoing research. In addition, our results suggest that, owing to scale separation in frequency, there is a large benefit to obtaining \bjm{and analyzing the scaling of} three-dimensional time-resolved spectral measurements.}
	
\moa{
\subsection{Effect of nonlinearity}
\label{sec.linear-nonlinear}
}

\moa{From a systems theory point of view, the nonlinear terms wrap a feedback loop around the linear sub-systems in the NSE and redistribute the energy. They determine the wall-normal shape and the magnitude/phase of the driving force for an individual mode. Therefore, the real flow is obtained by superposing the resolvent response modes that are weighted according to projection of the driving force on the resolvent forcing modes. 

We started by assuming that the nonlinear forcing is broadband in the wall-parallel directions and time and aligned in the principal resolvent forcing modes. It was shown that these simple assumptions can qualitatively produce different scaling regions of the streamwise energy spectra. Therefore, the proposed analysis effectively narrows down the scaling problem in wall-bounded turbulent flows to the problem of understanding the influence of nonlinearity on the inevitable scales that are admitted by the linear mechanisms\bjm{, i.e. determining which of those admitted modes will be required in real flows for the flow to be self-sustaining. A full description of the latter effects is beyond the scope of the present paper, but the subject of ongoing work.}}

\moa{A non-broadband forcing in time was accounted for by considering a weight function in the wave speed. We showed that ``training'' the \bjm{weights based on the} wave speed can result in streamwise energy intensities that quantitatively match DNS and experiments.} As the Reynolds number increases, the optimal weight functions increase for wave speeds in the inner region of the mean velocity. Representation of the optimal weights using similarity laws revealed that the amount of upward shift is linearly correlated with the weight function for wave speeds in the middle region of the mean velocity. In other words, the weight function increases with the energy intensity of the large scales and amplifies the universal inner-scaled energy density of the rank-1 model. Therefore, it implicitly captures the well-known coupling of small scales with the large scales and their subsequent amplification in real turbulent flows.  

\bjm{A consequence of the simplicity of the identified scaling in wavenumber-frequency domain is the success of the simple weighting based on convection velocity in post- and pre-dicting the variation of the streamwise velocity fluctuations with Reynolds number.} One of the main results of this study is that the rank-1 approximation, together with the optimal weight functions and the (well-known) mean velocity profile, is sufficient for predicting the streamwise energy intensity at high Reynolds numbers. 
\moa{Even though the weight function provides a rough intuition about the effect of nonlinearity, the explicit analysis of the nonlinear feedback on the velocity field remains a subject of ongoing research.}

\moa{
\subsection{Outlook of the present analysis as a predictive tool}
\label{sec.predictive}
}

\moa{The present study effectively divides the streamwise energy density of the rank-1 model with broadband forcing into inner- and outer-scaled universal regions with Reynolds number and a geometrically self-similar region with distance from the wall that bridges the gap between the inner and outer regions. This enables scaling of the streamwise energy density to arbitrary large Reynolds numbers. It was shown that the most energetic wave parameters and the corresponding scales roughly agree with the dominant near-wall motions in real turbulent flows. 

The identified self-similar resolvent modes facilitate analytical developments in the logarithmic region of the turbulent mean velocity and can result in significant simplifications in analysis of wall turbulence. In addition, the wall-normal locality of the self-similar modes in a given hierarchy suggests that the linear sub-systems in the NSE impose a direct correspondence between wall-parallel scales and wall-normal locations in the logarithmic region. In the classical cascade analogy, e.g. see the review paper by~\cite{jim12}, this is reminiscent of an inertial regime the study of which is a topic of ongoing research. Furthermore, ongoing research is focused on utilizing the identified scalings to better understand the structure and evolution of the hypothesized attached eddies.}

\moa{The available predictive models of wall turbulence, e.g. the attached eddy hypothesis~\citep{tow76,percho82} and the model of~\cite{markun03}, rely on physical intuition that is gained from DNS and experiments. For example, the method proposed by~\cite{markun03} is based on an assumption about the influence of outer-layer modes on the near-wall modes (their equation (2)), where the underlying functions are determined by empirical curve fits to the experimental data (their equations (3)-(5)). The present model is more fundamental as it directly uses the NSE for decomposing the flow into classes of modes that are uniquely scaled with the Reynolds number and distance from the wall. Since the wall-normal shape of these modes is one of the model outputs, the contribution of the present work goes beyond reporting an empirical fit to the model-based data, \bjm{namely by exploring the scaling of the modes admitted by the NSE}.}
    
\bjm{In essence, this work supports the efficacy of the low-rank model of wall turbulence proposed by~\cite{mcksha10} by demonstrating that it can be used both to determine self-similar mode scalings and to obtain a low-rank representation of the streamwise intensity, given appropriate, self-similar weighting of the modes.} Our ongoing research, to be reported elsewhere, is focused on analytical expression of the streamwise energy density for wave speeds in the logarithmic region of the mean velocity and \emph{a priori} derivation of the weight functions. Addressing the limitations and implications of the low-rank model for predicting the wall-normal and spanwise energy spectra as well as the Reynolds stresses is another topic of future research. 

\section*{Acknowledgments}
    \label{sec.Ack}

The support of Air Force Office of Scientific Research under grants FA 9550-09-1-0701 (P.M. John Schmisseur) and FA 9550-12-1-0469 (P.M. Doug Smith) is gratefully acknowledged.

\appendix

\section{Derivation of the inner scalings}
\label{sec.inner-scales}

\moa{
We show that the transfer function $H$ admits universal behavior for the modes with speeds $c \lesssim 16$. For these modes, following equation~(\ref{eq.U-Coles}), the $y\/$-dependent coefficients in the transfer function $H$, are either independent of $Re_\tau$, e.g. $U (y^+) - c$, or scale with $Re_\tau$, e.g. $U' (y^+)$. This allows for scaling the height of the resolvent modes with the viscous unit $\nu/u_\tau$. In addition, the balance between the viscous dissipation term, $(1/Re_\tau) \Delta$, and the mean advection terms, e.g. $\mri \kappa_x (U - c)$, in the resolvent in~(\ref{eq.RA}) requires scaling of the wall-parallel wavelengths with the viscous unit $\nu/u_\tau$
} 
	\be
	\lambda_x^+
	\, = \,
	{Re}_\tau \lambda_x,
	~~
	y^+
	\, = \,
	{Re}_\tau y,
	~~
	\lambda_z^+
	\, = \,
	{Re}_\tau \lambda_z.
	\non
	\ee
The differential operators in $y$ and the wavenumber symbols in the inner coordinates are
	\be
	\p/\p_{y^+}
	\, = \,
	{Re}_\tau^{-1} \p/\p_y,
	~~
	\kappa_x^+
	\, = \,	
	{Re}_\tau^{-1} \kappa_x,
	~~
	\kappa_z^+
	\, = \,	
	{Re}_\tau^{-1} \kappa_z,
	~~
	\Delta^+
	\, = \,
	{Re}_\tau^{-2} \Delta.
	\non
	\ee
Consequently in the inner coordinates, the operators $R_A$, $C$, and $C^{\dagger}$ in~(\ref{eq.A-C}) and~(\ref{eq.RA}) scale as
	\be
	\ba{rcl}
	R_A
	&\!\! = \!\!&
	\tbt
	{{Re}_\tau X_1}
	{0}
	{{Re}_\tau^2 X_3}
	{{Re}_\tau X_2}^{-1}
	\, = \,
	\tbt
	{{Re}_\tau^{-1} Y_1}
	{0}
	{Y_3}
	{{Re}_\tau^{-1} Y_2},
	\\[0.3cm]
	C
       &\!\! = \!\!&
	\thbt
       {C_1}{{Re}_\tau^{-1} C_2}
       {C_3}{0}
       {C_4}{{Re}_\tau^{-1} C_5},
       ~~
       C^{\dagger}
	\; = \;
	\tbth
	{C_1^{\dagger}}
       {C_3^{\dagger}}
	{C_4^{\dagger}}
	{{Re}_\tau C_2^{\dagger}}
	{0}
	{{Re}_\tau C_5^{\dagger}}.
	\ea
	\label{eq.RA-C-xp-yp-zp}
	\ee
For given $\kappa_x^+$ and $\kappa_z^+$, the operators $C_1$ to $C_5$ and their adjoints are independent of ${Re}_\tau$. On the other hand, the operators $X_1$ to $X_3$ and $Y_1$ to $Y_3$ contain spatially-varying coefficients, $U - {c}$ and its first two derivatives, that depend on ${Re}_\tau$. As discussed in the beginning of~\S~\ref{sec.Re-scaling-theory}, $U$ scales with $y^+$ and is independent of ${Re}_\tau$ for $y^+ \lesssim 100$. Therefore, for given $\kappa_x^+$, $\kappa_z^+$, and ${c} \lesssim U (y^+ = 100) = 16$, the operators $X_1$ to $X_3$ and $Y_1$ to $Y_3$ are independent of ${Re}_\tau$ when acting on functions whose supports are inside the interval $0 < y^+ \lesssim 100$. Since the principal resolvent modes are localized around the critical layer (i.e. the wall-normal location where the turbulent mean velocity equals ${c}$), the resolvent modes are negligible outside $y^+ \lesssim 100$ for ${c} \lesssim 16$ and all of the aforementioned operators are effectively independent of ${Re}_\tau$. It follows from~(\ref{eq.RA-C-xp-yp-zp}) that
	\be
	H
	\, = \,
	C R_A C^{\dagger}
	\, = \,
	\thbth
	{{Re}_\tau^{-1} H_{11}}{{Re}_\tau^{-1} H_{12}}{{Re}_\tau^{-1} H_{13}}
	{{Re}_\tau^{-1} H_{21}}{{Re}_\tau^{-1} H_{22}}{{Re}_\tau^{-1} H_{23}}
	{{Re}_\tau^{-1} H_{31}}{{Re}_\tau^{-1} H_{32}}{{Re}_\tau^{-1} H_{33}},
	\non
	\ee
where the operators $H_{ij}$ are effectively independent of ${Re}_\tau$ when acting on their principal resolvent modes. Therefore, the principal singular value of $H$ is proportional to ${Re}_\tau^{-1}$. In addition, the orthonormality constraints~(\ref{eq.orthonormal}) on $\hat{\bpsi}_1$ and $\hat{\bphi}_1$ require that these functions scale as ${Re}_\tau^{1/2}$. This is because the supports of $\hat{\bpsi}_1$ and $\hat{\bphi}_1$ are independent of ${Re}_\tau$ in inner units (hence, proportional to ${Re}_\tau^{-1}$ in outer units). In other words, $\hat{\bpsi}_1 (y)$ and $\hat{\bphi}_1 (y)$ become thinner and taller as ${Re}_\tau$ increases. Finally, the streamwise energy density $E_{uu} = \kappa_x^2 \kappa_z \sigma_1^2 \abs{u_1}^2$ scales with
	\be
	\big({Re}_\tau\big)^2 \,
	\big({Re}_\tau\big) \,
	\big({Re}_\tau^{-1}\big)^2
	\big({Re}_\tau^{1/2}\big)^2
	\, = \,
	{Re}_\tau^2.
	\non
	\ee

\section{Derivation of the outer scalings}
\label{sec.outer-scales}

\moa{For the modes with defect speeds $0 \lesssim U_{cl} - c \lesssim 6.15$, we show that the transfer function $H$ admits universal behavior with Reynolds number. For these modes, following equation~(\ref{eq.U-Coles}), the $y\/$-dependent coefficients in the transfer function $H$, e.g. $U (y) - c$, are independent of $Re_\tau$. This allows for scaling the height of the resolvent modes with $h$. Furthermore, the balance between the viscous dissipation term, $(1/Re_\tau) \Delta$, and the mean advection terms, e.g. $\mri \kappa_x (U - c)$, in the resolvent in~(\ref{eq.RA}) requires scaling of the spanwise coordinate with $h$ and the streamwise coordinate with $h {Re}_\tau$.} Therefore, the streamwise wavenumber symbol in the outer coordinates is given by $\kappa_x^- = {Re}_\tau \kappa_x$. The Laplacian
	\be
	~~
	\Delta
	\, = \,
	\p_{yy}
	\, - \,
	{Re}_\tau^{-2} (\kappa_x^-)^2
	\, - \,
	\kappa_z^2,
	\non
	\ee
is independent of ${Re}_\tau$ if $\kappa_z^2$ dominates ${Re}_\tau^{-2} (\kappa_x^-)^2$ for all values of ${Re}_\tau$. For fixed $\kappa_x^-$ and $\kappa_z$, it suffices that
	\be
	\dfrac{\lambda_x^-}{\lambda_z}
	\, = \,
	\dfrac{\kappa_z}{\kappa_x^-}
	\gtrsim
	\dfrac{\gamma}{{Re}_{\tau,\mathrm{min}}}.
	\label{eq.outer-in-y-const}
	\ee
In the outer coordinates, the operators $R_A$, $C$, and $C^{\dagger}$ in~(\ref{eq.A-C}) and~(\ref{eq.RA}) scale as
	\be
	\ba{rcl}
	R_A
	&\!\! \approx \!\!&
	\tbt
	{{Re}_\tau^{-1} \tilde{X}_1}
	{0}
	{\tilde{X}_3}
	{{Re}_\tau^{-1} \tilde{X}_2}^{-1}
	\, = \,
	\tbt
	{{Re}_\tau \tilde{Y}_1}
	{0}
	{{Re}_\tau^2 \tilde{Y}_3}
	{{Re}_\tau \tilde{Y}_2},
	\\[0.3cm]
	C
       &\!\! \approx \!\!&
	\thbt
       {{Re}_\tau^{-1} \tilde{C}_1}{\tilde{C}_2}
       {\tilde{C}_3}{0}
       {\tilde{C}_4}{{Re}_\tau^{-1} \tilde{C}_5},
       ~~
       C^{\dagger}
	\; \approx \;
	\tbth
	{{Re}_\tau^{-1} \tilde{C}_1^{\dagger}}
       {\tilde{C}_3^{\dagger}}
	{\tilde{C}_4^{\dagger}}
	{\tilde{C}_2^{\dagger}}
	{0}
	{{Re}_\tau^{-1} \tilde{C}_5^{\dagger}}.
	\ea
	\label{eq.RA-C-xm-y-z}
	\ee
For given $\kappa_x^-$ and $\kappa_z$ that satisfy the constraint~(\ref{eq.outer-in-y-const}), the operators $\tilde{C}_1$ to $\tilde{C}_5$ and their adjoints are approximately independent of ${Re}_\tau$. In addition, the defect velocity $U_{cl} - U (y)$ is independent of ${Re}_\tau$ for $y \gtrsim 0.1$. Therefore, for given $\kappa_x^-$, $\kappa_z$, and $U_{cl} - {c} \lesssim U_{cl} - U (y = 0.1) = 6.15$, the operators $\tilde{X}_1$ to $\tilde{X}_3$ and $\tilde{Y}_1$ to $\tilde{Y}_3$ are approximately independent of ${Re}_\tau$ when acting on functions whose supports are inside the interval $0.1 \lesssim y \lesssim 1$. From~(\ref{eq.RA-C-xm-y-z}), we have
	\be
	H
	\, = \,
	C R_A C^{\dagger}
	\, \approx \,
	\thbth
	{{Re}_\tau \tilde{H}_{11}}{{Re}_\tau^{2} \tilde{H}_{12}}{{Re}_\tau^{2} \tilde{H}_{13}}
	{\tilde{H}_{21}}{{Re}_\tau \tilde{H}_{22}}{{Re}_\tau \tilde{H}_{23}}
	{\tilde{H}_{31}}{{Re}_\tau \tilde{H}_{32}}{{Re}_\tau \tilde{H}_{33}}.
	\label{eq.H-outer-in-y}
	\ee
Owing to the locality of the principal resolvent modes around the critical layer, the operators $\tilde{H}_{ij}$ are approximately independent of ${Re}_\tau$ when acting on their principal resolvent modes. Therefore, the principal singular value of $H$ is proportional to ${Re}_\tau^{2}$. Since $\hat{\bpsi}_1$ and $\hat{\bphi}_1$ scale in the outer length scale, the orthonormality constraints~(\ref{eq.orthonormal}) require that these functions be independent of ${Re}_\tau$. Finally, the streamwise energy density $E_{uu} = \kappa_x^2 \kappa_z \sigma_1^2 \abs{u_1}^2$ scales with
	\be
	\big({Re}_\tau^{-1}\big)^2 \,
	\big(1\big) \,
	\big({Re}_\tau^{2}\big)^2
	\big(1\big)^2
	\, = \,
	{Re}_\tau^2.
	\non
	\ee

\moa{
\section{Derivation of the geometrically self-similar scalings}
\label{sec.self-similar-log-scales}
}

\moa{
The transfer function $H$ admits geometrically self-similar modes with speeds in the logarithmic region of the turbulent mean velocity. In this region, it follows from the discussion in~\S~\ref{sec.Req-self-similar-log} that the $y\/$-dependent coefficient in the transfer function $H$ can be expressed as $U (y) - c = (1/\kappa) \log (y/y_c)$, where $y_c$ is the critical wall-normal location corresponding to $c$, i.e. $c = U(y_c)$. Similarly, $U'$ and $U''$ are functions of $y/y_c$. This allows for scaling the height of the resolvent modes with $y_c$. Furthermore, the balance between the viscous dissipation term, $(1/Re_\tau) \Delta$, and the mean advection terms, e.g. $\mri \kappa_x (U - c)$, in the resolvent in~(\ref{eq.RA}) requires scaling of the spanwise wavelength with $y_c$ and the streamwise wavelength with $y_c^+ y_c$,
	\be
	\bar{\lambda}_x
	\, = \,
	\lambda_x/(y_c^+ y_c),
	~~
	\bar{y}
	\, = \,
	y/y_c,
	~~
	\bar{\lambda}_z
	\, = \,
	\lambda_z/y_c.
	\non
	\ee
The differential operators in $y$ and the wavenumber symbols in the $y_c\/$-scaled coordinates are
	\be
	\p/\p_{\bar{y}}
	\, = \,
	y_c (\p/\p_y),
	~~
	\bar{\kappa}_x
	\, = \,	
	(y_c^+ y_c) \kappa_x,
	~~
	\bar{\kappa}_z
	\, = \,	
	y_c \kappa_z.
	\non
	\ee
For given $\bar{\kappa}_x$ and $\bar{\kappa}_z$, the Laplacian
	\be
	\Delta
	\, = \,
	y_c^{-2} 
	\left(
	\p_{\bar{y} \bar{y}}
	\, - \,
	(y_c^+)^{-2} (\bar{\kappa}_x)^2
	\, - \,
	(\bar{\kappa}_z)^2
	\right),
	\non
	\ee
approximately scales with $y_c^{-2}$ if $(\bar{\kappa}_z)^2$ dominates $(y_c^+)^{-2} (\bar{\kappa}_x)^2$, i.e.
	\be
	\kappa_z/\kappa_x
	\, = \,
	\lambda_x/\lambda_z
	\, = \,
	y_c^+ (\bar{\lambda}_x/\bar{\lambda}_z)
	~\gtrsim~
	\gamma,
	\label{eq.log-region-const}
	\ee
where a conservative value for $\gamma$ is $\sqrt{10}$. Since the aspect ratio $\lambda_x/\lambda_z$ increases with $y_c^+$, the smallest value of $y_c^+$ for which~(\ref{eq.log-region-const}) is guaranteed is equal to $y_{c_1}^+ = \gamma (\bar{\lambda}_{z}/\bar{\lambda}_{x})$. Therefore, the smallest wave speed that satisfies the aspect ratio constraint and lies above the inner region is given by
	\be
	c_1
	\; = \;
	\mbox{max} 
	\big(
	16
	,~
	B
	\, + \, 
	(1/\kappa) \log y_{c_1}^+
	\big).
	\label{eq.c1}
	\ee
Then, the operators $R_A$, $C$, and $C^{\dagger}$ in~(\ref{eq.A-C}) and~(\ref{eq.RA}) scale as
	\be
	\ba{rcl}
	R_A
	&\!\! = \!\!&
	\tbt
	{\big(y_c^+ y_c\big)^{-1} \bar{X}_1}
	{0}
	{y_c^{-2} \, \bar{X}_3}
	{\big(y_c^+ y_c\big)^{-1} \bar{X}_2}^{-1}
	\, = \,
	\tbt
	{\big(y_c^+ y_c\big) \bar{Y}_1}
	{0}
	{(y_c^+)^2 \, \bar{Y}_3}
	{\big(y_c^+ y_c\big) \bar{Y}_2},
	\\[0.4cm]
	C
       &\!\! = \!\!&
	\thbt
       {(1/y_c^+) \, \bar{C}_1}{(y_c) \, \bar{C}_2}
       {\bar{C}_3}{0}
       {\bar{C}_4}{(1/{Re}_\tau) \, \bar{C}_5},
       ~~
       C^{\dagger}
	\; = \;
	\tbth
	{(1/y_c^+) \, \bar{C}_1^{\dagger}}
       {\bar{C}_3^{\dagger}}
	{\bar{C}_4^{\dagger}} 
	{(1/y_c) \, \bar{C}_2^{\dagger}}
	{0}
	{\big(y_c^+ y_c\big)^{-1} \bar{C}_5^{\dagger}}.
	\ea
	\label{eq.RA-C-xh-yh-zh}
	\ee
For given $\bar{\kappa}_x$ and $\bar{\kappa}_z$ that satisfy the constraint~(\ref{eq.log-region-const}), the operators $\bar{C}_1$ to $\bar{C}_5$ and their adjoints are approximately independent of $y_c$ and ${Re}_\tau$. In addition, the operators $\bar{X}_1$ to $\bar{X}_3$ and $\bar{Y}_1$ to $\bar{Y}_3$ are approximately independent of $y_c$ and ${Re}_\tau$ when acting on functions whose supports are localized in the interval $100/{Re}_\tau \leq y\leq 0.1$. From~(\ref{eq.RA-C-xh-yh-zh}), we have
	\be
	H
	\, = \,
	C R_A C^{\dagger}
	\, = \,
	\thbth
	{\big(y_c^+ y_c\big) \bar{H}_{11}}{\big(y_c^+\big)^2 (y_c) \bar{H}_{12}}{\big(y_c^+\big)^2 (y_c) \bar{H}_{13}} 
	{(y_c) \bar{H}_{21}}{\big(y_c^+ y_c\big) \bar{H}_{22}}{\big(y_c^+ y_c\big) \bar{H}_{23}} 
	{(y_c) \bar{H}_{31}}{\big(y_c^+ y_c\big) \bar{H}_{32}}{\big(y_c^+ y_c\big) \bar{H}_{33}},
	\non
	\ee
where the operators $\bar{H}_{ij}$ are effectively independent of $y_c$ and ${Re}_\tau$ when acting on their principal resolvent modes. Therefore, the principal singular value of $H$ is proportional to $(y_c^+)^2 (y_c)$. In addition, the orthonormality constraints~(\ref{eq.orthonormal}) on $\hat{\bpsi}_1$ and $\hat{\bphi}_1$ require that these functions scale with $(y_c)^{-1/2}$. This is because the supports of $\hat{\bpsi}_1$ and $\hat{\bphi}_1$ expand with $y_c$. Finally, the streamwise energy density $E_{uu} = \kappa_x^2 \kappa_z \sigma_1^2 \abs{u_1}^2$ for the waves that belong to the same hierarchy scales with 
	\be
	\big(y_c^+ y_c\big)^{-2} \, 
	\big(y_c\big)^{-1} \,
	\big((y_c^+)^2 (y_c)\big)^2
	\big(y_c\big)^{-1}
	\, = \,
	{Re}_\tau^2.
	\non
	\ee
}

\bibliographystyle{jfm}
\bibliography{../bib/couette,../bib/mj-complete-bib,../bib/periodic,../bib/covariance,../bib/control-pde,../bib/ref-added-rm}

\end{document}

%% file: commands.tex
\newcommand{\beq}{\begin{equation}}
\newcommand{\eeq}{\end{equation}}
\newcommand{\bseq}{\begin{subequations}}
\newcommand{\eseq}{\end{subequations}}
\newcommand{\beqn}{\begin{eqnarray}}
\newcommand{\eeqn}{\end{eqnarray}}
\newcommand{\ba}{\begin{array}}
\newcommand{\ea}{\end{array}}
\newcommand{\bct}{\begin{center}}
\newcommand{\ect}{\end{center}}
\newcommand{\btmz}{\begin{itemize}}
\newcommand{\etmz}{\end{itemize}}
\newcommand{\benum}{\begin{enumerate}}
\newcommand{\eenum}{\end{enumerate}}

\newcommand{\norm}[1]{\| #1 \|}                 

\newcommand{\hv}{\hat{v}}

\newcommand{\hu}{\hat{u}}
\newcommand{\hw}{\hat{w}}

\newcommand{\matbegin}{
        \left[
}
\newcommand{\matend}{
        \right]
}

\newcommand{\tbt}[4]{
  \matbegin \begin{array}{cc}
       #1 & #2 \\ #3 & #4
       \end{array} \matend }
\newcommand{\thbt}[6]{
  \matbegin \begin{array}{cc}
       #1 & #2 \\ #3 & #4 \\ #5 & #6
       \end{array} \matend }
\newcommand{\tbth}[6]{
  \matbegin \begin{array}{ccc}
       #1 & #2 & #3\\ #4 & #5 & #6
       \end{array} \matend }
\newcommand{\thbth}[9]{
 \matbegin \begin{array}{ccc}
                #1 & #2 & #3 \\
                #4 & #5 & #6 \\
                #7 & #8 & #9
                \end{array}\matend}

\newcommand{\be}{\begin{equation}}
\newcommand{\ee}{\end{equation}}

\newcommand{\cplxs}{ C\kern -.35em \rule{0.03 em}{.7 ex}~   }

\def\complex{\hbox{C\kern -.45em \rule{0.03 em}{1.5 ex}}~}

\newcommand{\bi}{\begin{itemize}}
\newcommand{\ei}{\end{itemize}}